\documentclass[journal]{IEEEtran}

\usepackage{changepage}
\usepackage{svg}
\usepackage{array}
\usepackage{booktabs}
\usepackage{textcomp}
\usepackage{comment}
\usepackage{lineno}
\usepackage{multirow}
\def\textbf#1{{\bfseries #1}}
\usepackage{multicol}
\usepackage{graphicx}
\usepackage{epstopdf}
\usepackage[cmex10]{amsmath,mathtools}
\usepackage{bm}
\usepackage{mdwtab}
\usepackage{amssymb}
\usepackage{stfloats}
\usepackage{algorithmic}
\usepackage{algorithm}
\usepackage{float}
\usepackage{esint}
\usepackage{diagbox}
\usepackage{url}
\usepackage{empheq}
\usepackage{color}
\usepackage{cite}
\usepackage{xcolor}
\usepackage{verbatim}
\usepackage{pifont}
\usepackage{rotating}
\usepackage{hhline} % added by wyl
\usepackage{ifpdf}
\usepackage{tabularx,booktabs}
\usepackage{array}
\usepackage{tikz}
\usepackage{colortbl}
\usepackage{placeins}
\usepackage{hyperref} 
\usepackage{threeparttable}
\usepackage[T1]{fontenc}
\usepackage{tgtermes}
\newcommand*{\blue}{\color{blue}}

\definecolor{mygreen}{RGB}{32,178,170}

\newcommand*{\chat}{\color{black}}

\newcolumntype{P}[1]{>{\centering\arraybackslash}p{#1}}

\ifCLASSINFOpdf
\else
\fi
\title{
Failure Analysis in Next-Generation Critical Cellular Communication Infrastructures
}

\author{Siguo Bi, Xin Yuan,~\IEEEmembership{Senior Member,~IEEE,} Shuyan Hu,~\IEEEmembership{Member,~IEEE,} Kai Li,~\IEEEmembership{Senior Member,~IEEE,} \\
Wei Ni,~\IEEEmembership{Fellow,~IEEE,} 
Ekram Hossain,~\IEEEmembership{Fellow,~IEEE,}
and Xin Wang,~\IEEEmembership{Fellow,~IEEE} 
\IEEEcompsocitemizethanks{
\IEEEcompsocthanksitem S. Bi, S. Hu, and X. Wang are with Fudan University, Shanghai 200433, China (e-mail: \{fdbsg,\,syhu14,\,xwang11\}@fudan.edu.cn). 
\IEEEcompsocthanksitem X. Yuan and W. Ni are with Data61, Commonwealth Scientific and Industrial Research Organization (CSIRO), Sydney, NSW 2122, Australia, and the School of Computer Science and Engineering, University of New South Wales, Sydney, NSW 2052 (e-mail: \{xin.yuan, wei.ni\}@data61.csiro.au).
\IEEEcompsocthanksitem K.~Li is with the Department of Engineering, University of Cambridge, CB3 0FA Cambridge, U.K., and also with Real-Time and Embedded Computing Systems Research Centre (CISTER), Porto 4249--015, Portugal (e-mail: kaili@ieee.org).
\IEEEcompsocthanksitem E. Hossain is with the Department of Electrical and Computer Engineering, University of Manitoba, Canada (e-mail: ekram.hossain@umanitoba.ca).
}
}
\begin{document}

\IEEEcompsoctitleabstractindextext{%
\begin{abstract}
The advent of communication technologies marks a transformative phase in critical infrastructure construction, where the meticulous analysis of failures becomes paramount in achieving the fundamental objectives of continuity, security, and availability. This survey enriches the discourse on failures, failure analysis, and countermeasures in the context of the next-generation critical communication infrastructures. Through an exhaustive examination of existing literature, we discern and categorize prominent research orientations with focuses on, namely resource depletion, security vulnerabilities, and system availability concerns. 
We also analyze constructive countermeasures tailored to address identified failure scenarios and their prevention. 
Furthermore, the survey emphasizes the imperative for standardization in addressing failures related to Artificial Intelligence (AI) within the ambit of the sixth-generation (6G) networks, accounting for the forward-looking perspective for the envisioned intelligence of 6G network architecture. 
By identifying new challenges and delineating future research directions, this survey can help guide stakeholders toward unexplored territories, fostering innovation and resilience in critical communication infrastructure development and failure prevention. 

\end{abstract}

}

\maketitle

\IEEEdisplaynotcompsoctitleabstractindextext
\IEEEpeerreviewmaketitle

\section{Introduction}
\label{sec_introduction}
\begin{table*}
\label{table0}
\caption{List of acronyms and abbreviations.}
\begin{tabular*}{\linewidth}{p{2cm}p{5.5cm}p{2cm}p{5.5cm}}
\toprule
 3GPP                            & 3rd Generation Partnership Project &    &  \\
5G                    & The fifth-generation         &IEG& Independent Evaluation Group \\
 &  &IIoT& Industrial Internet-of-Things \\
5G-R                           & 5G Mobile Networks for Railway &IL& Incremental Learning \\
5G-R                           & 5G Mobile Networks for Railway &ILP& Integer Linear Programming \\
6G                    & The sixth-generation         &IoMT                   &Internet of Medical Things \\
AI                    & Artificial Intelligence      &IoT                      & Internet of Things  \\
AKA& Authentication and Key Agreement&IoV& Internet of Vehicle \\
AP & Access Point&IPFA&International Symposium on Physical and Failure Analysis of Integrated Circuits \\
ATIS& Alliance for Telecommunications Industry Solutions&IP                & Infrastructure Provider   \\
B5G                   & Beyond 5G  &   &   \\
 &  &IT                    & Information Technology \\
BPDC& Blockchain-based Privacy-aware Distributed Collection&ITU& International Telecommunications Union \\ 
CA& Certificate Authority&KC-FS & $k$-connected Function Slicing  \\
CAV& Connected and Autonomous Vehicle &KC-NS                 & $k$-connected Network Slicing \\
  &  &KC-SLG                            &$k$-Connected Service Function Slices Layered Graph  \\
CGC& Centralized Graph Coloring&KGCs& Key Generation Centers \\
C-H& Controller-Hypervisor  &                                &   \\
CNN& Convolutional Neural Network&   &     \\
CPS& Cyber-physical System&LiDAR& Light Detection and Ranging \\
CRAN& Cloud RAN&LLM& Large Language Model \\
 &     &LoS                           &Line-of-sight  \\
CRN&  Cognitive Radio Network&LSTM                              & Long Short-Term Memory \\ 
 &   &M2M&Machine-to-Machine \\
C-V2X           &Cellular V2X &    &    \\
D2D                   & Device-to-Device  &  &   \\
DDoS&Distributed Denial of Service&MCS& Modulation and Coding Scheme \\
DITEN&Digital Twin Edge Networks&MDA& Management Data Analytics \\
DME& Distance Measuring Equipment&MEC                   & Multi-access Edge Computing \\
DNN& Deep Neural Network&MEMR& Miniaturized Electromechanical Relays \\
DNS                   & Domain Name System   &ML& Machine Learning \\
DoS& Denial of Service&MMALCCA& Multiple Machine Access Learning with Collision Carrier Avoidance \\
DQN&  Deep Q-Network &mMIMO& Massive Multiple-Input Multiple-Output \\
DRL& Deep Reinforcement Learning  &mmWave& Millimeter-Wave \\
DRX & Discontinuous Reception &MTC                    & Machine-type Communication \\
DPA& Destructive Physical Analysis&NEMO-BS                   &Network Mobility Basic Support \\ 
DT& Digital Twin&NF& Network Function \\
          & &NFV &  Network Function Virtualization \\   
EC                        & Edge Computing &NGMN& Next Generation Mobile Networks \\
ECU &  Electronic Control Unit&    &  \\
&   &NIST& National Institute of Standards and Technology \\
ENI                        & Experiential Networked Intelligence &NLoS& Non-Line-of-Sight \\
ETSI                        & European Telecommunications Standards Institute  &NOMA                   &Non-Orthogonal Multiple Access \\
   &    &NR&New Radio  \\
   &  &NRF& Non-Radio-Frequency \\
Fast-CRO&Fast Chemical Reaction Optimization &                        & \\
FDMA        & Flow-Enabled Distributed Mobility Anchoring  &NSGA& Non-dominated Sorting Genetic Algorithm \\
FL& Federated Learning&NWDAF& Network Data Analytics Function \\
FLISR                    &  Fault Location, Isolation and Service Recovery &OBC& Onboard Charging \\
FPGA& Field Programmable Gate Array&ODN& Optical Distribution Network \\
FSO& Free-space Optics&OFTLPA& Overlapping Fault-tolerant Large Passenger Aircraft \\
FTA& Fault Tree Analysis &PAC                 & Protection, Automation and Control  \\
FT-SFGE               & Fault-Tolerant Service Function Graph Embedding &   &  \\
GAN& Generative Adversarial Network&PBFT& Practical Byzantine Fault Tolerance \\
&   &PCB& Printed Circuit Board \\
GBS & Ground Beacon Station&PCRF& Policy and Charging Rules Function \\
                     &      &                         &   \\
GNN& Graph Neural Network&PHY&  Physical Layer\\
GPS& Global Positioning System&PNEMO       & Proxy NEMO \\
  &  &PON& Passive Optical Network \\
% HF-ANN& Hybrid Fuzzy with Artificial Neural Network
& &PoS& Proof of Stake \\
  &   &PoW& Proof of Work \\
HSR& High-Speed Rail&PTP& Precision Time Protocol \\
HSS& Home Subscriber Server&QLC& Q-Learning for Cooperation \\
IAB                       & Integrated Access and Backhaul &QoS                        & Quality of Service  \\
% \bottomrule
\end{tabular*}
% }
\end{table*}

\begin{table}
\label{table00}
\begin{tabular}{p{2cm} p{5.5cm}}
RAN                   & Radio Access Network \\
RCA& Root Cause Analysis \\
RedCap&Reduced Capability \\
RF                    & Radio Frequency  \\
RIS                   & Reconfigurable Intelligent Surfaces \\
RLQ & Radio Link Quality  \\
RPFM& Reverse Path-Flow Mechanism \\
RRM &Radio Resource Management \\
RSSI& Received Signal Strength Indicator \\
RTT& Round-trip-time \\
SBS& Small Base Station \\
SDN                   & Software-Defined Networking  \\
SD-RAN&Software-Defined RAN  \\
SFC                    & Service Function Chain \\
SFF                   & Service Function Forwarder \\
SF                    & Service Function  \\
SINR                            &Signal-to-Interference-plus-Noise Ratio \\
SMF & Single Mode Fiber \\
STECN                    & Satellite-terrestrial Edge Computing Network \\ 
SVFMF & Service Virtualization and Flow Management Framework \\
SVM  & Support Vector Machine \\  
THz & Terahertz \\
TTF & Time to Failure \\
UAV                          &  Unmanned Aerial Vehicle   \\
UDM& Unified Data Management \\
UE & User Equipment  \\
URLLC                   & Ultra Reliable and Low-Latency Communications  \\
UWB                        & Ultra-Wide Band \\
V2X & Vehicle-to-Everything \\
VANET                & Vehicular Ad-hoc Network   \\
vIMS& Virtual IP Multimedia Subsystem \\
VL &  Visible Light \\
VNS              &Virtual Network Services \\
vSDN & Virtualized Software-defined Network \\ 
WWRF& Wireless World Research Forum \\
ZSM & Zero Touch Management \\
\bottomrule
\end{tabular}
% }
\end{table}

The imminent advent of the sixth-generation (6G) communication standard heralds a transformative where adaptability and intelligence seamlessly intertwine, fundamentally reshaping the very fabric of critical infrastructures~\cite{8412482}. At the heart of this transformative wave is the central role of communication and network systems, which threads through essential sectors, including but not limited to telecommunications~\cite{wang2015}, electric power systems~\cite{hu20tgcn}, banking and finance~\cite{8808168}, transportation~\cite{hu22tvt}, water supply systems~\cite{8766143}, supply chain~\cite{8792135}, government services~\cite{8760275}, and emergency services~\cite{hu23visual}. 
One notable case is Hurricane Katrina, which struck the Gulf Coast of the United States in 2005. The hurricane resulted in widespread destruction, flooding, and a breakdown of critical infrastructure, highlighting the essential role of communication systems in emergency response and recovery~\cite{notable_fail_2024_01_10}.
In this impending epoch, 6G is poised to ascend to a pivotal position, advancing beyond the achievements of its predecessor, i.e., the fifth-generation (5G) communication system.

\begin{figure}
\centering
\includegraphics[width=6cm]{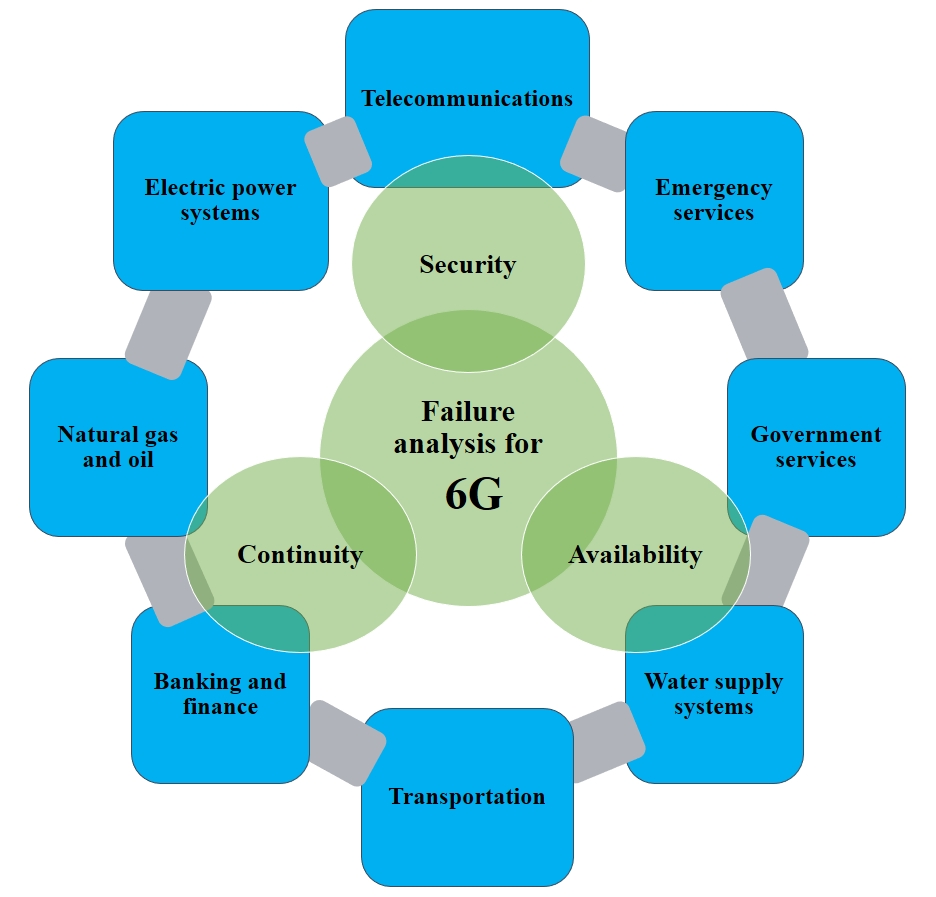}
\caption{\label{1} A schema of failure analysis in 6G for enhancing and promoting the construction of critical infrastructure.}
\label{F:schema1}
\end{figure}

The implementation of 5G, characterized by its unwavering focus on security, continuity, and availability, has already left an indelible mark on critical infrastructures. Not merely confined to incremental progress, 5G technology has orchestrated remarkable feats in diverse domains. In the area of smart grid management, the precision and efficacy of 5G have ushered in a new era of energy resource exploitation\cite{Bi_power_2019}, demonstrating unparalleled high flexibility and stringent security measures~\cite{5g_smart_grid,5g_gas, hu16, hu18}. The financial and banking sectors, driven by an insatiable demand for low-latency and high-security services, have found a reliable ally in 5G, ensuring efficient operations and robust governmental monitoring services~\cite{5g_bank,5g_gov}. The transportation and emergency services sectors, both inherently reliant on instantaneous and secure communication, have witnessed transformative benefits by empowering 5G technology~\cite{5g_transport, 5g_emergency}. Even the processes within water supply systems have been witnessed, with 5G-enabled intelligent sensors orchestrating the efficient monitoring and management of clean water distribution and wastewater treatment~\cite{5g_water}.

\subsection{Motivation}

The ascension to 6G is not merely an evolution but a revolution, promising a quantum leap in intelligence, reliability, and flexibility. Playing the central role of critical infrastructures, 6G is poised to deepen its roots in telecommunications, fortify the resilience of power grids, enhance the security fabric of financial systems and supply chains, revolutionize transportation networks, refine water supply management, and amplify the efficiency of emergency services. The discussion on the central role of 6G is timely and critical due to the indispensable nature of these critical infrastructures in shaping and safeguarding the very foundations of modern society, as illustrated in Fig.~\ref{F:schema1}.

As we transition to 6G, it becomes imperative to anticipate and address potential failures that could impede its pivotal role in sustaining critical infrastructures~\cite{8782879}. The extensive integration of components, services, and applications within future 6G systems introduces unprecedented complexity, heightening the risk of failures. This complexity, coupled with the interconnected nature of numerous components, modules, and subsystems, poses challenges for effective failure analysis in 6G~\cite{8869705}.

The recent surge in telecom failures on a global scale has brought to light the critical challenges faced by telecommunication networks, impacting millions and exposing vulnerabilities in their infrastructure.
In 2020, a significant portion of telecom downtime, equivalent to 346 million hours, was attributed to software glitches~\cite{notable1}. Severe weather events have also taken a toll, as witnessed in February 2021, when a snowstorm in the central United States led to power outages and caused widespread telecommunication service failures~\cite{flevent3}. A global outage in June 2021, originating from a network software failure triggered by an inappropriate update, affected critical websites and apps across continents~\cite{notable2}. Incidental outages in the same month wreaked havoc on major international services due to edge DNS failures~\cite{notable3}. Subsequent incidents in 2022 and 2023, ranging from operational disruptions in London data centers to cyberattacks on T-Mobile and hardware failures in Hong Kong, underscore the diverse challenges faced by telecom networks~\cite{notable5, flevent4, flevent5, flevent8, notable4, flevent9, flevent7, notable6, flevent6}. 

These incidents reveal the intricate interplay of environmental and technological factors, necessitating robust resilience strategies~\cite{9040264}. As telecommunication networks grapple with evolving risks, comprehensive risk management and international cooperation for standards and response protocols become imperative~\cite{9040431}. To this end, the upcoming 6G systems demand a meticulous understanding of potential failure points and preemptive strategies for their prevention and mitigation~\cite{9144301}.

The urgency of devising strategies to safeguard 6G against catastrophic failures underscores the importance of thorough preemptive failure analysis. As a meticulous engineering discipline, failure analysis aims to identify and trace the underlying mechanisms of failures, offering actionable insights to enhance system robustness and prevent severe repercussions~\cite{9145564}. While considerable efforts have been directed towards understanding failures in 6G, the challenges arising from system complexity have prompted researchers to delve deeper into this area~\cite{9023459}.
% In response to the complexity and challenges inherent in earlier generations of communication systems, including 5G systems, and arising from the upcoming 6G systems, 
A substantial body of work has emerged, proposing methodologies and solutions for effective failure analysis~\cite{9349624}. Researchers from diverse disciplines have contributed to this collective effort, each bringing a unique perspective to bolster the reliability of future 6G systems and applications~\cite{9397776}. Despite these valuable contributions, there exists a gap in the form of a comprehensive survey and tutorial that systematically organizes and articulates the key findings and focal points of these studies.

\subsection{Contribution}
\begin{figure}
\centering
\includegraphics[scale=0.2]  {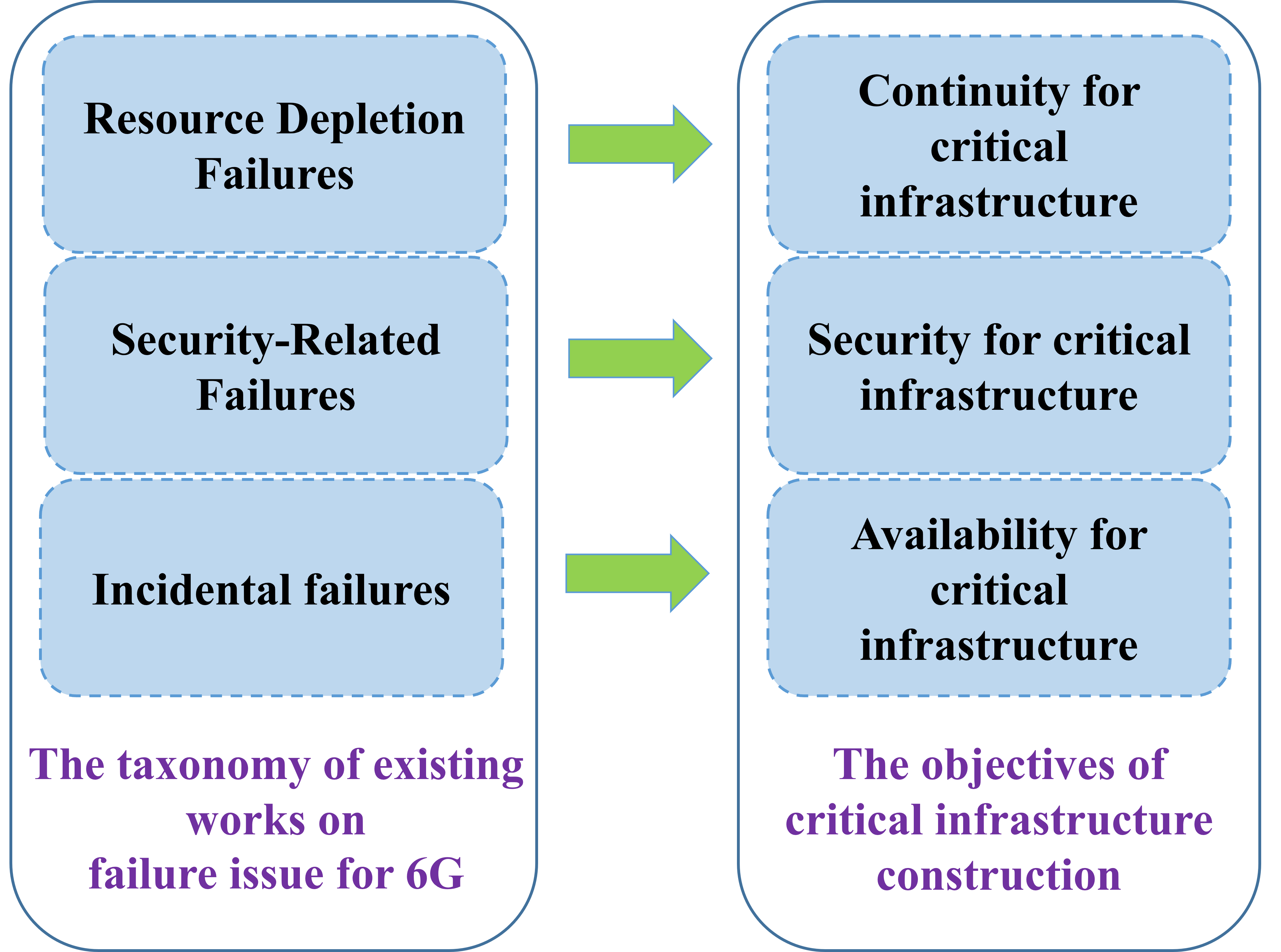}
 \caption{\small The taxonomy of existing works on failure in 6G corresponding to the three objectives of critical infrastructure.}
 \label{F:corresp}
 \end{figure}
By reviewing existing literature, this study aims to fill this gap by focusing on failure issues within critical 6G systems and applications and their broader impact on critical infrastructure. This survey examines failure analysis in 6G, providing formal definitions, exploring investigated failures in 5G and 6G systems, addressing security-induced system failures, exploring incidental failures, and discussing lessons learned. The survey concludes with opportunities and future research directions in 6G failure analysis, contributing to the ongoing debate.

This survey makes several novel contributions, enriching the discourse on failures, failure analysis, and countermeasures in 6G and critical communication infrastructures:
\begin{itemize}
\item 
% An exhaustive literature review reveals and categorizes the prominent research orientations in 6G failures. Specifically, 
We identify resource depletion challenges, security vulnerabilities, and concerns about system availability in 6G. It aligns with the triad of objectives for building robust and resilient critical infrastructures, see Fig. \ref{F:corresp}.

\item
% This survey delves into a comprehensive review of representative works. Specifically, 
We categorize typical failures in 6G systems and applications. We also summarize constructive countermeasures tailored to address these identified failure scenarios, contributing to the growing knowledge in 6G failure analysis and prevention.

\item
A general and standardized procedure for dealing with critical and typical failures is developed with insights from the extensive review of existing works. It can serve as a practical guide for researchers and practitioners involved in 6G failure analysis and prevention.

\item
In the context of 6G networks, we emphasize the need for standardization in addressing AI failures. A forward-looking perspective is required for 6G network evolution, especially in terms of failure standardization, given the envisioned intelligence of 6G.

\item 
In addition to consolidating existing knowledge, our survey identifies new challenges and outlines future research directions. In the context of 6G failures and critical infrastructure development, this forward-looking perspective fosters innovation and resilience.

\end{itemize}
% These contributions collectively contribute to the establishment of a comprehensive and forward-thinking framework for understanding, addressing, and advancing the discourse on failures and their implications in the dynamic landscape of 6G and critical communication infrastructures.

The rest of this survey is organized as follows.
Section II showcases many recent real-world failures in communication systems.
Section III reviews existing surveys on 6G systems, architecture, and applications, and reveals a significant gap in the failures and failure analysis of the systems.
Section IV defines the general types of failures in 6G systems.
Sections V through VII provide examples of preventing and mitigating resource depletion failures, security-related failures, and incidental failures.
In Section VIII, we discuss efforts to standardize and mitigate failures in 6G communications and closely related AI systems.
Section IX summarizes the lessons learned, challenges, and open issues.
The paper is concluded in Section X.
Table I collates the acronyms and abbreviations used in this survey.

\begin{figure} %[htbp]
\includegraphics[width=0.35\textheight]  {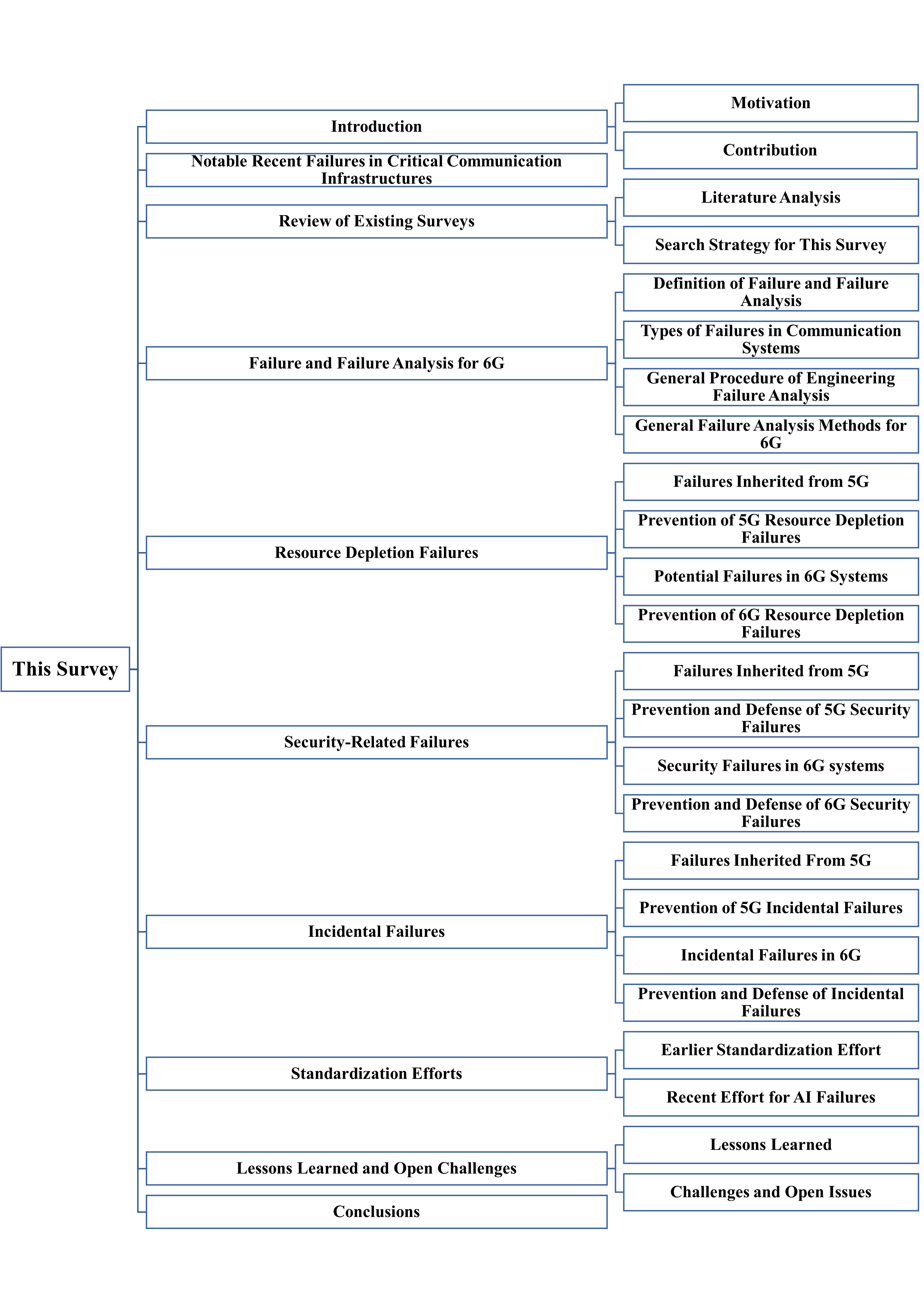}
 \caption{\label{1} The structure of this survey.}
 \label{F:timeline}
 \end{figure}

\section{Notable Recent Failures in Critical Communication Infrastructures}

% As illustrated in Fig.~\ref{F:timeline},
In 2020, an EU annual report on telecom security incidents highlighted that 40\% of time loss, that is, 346 million hours, was attributed to telecom failures caused by faulty software changes or updates~\cite{notable1}. In February 2021, a severe snowstorm in the central United States led to power outages, causing communication service failures and significantly impacting the communication needs of millions of people~\cite{flevent3}.

June 2021 witnessed a catastrophic outage failure affecting critical websites and apps globally, spanning the Americas, Europe, Asia, and South Africa. The incident resulted from a network software failure triggered by an inappropriate software update~\cite{notable2}. An incidental outage in the same month damaged an international IT organization's edge DNS service. Airlines, subways, banks, and international businesses were affected by this failure~\cite{notable3}. Telecommunications and power systems in the United States were threatened by devastating floods~\cite{flevent5}.

In July 2022, serious service failures in data centers in London, UK, led to operational disruptions for major IT companies like Google and Oracle. The core reason behind the failure was extreme weather affecting the normal operation of the refrigeration system~\cite{notable5}. Concurrently, millions of users in Japan faced disconnections due to outages caused by equipment failure, affecting finance and transportation operations~\cite{flevent4}. 
November 2022 witnessed a security failure resulting from a hacker attack, leading to severe information leakage for millions of T-Mobile customers~\cite{flevent8}. There were extensive breakdowns in Macau, China, in December 2022 due to a service failure in Hong Kong. The reason was an incidental hardware failure in the cooling system~\cite{notable4}.

August 2023 brought severe telecom service failures to Maui, USA, due to wildfires causing power failures in telecom service infrastructures~\cite{flevent9}. In the same month, a large number of users in Haiti experienced severe service failures due to the damage of fiber optic cables~\cite{flevent7}. The data communication system for nationwide bank statements in Japan failed in October 2023, affecting massive banks and financial institutions. The initial report suggested the failure was possibly due to the updating process of the relay processor reaching its operational limit~\cite{notable6}. In November 2023, the telecom outage failure of Optus, the second-largest provider in Australia, led to disconnections in phone and internet services, impacting various sectors with the cause not immediately identified~\cite{flevent6}.

Recent telecom failures, spanning diverse incidents globally, highlight the multifaceted challenges that telecom networks face. Service disruptions are frequently caused by faulty software updates, extreme weather conditions, natural disasters, and cyberattacks. 
It is important to emphasize robust resilience strategies when considering the interconnection of telecommunication infrastructure with environmental and technological factors. 
Nevertheless, comprehensive risk management is essential in light of the diversity of incidents, including hardware failures and software vulnerabilities. It becomes increasingly challenging to ensure these critical systems are reliable and secure. These failures also highlight the need for international cooperation to develop standards, response protocols, and mitigation strategies.

\section{Review of Existing Surveys}
\label{sec_II}

 \begin{figure}%[!h]
 % \captionsetup{font={\large}}
\center{\includegraphics[width=6cm]  {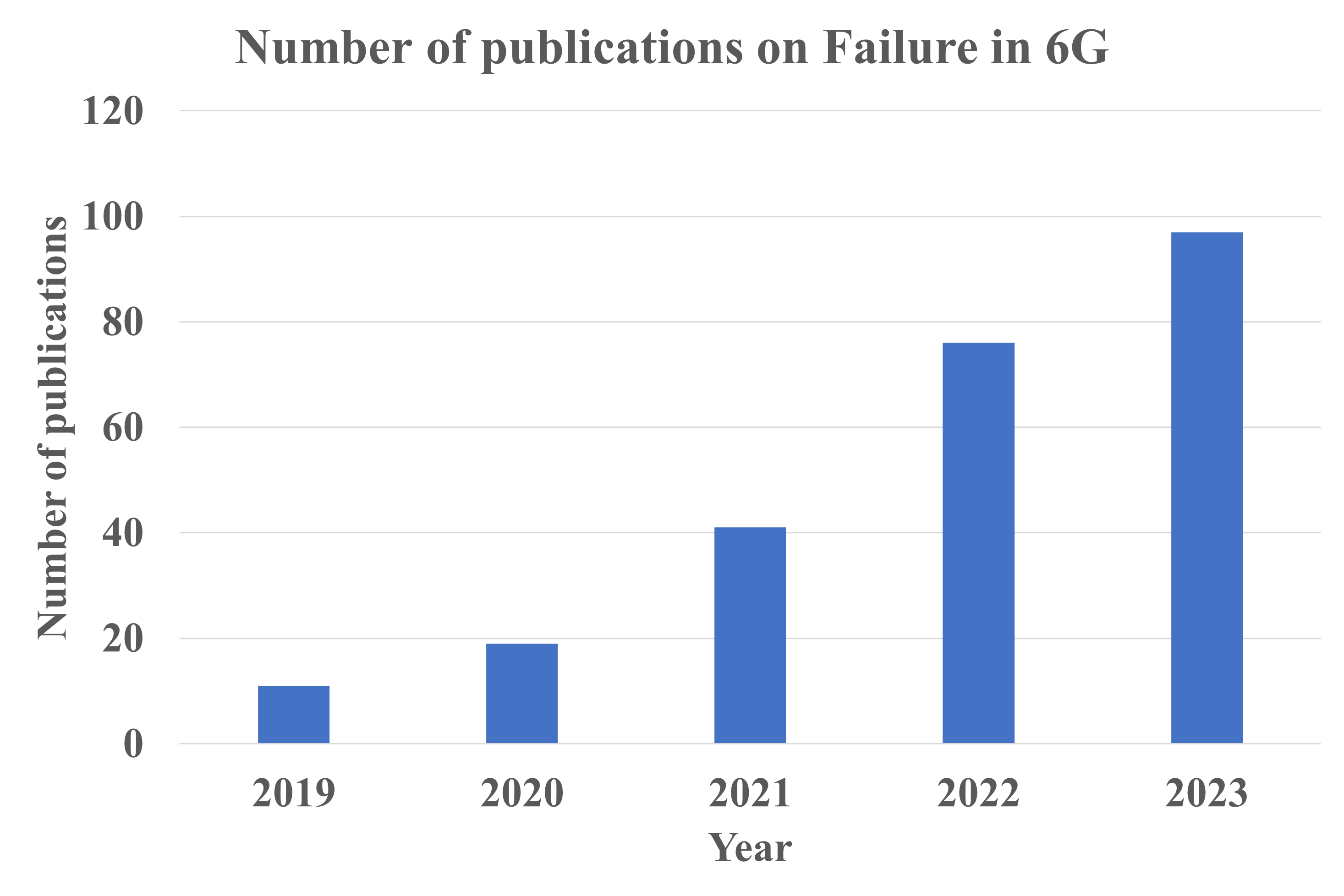}}
 \caption{\small Recent publications on failures in 6G based on the search result from the Scopus database as of 5 January 2024, where the search keywords are ( ``6G"~or~``6th~generation"~or~``sixth~generation")~and~(``failure"~or~``fault"), and the range is specified to be within the recent five years.}
 \label{F:TREND}
 \end{figure}

\begin{figure}
\centering
\includegraphics[width=8cm]  {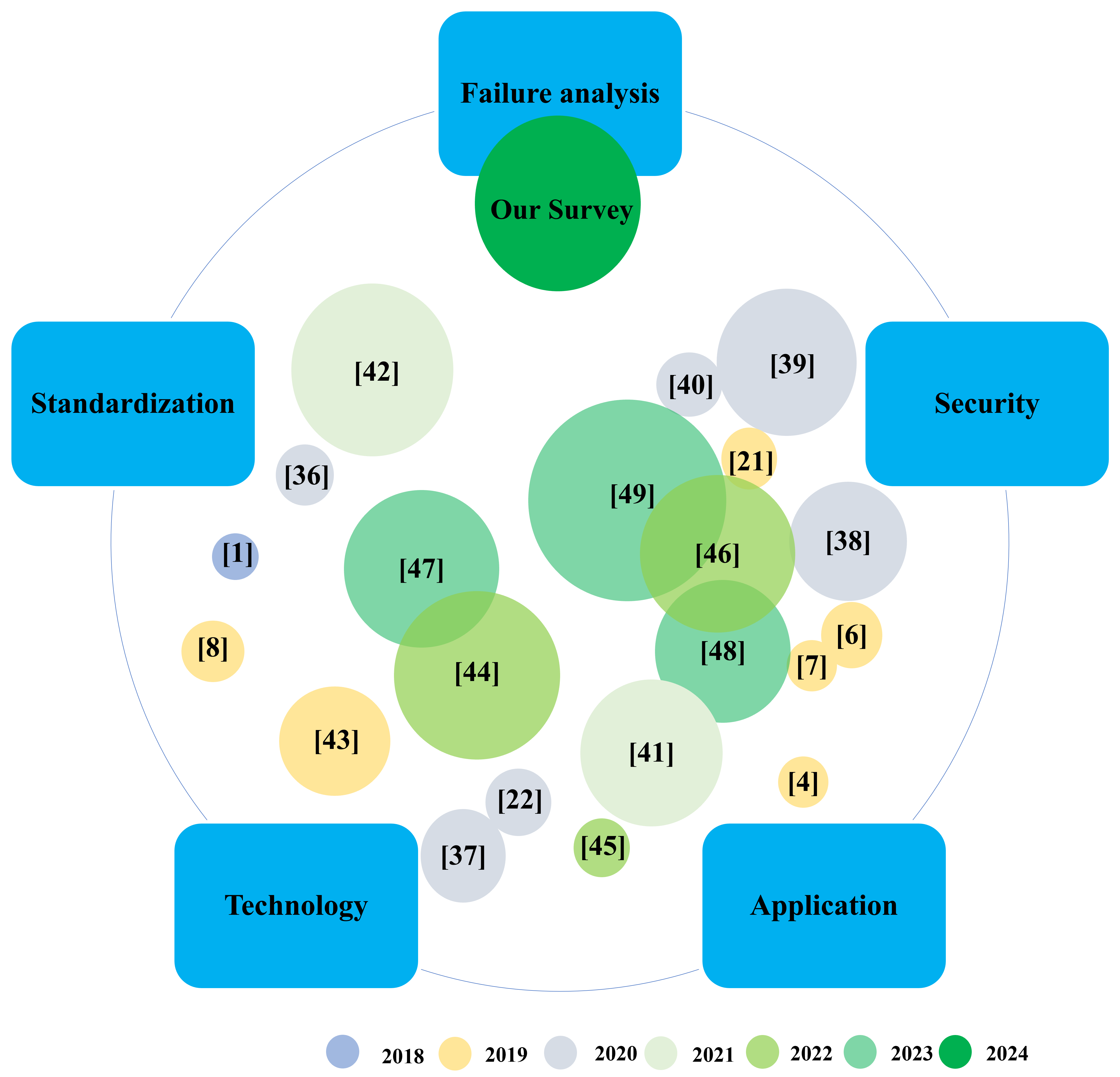}
\caption{\small The positioning of this survey with respect to the existing surveys, where the size of a circle indicates the number of papers reviewed in the corresponding survey. Different colors indicate different years of publication.}
\label{F:survey_topic}
\end{figure}

\subsection{Literature Analysis}
In the evolving landscape of 6G systems and applications, the rapid advancements in 6G technology and failure analysis have spurred our continuous monitoring of these domains. 
A pivotal aspect of our research involves a meticulous review of existing surveys on 6G, where we systematically distill the characteristics of these works, identify gaps, and underscore the critical significance of our investigation.
We judiciously select representative works that have garnered substantial citations, as shown in Fig.~\ref{F:survey_topic}.

One of the focal points explored across several surveys, such as \cite{8766143,9144301,9023459,9349624,8922617,8792135}, revolves around the architectural vision of 6G. These works propose innovative architectures designed to address the escalating demands, building upon the challenges posed by the existing 5G architecture. The transition from 5G to 6G is characterized by heightened transmission rates and an increased emphasis on security, intelligence, reliability, and flexibility. 
% These surveys introduce architectural paradigms that encapsulate the envisioned capabilities and features of 6G.
In addition to architectural considerations, techniques, standardization solutions, requirements, and 6G-enabled applications have been investigated, e.g., in \cite{8869705,8808168,9040264,8782879,9145564,8412482,8760275,9397776,9040431}. 
% These surveys delve into a spectrum of proposed techniques, ranging from terahertz communications to quantum communications and network slicing. 
% These techniques showcase robustness and sophistication compared to their predecessors in previous generations. 
Despite the advancements ranging from terahertz communications to quantum communications and network slicing, none explicitly captures 6G failure analysis and the construction of critical infrastructures.

While some surveys allude to concepts like critical infrastructure and failure analysis, our survey delves into the impact of applying failure analysis within 6G, examining how it influences and contributes to the construction of critical infrastructure. 
% This approach allows us to unveil the nuanced and intricate dynamics between 6G technology, failure analysis methodologies, and the overarching framework of critical infrastructure construction. 
Our survey 
% aims to provide a comprehensive understanding of these interdependency, shedding 
sheds light on potential challenges, opportunities, and the transformative potential of integrating failure analysis into the fabric of 6G systems.

\vspace{-3 mm}
\subsection{Search Strategy for This Survey}
In pursuing an extensive exploration of the existing literature on communication failures, we employed Scopus\cite{scopus1}, a renowned academic search engine, as the primary source for our initial data collection. 
We conducted targeted searches on the subjects of 5G and 6G. This focused approach allows us to discern and identify the ongoing and enduring nature of failures within these critical and contemporary communication infrastructures. The following search strings were used for our initial 5G and 6G collection.
\begin{itemize}
    \item \textit{TITLE-ABS-KEY~(( ``5G"~OR~``5th~generation"~OR~``fifth~generation")~AND~(``failure"~OR~``fault"))}

    \item \textit{TITLE-ABS-KEY~(( ``6G"~OR~``6th~generation"~OR~``sixth~generation")~AND~(``failure"~OR~``fault"))}
\end{itemize}

We focused on title, abstract, and keywords (as opposed to using ALL) as the filtering field to ensure a more precise and relevant collection. Our choice was driven by the objective of obtaining accurately matched articles. Additionally, we included the term ``fault" alongside ``failure" because parts of failures that are subject to repair are generally referred to as ``faults"\cite{fa_book_1}. This approach allows us to cover related research comprehensively.

Following our initial collection on communication failures, we identified over 1,400 articles on failures in 5G and 240 articles on failures in 6G from 2019 to 2023. An observation from the collected data is that 
the research on failure in 6G shows a steady upward trend, aligning with the global development and promotion of 6G; see Fig.~\ref{F:TREND}.

We also meticulously refined our dataset. The majority of the resulting papers originate from prestigious journals, such as IEEE Transactions\cite{ieee_home}, Elsevier\cite{else1}, and Springer\cite{spring1}. These carefully selected papers serve as the focal point for our review, prioritizing a comprehensive exploration of the most promising and in-depth patterns related to communication failures.

\section{Failure and Failure Analysis for 6G}
\label{sec_concept}

This section elucidates the symbiotic relationship between 6G and critical infrastructure, underscoring how failure analysis becomes a linchpin for optimizing 6G deployment in critical infrastructure construction. This unfolds with precise formal definitions of failure and failure analysis tailored specifically to the landscape of 6G. Following this groundwork, we present a systematic engineering failure analysis procedure, offering concrete insights into its application within the unique ambit of 6G scenarios.

\vspace{-3 mm}
\subsection{Definition of Failure and Failure Analysis}
\subsubsection{Definition of Failure}
\begin{figure}%[h]
\centering
\includegraphics[scale=0.4]  {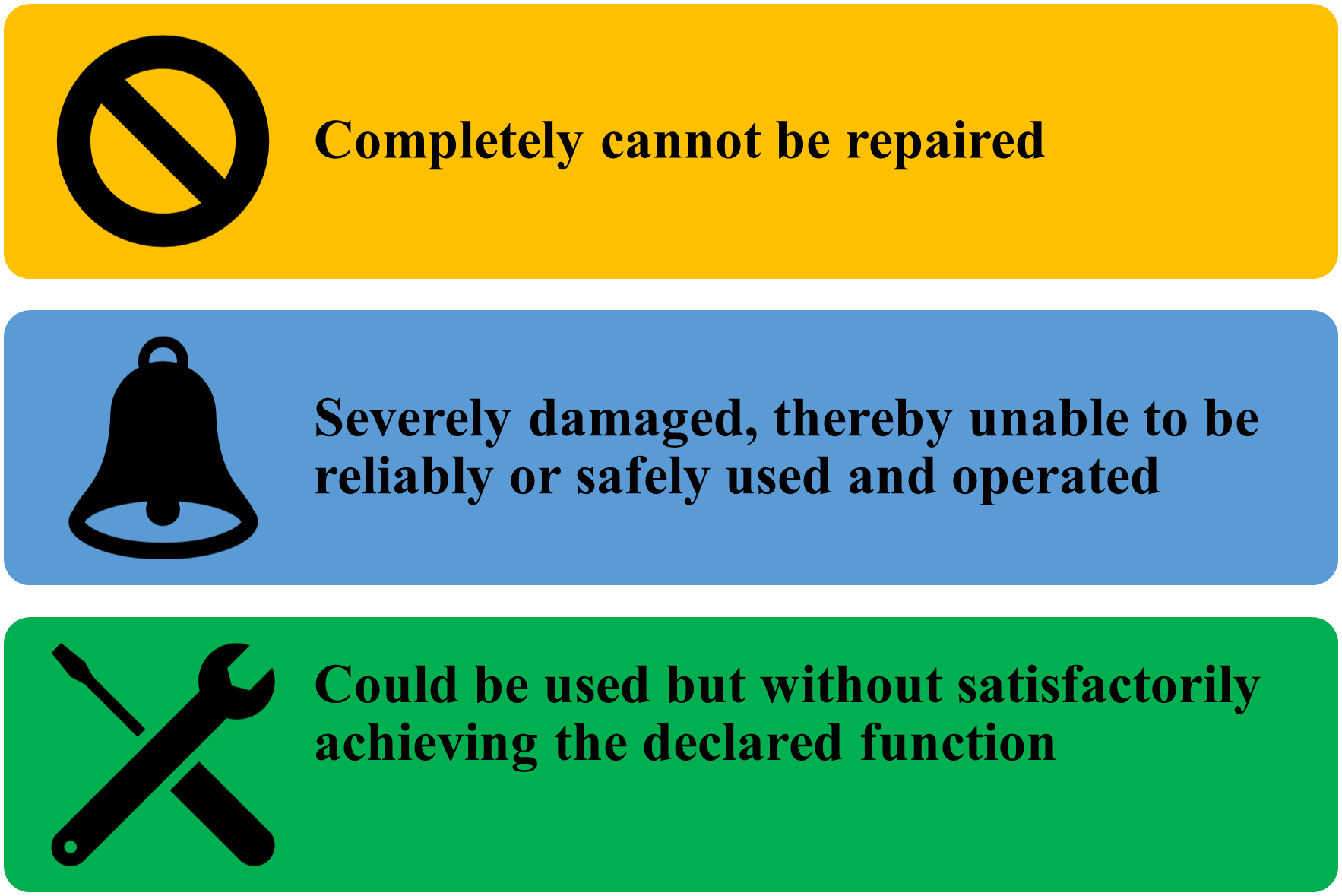}
 \caption{\small Three failure states, where the failure state in the orange block lays stresses on devices or systems and the failures must be replaced for recovery, while the rest emphasize on devices or systems and deserve special interventions.}
 \label{F:failure_def}
 \end{figure}

The notion of failure can be illustrated when one system, product, device, or component malfunctions due to the deterioration of the exterior appearance or interior structure, thereby depriving the original declared function. Generally, the product, device, or component can be considered under failure once it is in accord with being under any of the three states\cite{fa_book_1}, as shown in Fig. \ref{F:failure_def}. It is worth noting that even when a system, product, device, or component is in a comparatively mild state, i.e., barely enough working or just without achieving the declared function, it has already been at potential risk of severe crashes. It is no longer available for sustaining operations and must be instantly terminated to avoid more disastrous consequences.

\subsubsection{Failure Analysis}
As an efficient countermeasure to handle failures, 
% avoiding further exacerbating or preventing the system from the failure states in advance, 
failure analysis is a crucial scientific subject specifically used for identifying the cause of failure, the underlying failure mechanism based on the failure mode~\cite{failure_mode}. 
% In practice, failure analysis and prevention are of utmost importance to both economy and society since almost every sector of society is involved with the underlying risk of failures and the need for safety and reliability, especially in constructing critical infrastructures.
Failure analysis is widely acknowledged as a rigorous and formal scientific process that relies on multidimensional data collected from various sectors of production and operation. This process involves conducting comprehensive analyses to trace the in-depth mechanisms and root causes leading to failures. 
% Failure analysis is generally considered a rigorous and formal scientific process relying on the multidimensional data collected in each sector of production and operation to conduct comprehensive analysis for tracing the in-depth mechanism and cause leading to failure and further to provide constructive countermeasures and tactics for effective failure prevention. 

\subsection{Types of Failures in Communication Systems}
\begin{table*}[h]
\caption{Comparative Summary of Failure Types}
    \label{tab:failure_types}
    \centering
    \begin{tabular}{>{}p{2cm}>{}p{6.5cm}>{\arraybackslash}p{7.5cm}}
        \toprule
        \textbf{Failure Type} & \textbf{Definition} & \textbf{Main Causes} \\
        \midrule
        Transmission Failure & Task failing in fulfilling designated transmission requirements; issues in channel transmission and link access. & Channel fluctuation, channel interference, handover involving user mobility, hardware malfunction. \\
        \midrule
        Service Failure & Disruptions in core services like cloud, edge, storage, computation, and virtualization. & Inefficient management of dynamic requirements, imbalanced service allocation, inadequacies in managing service demand. \\
        \midrule
        Network Failure & Congestion, blockage, conflicts, and collisions; cascading failure; traffic reliability issues in network components. & Incidental disruptions, inappropriate traffic management, improper virtual network placement, resource conflicts. \\
        \midrule
        Power Failure & Malfunction in communication networks within power grids. & Connectivity and energy consumption issues, incidental equipment damage, unreliability of Device-to-Device connections. \\
        \midrule
        Component Failure & Malfunction or fault of critical components or modules in critical infrastructure. & Electromagnetic device deterioration, aging, and fracture, incidental disruption from natural disasters, environmental corrosion. \\
        \midrule
        Authentication Failure & Mistaken granting or denial of permissions; critical in 6G's dynamic, AI-driven environment. & Complexity in dynamic 6G environments, decentralized nature, and traditional authorization mechanisms may struggle. \\
        \midrule
        Task Integrity Failure & Compromised or altered tasks crucial for maintaining trustworthiness in 6G systems. & Advanced cyberattacks, interference in interconnected devices, internal system errors. \\
        \midrule
        Physical Security Failure & Unauthorized interference or tampering with physical components in 6G systems. & Device theft, tampering, sabotage of critical infrastructure nodes, compromising localized networks. \\
        \bottomrule
    \end{tabular}    
\end{table*}

Table \ref{tab:failure_types} provides a comprehensive overview of various critical issues affecting the robustness and reliability of 6G technology. The identified failure types encompass transmission failure, service failure, network failure, power failure, component failure, authentication failure, task integrity failure, and physical security failure. 

\subsubsection{Transmission Failure}
In both 5G and 6G systems, transmission failure refers to the transmission task failing or malfunctioning, thereby not fulfilling the designated transmission performance requirements. The transmission failure mainly involves the processes in channel transmission and link access\cite{aemo}. The main causes for transmission failure consist of channel fluctuation, channel interference, handover involving user mobility, and malfunction of device hardware. Specifically, hardware malfunction, for instance, radio frequency (RF) signal transceiver, antenna, modulation, demodulation, and so forth, is generally caused by general electrical component failures.

\subsubsection{Service Failure}
Service failures primarily pertain to disruptions in operating services critical for supporting core system and application functions. These services, which encompass cloud services, edge services, resource storage services, computation services, and virtualization-based services, cease to operate normally or consistently, impacting their ability to fulfill designated functions. The key contributors to service failures are inefficient management of dynamic requirements, imbalanced service allocation, and inadequacies in managing service demand. 
% Addressing these challenges is crucial for ensuring the reliability and stability of the advanced functionalities provided by these systems.

\subsubsection{Network Failure}
Network failures are characterized by congestion, blockage, conflicts, and collisions within the network, primarily stemming from inefficient management of network traffic resources~\cite{zhang2020analysis,kaiser2021network,9342800}. Network failures predominantly manifest as traffic reliability issues in Backhaul, Software-Defined Networking (SDN), Multi-access Edge Computing (MEC), Radio Access Network (RAN), and network slices. The key factors contributing to network failures include incidental disruptions, inappropriate dynamic traffic management, improper dynamic placement of virtual networks, inadequate orchestration, insufficient BS management, and resource conflicts. 
% Addressing these factors is essential for ensuring the seamless operation and performance of the network components in these advanced communication systems. 

\subsubsection{Power Failure}
The reliability of power systems is paramount in developing and constructing critical infrastructure, particularly in light of the growing energy crisis. Power failures in the context of communication networks within power grids signify malfunctions that can have significant repercussions. The primary causes of power failures include connectivity and energy consumption issues, incidental equipment damage, and the unreliability of Device-to-Device (D2D) connections. 
% Given the critical role of power in supporting communication infrastructure, addressing these causes is crucial for ensuring the robustness and uninterrupted operation of power-dependent networks.

\subsubsection{Component Failure}
5G has been pervasively deployed in critical industrial automation and intelligent transportation. 6G is envisioned to play a much more versatile role\cite{6G_wang_2023}. In critical infrastructure, the failure of critical components or modules can generally cause disastrous consequences.  
% Component failure is mainly referred to as the malfunction or fault of critical components or even modules in all realms of critical infrastructure. 
The main causes for component failure are electromagnetic device deterioration, device aging, device fracture, incidental disruption from natural disasters\cite{harsh_fail_1}, and environmental corrosion to the location for deployment.

\subsubsection{Authentication Failure}
% In 6G, characterized by ultra-reliable and low-latency communication, extensive edge computing, and a vast array of user devices, an 
An authorization failure represents a critical flaw where a device or application is mistakenly granted or denied permission to access resources, execute tasks, or perform actions within the network. Such failures could arise from complexities introduced by the highly dynamic, AI-driven, and decentralized nature of 6G environments, where traditional centralized authorization mechanisms may struggle to keep pace~\cite{9430901,9838599}. 
% An authorization failure in 6G could lead to unauthorized access to sensitive data, execution of malicious tasks, or even disruption of critical services, emphasizing the need for robust, adaptive, and context-aware authorization strategies.

\subsubsection{Task Integrity Failure}
In 6G, a task integrity failure denotes a scenario where a task, including its data transmission, processing, or any network operation, is compromised or altered, either maliciously or inadvertently. This integrity breach could result from advanced cyberattacks targeting AI functions, interference in the vast web of interconnected user devices, or internal system errors. Given that 6G promises operations at unparalleled scales and speeds, even minor task alterations can cascade into significant disruptions, ensuring task integrity is crucial.

\subsubsection{Physical Security Failure}
6G systems are envisioned to converge intricate digital infrastructures and a proliferation of physical devices, including Internet of Things (IoT) sensors, edge servers, and radio transmitters. A physical security failure pertains to unauthorized physical interference or tampering with these critical components. 
% Given the heightened reliance on distributed edge nodes and pervasive devices in 6G, such failures could compromise localized networks and propagate disruptions across the broader interconnected applications. 
From device theft and tampering to sabotage of critical infrastructure nodes, physical security breaches can introduce catastrophic vulnerabilities. Integrating robust physical safeguarding measures is paramount, complementing the advanced digital security protocols they uphold.

\begin{table*}[h]
\centering
\caption{Interplay of Software, Hardware, and System Failures in 6G.}
\label{tab:failures_6g}
\begin{tabular}{l p{6cm} p{7cm}}
\hline
\textbf{Component} & \textbf{Possible Failures} & \textbf{Interplay and Impact} \\
\hline
\textbf{Hardware} &
\begin{itemize}
    \item Malfunctioning BS
    \item Antenna Damage
    \item Network Component Failure
    \item Server Breakdown
\end{itemize} &
\begin{itemize}
    \item BS failure impacting communication
    \item Antenna issues affecting network connectivity
    \item Network component failure causing data processing
    \item Server breakdown leading to system downtime
\end{itemize} \\
\hline
\textbf{Software} &
\begin{itemize}
    \item Virtualization Software Errors
    \item AI Algorithm Deficiencies
    \item Digital Twin Inaccuracies
\end{itemize} &
\begin{itemize}
    \item Virtualization errors affecting system functions
    \item AI algorithm issues impacting network management
    \item Digital twin inaccuracies affecting data processing
\end{itemize} \\
\hline
\textbf{System Functions} &
\begin{itemize}
    \item Disruptions in Communication Protocols
    \item Network Downtime
    \item Data Processing Errors
\end{itemize} &
\begin{itemize}
    \item Communication protocols affected by hardware failure
    \item Network downtime due to software errors
    \item Data processing errors resulting from system failures
\end{itemize} \\
\hline
\end{tabular}
\end{table*}

% \vspace{-5 mm}
\subsection{General Procedure of Engineering Failure Analysis}

\begin{figure}
\centering
\includegraphics[width=0.33\textheight]  {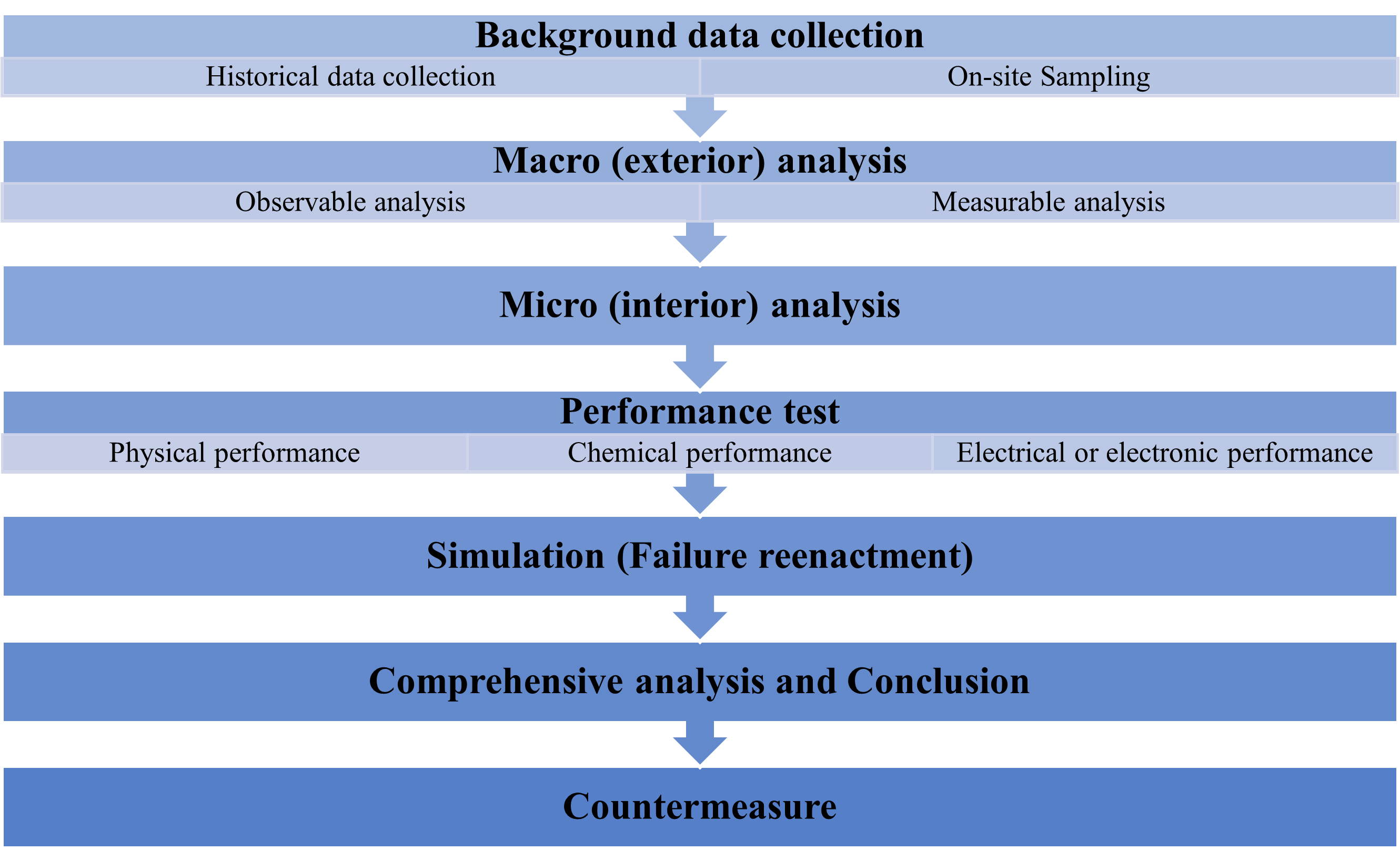}
 \caption{\small The general procedure of engineering failure analysis.}
 \label{F:failure_proced}
 \end{figure}

A standardized procedure for engineering failure analysis serves as a systematic guide, as illustrated in Fig. \ref{F:failure_proced}. To address the specific challenges of failure analysis in 6G, we delve into each stage of the procedure: 
\begin{itemize}
\item \textbf{Background Data Collection:} This initial stage gathers comprehensive data, including historical blueprints, parameter variations, and on-site samples. In the case of 6G, this entails capturing data related to the original network architecture and detailed parameter surveillance, such as peak signal transmission values and traffic throughput.

\item \textbf{Macro-analysis:} This stage is divided into observable and measurable analyses. Observable analysis evaluates failure modes, such as fracture, corrosion, and wear, from an exterior perspective. Measurable analysis assesses failure modes based on measurable aspects like distortion and attenuation. In 6G, issues involve hardware and signal analysis, including fracture assessment, electronic circuit corrosion, and signal distortion/attenuation in components.

\item \textbf{Micro-analysis:} At this stage, the focus shifts to understanding variations in constituents and configurations related to the identified failure mode, e.g., time-series electronic or electrical work by setting up a testbed with on-site sensors in 6G hardware to identify procedures leading to failure.

\item \textbf{Performance Test:} This stage involves checking, identifying, and testing the physical, chemical, and electrical or electronic aspects of the system. In the 6G scenario, the issues encompass evaluating the physical, electrical, and electronic aspects of 6G hardware.

\item \textbf{Simulation:} In this stage, the goal is to reenact the failure procedure based on analyses from earlier steps. This involves setting up simulation platforms and conducting analyses, which can be accomplished by establishing hardware or software simulation platforms, e.g., using 
% previously collected analysis data. In particular, AI-aided solutions, such as 
ML combined with 6G Digital Twin. 
% are increasingly applied to simulate and determine failure issues in 6G systems.

\item \textbf{Comprehensive Analysis and Conclusion:} This stage synthesizes insights into the causes of failure by considering all collected data, macro and microanalyses, performance tests, and simulations. In 6G, the typical causes of failures are resource depletion, security vulnerability, and accidental failure. 
% providing a foundation for detailed discussion in the subsequent section.

\item \textbf{Countermeasure:} The final stage focuses on proposing efficient and effective solutions to prevent the identified failures from recurring. 
% In the 6G context, specific countermeasures are discussed for general failure modes across different scenarios.

\end{itemize}

This structured approach ensures a thorough examination of failure scenarios in 6G, facilitating a comprehensive understanding and effective mitigation strategies.

\subsection{General Failure Analysis Methods for 6G}

\begin{figure} %[h]
\centering
\includegraphics[width=6cm]  {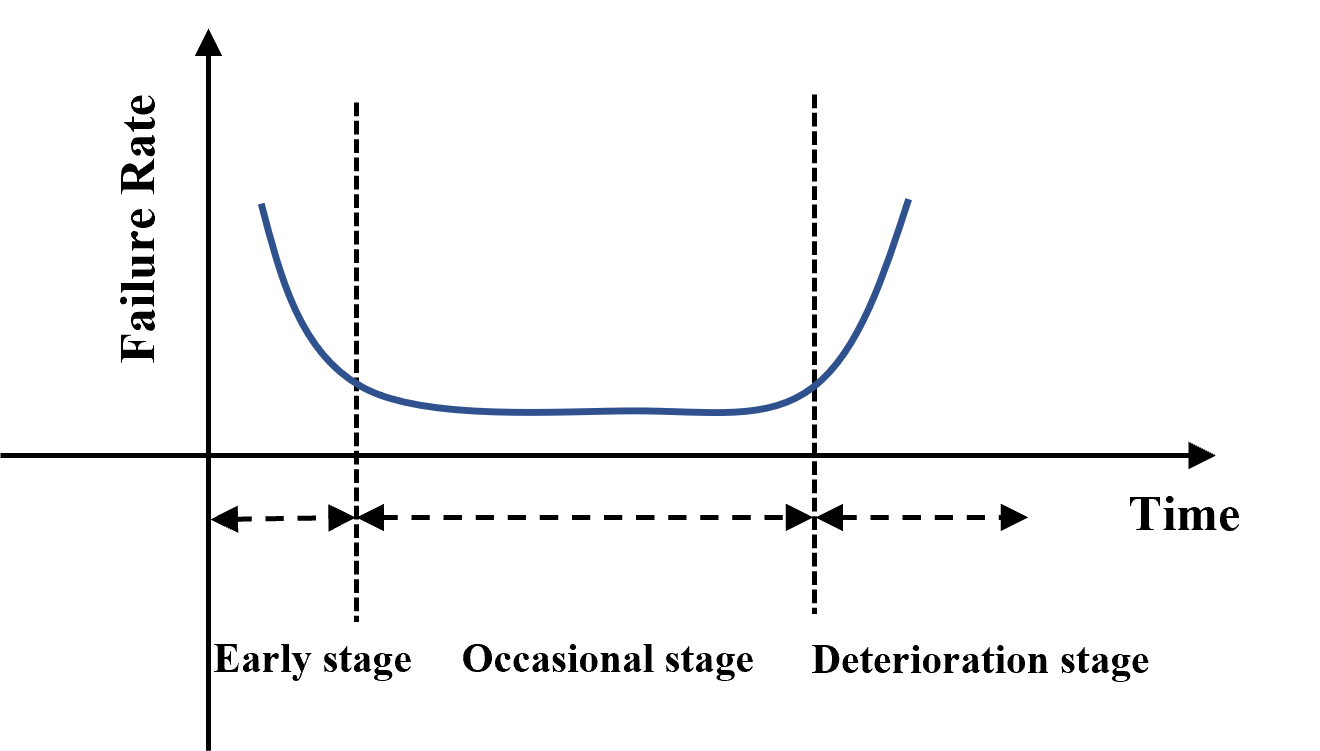}
 \caption{\label{1} The variation of the failure rate over time for general electrical device products.}
 \label{F:failure_rate}
 \end{figure}

 \begin{figure}[!h]
\centering
\includegraphics[scale=0.28]  {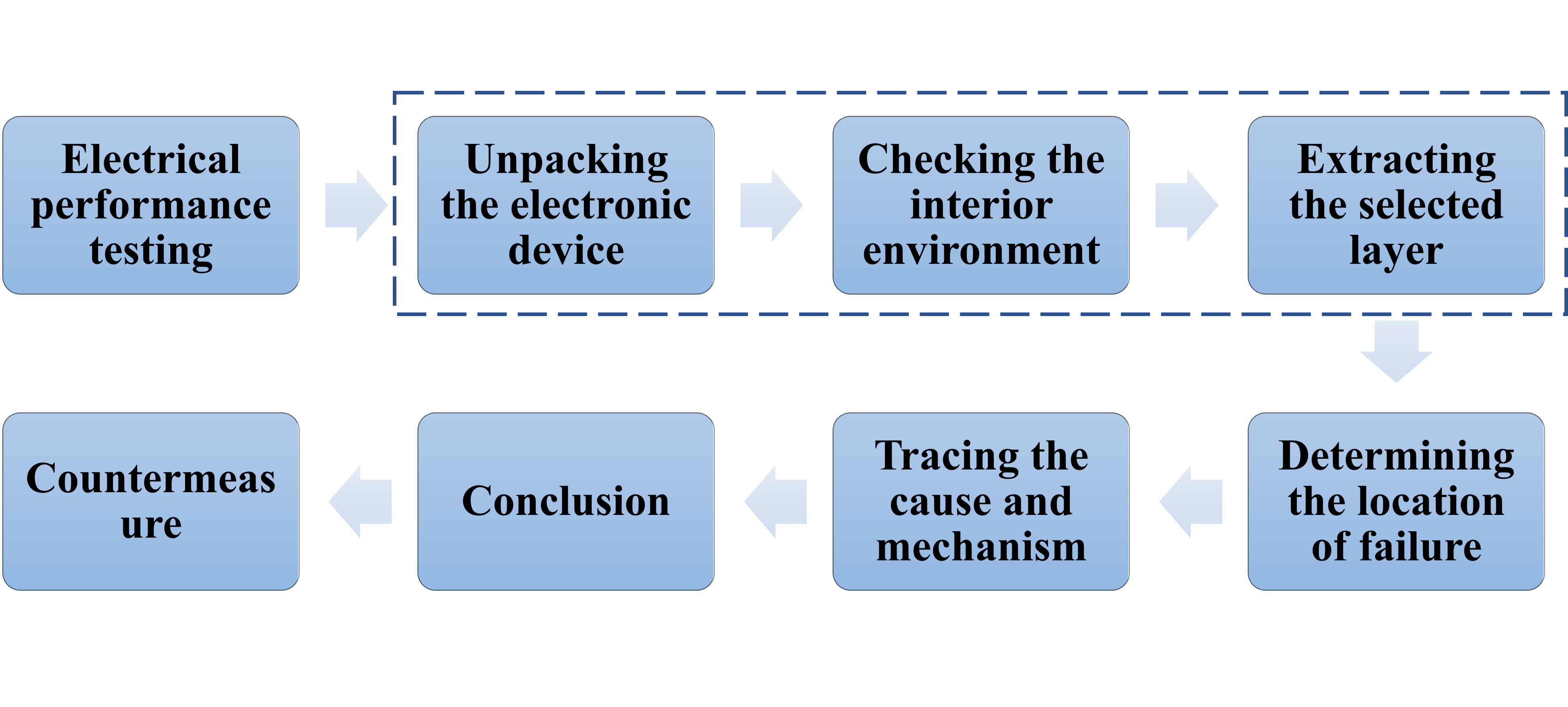}
 \caption{\small The general failure analysis procedures for 6G hardware, where the principle for this procedure is to timely, efficiently, and effectively identify and locate a failure and to trace further the very cause inducing the failure. The part of the procedure within the dashed block is also called DPA, which can be replaced by other non-destructive solutions.}
 \label{F:hardware}
 \end{figure}

 \begin{figure}[!h]
\centering
\includegraphics[width=6cm]  {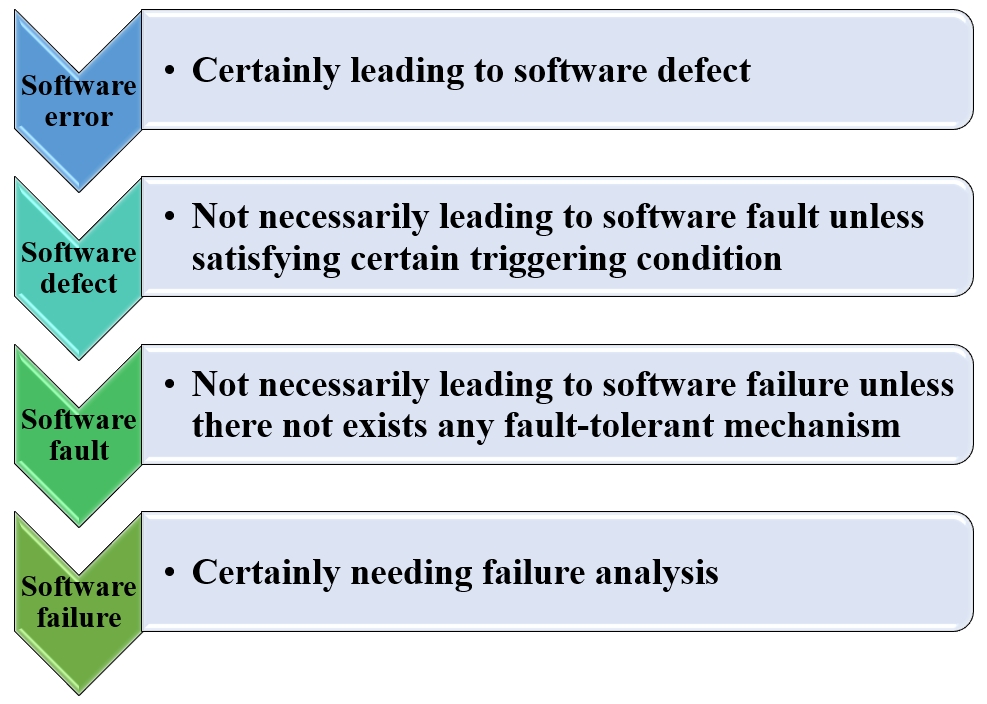}
 \caption{\label{1} The relationship among the four critical concepts of software failure.}
 \label{F:softrelation}
 \end{figure}

 \begin{figure}
\centering
\includegraphics[width=0.34\textheight]  {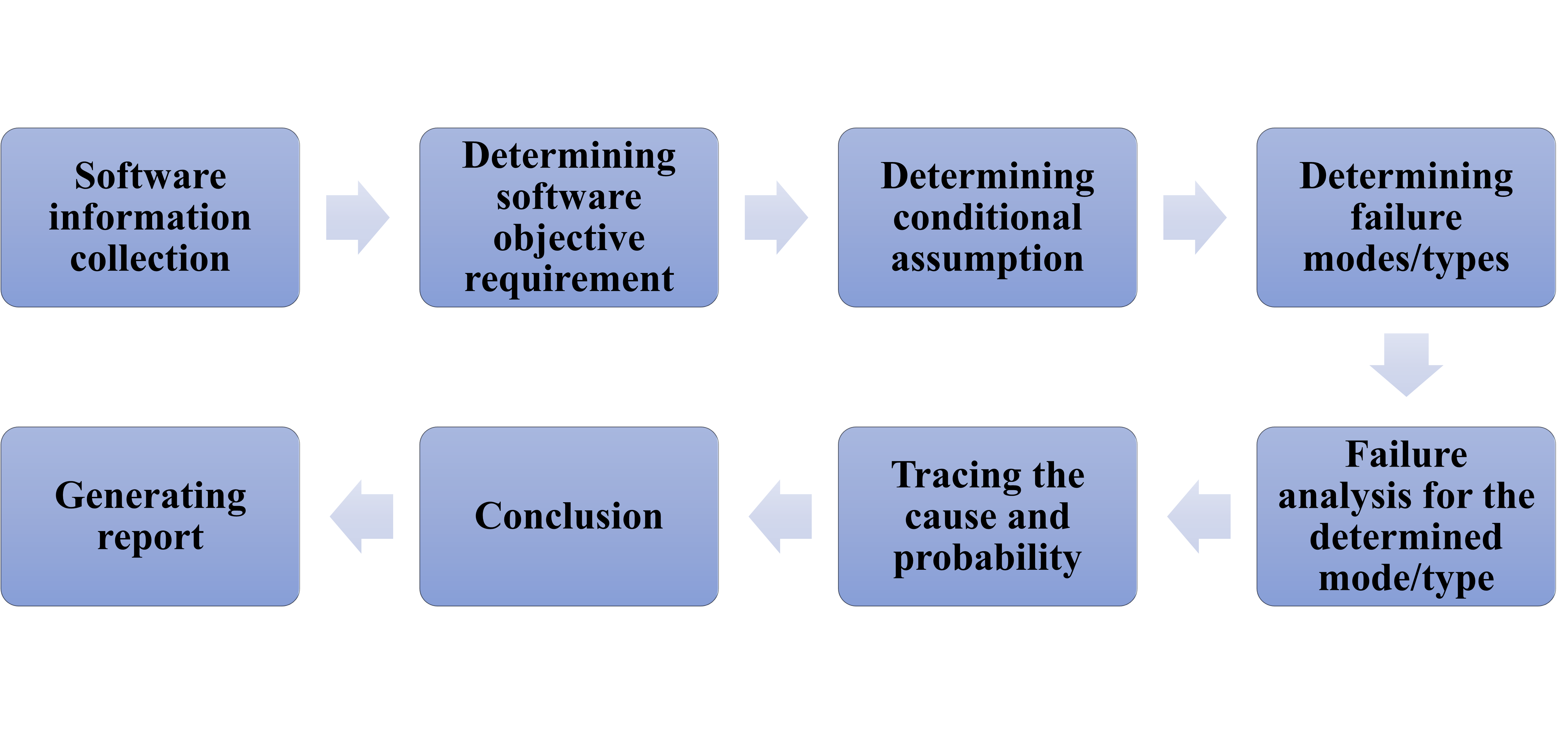}
 \caption{\small The general procedure of failure analysis for 6G software, whose principle is to timely, efficiently, and effectively identify software errors, software defects, software faults, and software failures, trace the very cause potentially resulting in failures, and verify whether code meets the specific objective stated in the requirements.}
 \label{F:software}
 \end{figure}

A direct application of general procedural principles to the distinctive context of 6G is deemed unrealistic. Given the inherent complexity of 6G, a more tailored approach is imperative. We provide a comprehensive exploration encompassing hardware failure, software failure~\cite{soft_fail_1}, and system/network failure, considering the unique challenges posed by 6G, as summarized by Table~\ref{tab:failures_6g}.

 \subsubsection{Hardware Failure Analysis}
 \label{hard_fail}

When a physical entity fails to fulfill its original purpose, it is called a hardware failure. Hardware is essential for critical infrastructure.
In 6G, hardware predominantly refers to key devices, modules, and components crucial for emerging technologies. In THz communication, holographic beamforming, and RIS, digital baseband processing units, RF transceivers, and array antennas are indispensable.

Specific hardware failures result from physical damage, defect, or aging of devices, modules, or components. Fig.~\ref{F:failure_rate} depicts the variation of the failure rate over time for general electrical devices. The majority of 6G hardware failures are due to electrical or electronic issues, such as open circuits, short circuits, abnormal leakage currents, and electric breakdowns. 
The failure analysis process for 6G hardware can generally be outlined sequentially, as illustrated in Fig.~\ref{F:hardware}. 
The procedure within the dashed block of Fig. \ref{F:hardware} is also called Destructive Physical Analysis (DPA), primarily used for better determining the defective devices potentially leading to severe consequence~\cite{zhou_dpa}. In general scenarios, DPA can be flexibly replaced by resorting to flaw-detection tools, e.g., X-ray or ultrasonic waves, especially considering the powerful AI specialized in processing imaging~\cite{zhao_ndt_ai}.

\subsubsection{Software Failure Analysis}

A critical, logical entity intricately connected with hardware is crucial in delivering essential services for constructing and developing critical infrastructure. 
A software failure occurs if any virtual entity fails to achieve its originally declared function due to the inappropriate resolution of software faults. 

Software errors, software defects, or software faults often cause software failures, as shown in Fig.~\ref{F:softrelation}. 
Software errors are unintentional defects in code or script introduced by software developers. Errors can occur at each stage of a software life cycle, including requirement analysis, and high- and low-level designs.
Software errors do not necessarily result in software failures. Software defects are errors or bugs in software. Software defects pose a serious risk if they are not addressed appropriately, even if they seem normal.
In the event the defective software at risk continues to operate, it may progress to a state of software fault. However, the existence of a fault-tolerant mechanism does not guarantee that a software fault leads to software failure. Different software faults can result from the same software defect under specific conditions.

In the context of 6G, software primarily refers to emerging technologies such as 6G virtualization, 6G AI, 6G-enabled digital twin, etc. All these software-based technologies are typical characteristics of general software, relying on the data and predefined instruction set to conduct training, computation, and operation, as illustrated in Fig.~\ref{F:software}.

\subsubsection{System Failure Analysis}
 
System failure emerges as a critical challenge in 6G, constituting a highly sophisticated and interconnected system amalgamating hardware and software components. This type of failure is discerned by the inability of the system to fulfill its initially asserted functions, a predicament that can be attributed to a myriad of factors, including hardware failures, software failures, and the intricate interplay of both. 

The system failure of 6G is not merely a conventional breakdown but a multifaceted challenge that demands a holistic understanding of the interplay between hardware and software. As 6G continues to evolve and unfold, it is critical to comprehend and address the system failures, contributing to the resilience and reliability of 6G.

% \begin{table*}[ht]
% \centering
% \caption{Comparison between System Failures and Application Failures in 6G.}
% \label{tab:failures_comparison}
% \begin{tabular}{l | p{7cm} | p{7cm}}
% \hline
% \textbf{Aspect} & \textbf{Operational Failures} & \textbf{Application Failures} \\
% \hline
% \textbf{Definition} & Failures that involves the overall functionality and performance of the 6G network system and derivative key technologies. & Failures specific to individual applications/tasks running on the 6G network. \\
% \hline
% \textbf{Impact} & Wide-reaching impact on multiple services and components within the 6G infrastructure. & Impact limited to the functionality of a particular 6G empowered application. \\
% \hline
% \textbf{Examples} & \begin{itemize}
%     \item Network protocol failures.
%     \item Hardware malfunctions.
%     \item Centralized server crashes.
% \end{itemize} & \begin{itemize}
%     \item  Advanced cyberattacks targeting
%     AI-driven function.
%     \item Handover failure within multiple services. 
% .
%     \item User mobility challenging offloading in MEC.
% \end{itemize} \\
% \hline
% \textbf{Countermeasures} & \begin{itemize}
%     \item Redundancy in network components.
%     \item Regular hardware maintenance.
%     \item Automated system monitoring/resilience recovery.
% \end{itemize} & \begin{itemize}
%     \item Blockchain-centric/Trustworthy  cipher mechanism.
%     \item Flexible  and intelligent BSs allocation scheme.
%     \item Digital Twin Edge Network based mobile offloading scheme.
% \end{itemize} \\
% \hline
% \end{tabular}
% \end{table*}

\section{Resource Depletion Failures}
\label{sec_IV}

This section delves into the challenges surrounding 6G technology, particularly focusing on failures arising from inadequate resources.
Despite advancements in 5G, the risk of failures persists due to inefficient resource allocation exacerbated by unpredictable factors. Specific instances of failures within 5G systems are explored to offer insights into these challenges.
The importance of optimal coordination, allocation, and scheduling is highlighted in the face of failures in 6G scenarios.

\subsection{Failures Inherited from 5G }

Despite the advancements of 5G, risks of failure that can substantially hinder the practical deployment of these systems persist. These failures often stem from the ineffective and inefficient allocation of limited communication resources, exacerbated by unforeseeable factors like environmental fluctuations and sudden traffic surges. 

% As depicted in Table~\ref{tab:failures_comparison}, 
Typical failures undergone in communication systems can be further categorized into operational or functional failures, and application failures, by considering their distinct impacted aspects of the systems.

\subsubsection{Operational Failures}

The ensuing discussion delves into specific instances of functional and operational failures resulting from the shortage of power, spectrum, and other resources within 5G systems.

\begin{figure}[htbp]
\centering
\includegraphics[width=8.5cm]  {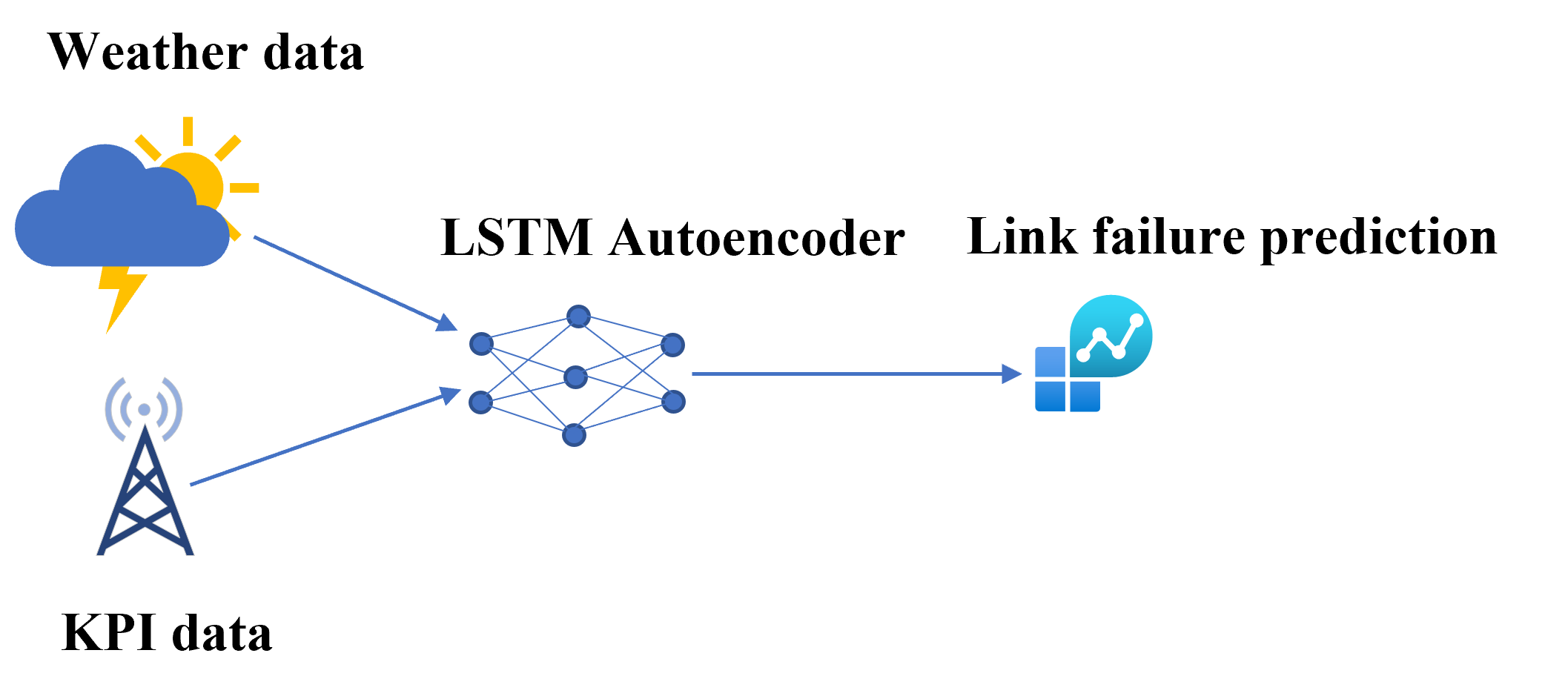}
 \caption{ \small Incorporating spatial and temporal correlations between radio and weather data, the model proposed in~\cite{Islam20231140} leverages weather forecasts for enhanced failure prediction. Integrating weather conditions anticipated during link operation aligns with observed performance improvements over models using current weather data.}
 \label{F:48png}
 \end{figure}

$\bullet$ \textbf{Transmission Failures}

As part of 5G infrastructure, RANs comprise radio BSs that establish wireless radio links. Environmental fluctuations, such as adverse weather conditions, can disrupt these communications. Critical applications are especially vulnerable to interruptions. To preemptively address these issues, a viable approach {\chat proposed by \cite{Islam20231140}} predicts potential failures and adjusts radio links resource allocation accordingly, as illustrated in Fig.~\ref{F:48png}.

 For applications with the highest quality-of-service (QoS) requirements, ultra-reliable and low-latency communications (URLLC) were introduced for 5G. Nonetheless, in~\cite{Li20232176}, the authors investigated the challenge posed by the dynamic nature of wireless channels. The high transmission power required by URLLC's stringent QoS requirements may conflict with the practical power constraints of real-world systems, further increasing transmission failure risks.
The authors of~\cite{Li20232176} enhanced energy efficiency in URLLC by optimizing joint uplink and downlink resource allocation. Frequency-hopping and proactive dropping were designed to reduce failure rates in deep fading scenarios, offering a solution to avoid system failures.

URLLC services are delivered using a sequence of software-based network functions, commonly called a service function chain (SFC). Ensuring fault tolerance in deploying an SFC is a complex endeavor, as protection mechanisms must simultaneously address transmission failures~\cite{Zheng20221}.
Network Function Virtualization (NFV) employs SFCs comprising service functions (SFs) and service function forwarders (SFFs) to provide services. However, the authors of \cite{Peng2023} indicated that SFFs within an SFC may encounter transmission failures while forwarding traffic to specified SF instances.

Non-orthogonal multiple access (NOMA) in 5G systems can be categorized into grant-based and random access. The letter empowers UEs to send information packets directly using uplink resources without requiring grant information~\cite{9239911}.
The authors of \cite{Zhang20235490} claimed that grant-free NOMA is well-suited for IoT services with small packets. Power collisions can result in severe transmission failures in machine-type communication (MTC) scenarios when uncoordinated resource selection occurs
% heightened access delay, and increased transmissions.

\begin{table*}[htbp]
\centering 
\caption{Operational failures in 5G caused by network resource depletion} 
\begin{tabular}{ p{1.75cm}<{\centering} m{2cm} m{4.5cm} m{4cm} m{1.5cm}<{\centering} } 
\hline 
\textbf{Failure Type}  & \textbf{5G Technology} & \textbf{Cause for Failure}  & \textbf{Countermeasure for Failure} & \textbf{Perpetuate in 6G} \\ \hline

\multirow{5}{*} & RAN\cite{Islam20231140}   & Limited radio resources for allocation\cite{Islam20231140}  & LSTM-autoencoder based radio resource allocation scheme\cite{Islam20231140}  & \checkmark \\ \cline{2-5}

&  URLLC \cite{Li20232176} & Stringent QoS needs\cite{Li20232176} & Energy-efficient packet
delivery mechanism \cite{Li20232176}  & \checkmark \\ \cline{2-5} 

 Transmission failure &  SFC \cite{Zheng20221} & Stringent QoS needs \cite{Zheng20221} & K-heterogeneous-faults-tolerance mechanism \cite{Zheng20221} & \checkmark \\ \cline{2-5} 

&  NFV \cite{Peng2023} & Forwarding traffic
instances \cite{Peng2023}& Auxiliary backup  transferring mechanism \cite{Peng2023} & \checkmark \\ \cline{2-5}

&  NOMA \cite{Zhang20235490} & Power collision interference \cite{Zhang20235490}& Limited Interference Resolution signaling \cite{Zhang20235490} & \checkmark \\ \cline{2-5}

   & Radio link
management \cite{Hasan20228895} & Weak signals and signaling overhead \cite{Hasan20228895} &  Resource-efficient
Flow-Enabled Distributed Mobility Anchoring mechanism \cite{Hasan20228895}& \checkmark \\ \cline{1-5}

 &  New radio \cite{Selim2022}   & Small cell backhauling dilemma \cite{Selim2022} & Self-healing scheme \cite{Selim2022}& \checkmark \\ \cline{2-5}

&  STECN\cite{Esmat202314621}   & Inappropriate traffic management and orchestration\cite{Esmat202314621} & Autonomous reconfiguration mechanism \cite{Esmat202314621}& \checkmark \\ \cline{2-5}

 & MEC\cite{Haber20216838}
   & Inflexibility of ground-based MEC \cite{Haber20216838}& UAV-aided ultra-reliable low-latency computation
offloading mechanism \cite{Haber20216838}& \checkmark \\ \cline{2-5}

 Network failure              &  RAN\cite{DiCicco2022}   & Inappropriate dynamic traffic management\cite{DiCicco2022} & Optimal virtual function
placement mechanism \cite{DiCicco2022}& \checkmark \\ \cline{2-5}

 &   NFV \cite{Dandachi2022}  & Inappropriate  dynamic virtual networks placement \cite{Dandachi2022}& DQN based self-adaptive  strategy \cite{Dandachi2022}& \checkmark \\ \cline{2-5}
& MEC based V2X system \cite{Wang2023ssssss}  & Rapidly varying computing and energy loads \cite{Wang2023ssssss} &  
MEC-based hierarchical resource management framework \cite{Wang2023ssssss}& \checkmark \\ \cline{2-5}

 & NFV based VANETs \cite{Cao202222492} & Incidental disruption \cite{Cao202222492} & Dynamic virtual resource
allocation mechanism \cite{Cao202222492}& \checkmark \\ \cline{1-5}

Power failure       & SD-RAN based smart grid \cite{NaitBelaid20225874}  & Incidental connectivity and energy consumption problem \cite{NaitBelaid20225874} &   Joint
routing and link scheduling for failure \cite{NaitBelaid20225874}& \checkmark \\ \cline{2-5}

 & IoT based smart grid \cite{Ilahi2021} & Incidental equipment damage \cite{Ilahi2021} & Non-intrusive detection for
Partial Discharge mechanism \cite{Ilahi2021}& \checkmark \\ \cline{1-5}

\end{tabular} 
\label{t.op.res}

\end{table*}

Mobile devices within the Internet of Medical Things (IoMT), including ambulances, medical drones~\cite{li2022data}, and emergency movable medical device, encounter significant signal distortions characterized by interference, packet loss, delay, and reduced throughput when moving.
A Network Mobility Basic Support (NEMO BS) Protocol was introduced, leveraging an IP-based Wi-Fi resolution.
In \cite{Hasan20228895}, it was revealed that weak signals, additional signaling overhead, and increased delays are hindering the handover process. This situation can lead to radio link failures.

$\bullet$ \textbf{Network Failures}

A critical issue in 5G is cascading failures. These failures are typically initiated by a small subset of network nodes. The redistributed data flow exceeds the capacity of other links and routers, resulting in a network outage \cite{cas_fail_power_review}. As 6G approaches, data flows may be congested due to diverse demands on network resources.
Table~\ref{t.op.res} summarizes 5G operational failures caused by network resource shortage.

The majority of existing solutions focus on model-based simulations or reenactments to anticipate cascading failures \cite{zhang2020analysis,9342800,cas_fail_network_1,cas_fail_network_2}. Alternative approaches include timely isolation countermeasures to prevent failures from spreading \cite{kaiser2021network} and optimizing resource scheduling and routing algorithms for efficient resource allocation and transfer in the network \cite{cas_fail_power_3}. 
Correspondingly, promising technologies such as network slicing~\cite{8994208}, featuring logically isolated network resources, and resource orchestration~\cite{8793221}, facilitating optimal resource distribution~\cite{9187796}, have emerged. In 6G, these technologies align with the vision set by the European Telecommunications Standards Institute (ETSI) Experiential Networked Intelligence (ENI). Fully automated network slices and resource orchestration, integrating AI and context-aware policies, are required to achieve this vision.
% These elements enable service delivery to adapt to time-varying user requirements, environmental conditions, and task objectives.
In \cite{Dandachi2022}, the authors contributed to this objective by dealing with the challenge of optimally placing dynamic virtual architectures via a self-adaptive learning-reliant policy.
In this evolving system, random and high-dimensional state and action spaces may not always align with real-time implementations, increasing the risk of network failures.

For coverage and capacity, 5G network operators are exploring small cells. It was found in \cite{Selim2022} that a challenge is cost-effectively backhauling the traffic from many gNBs to the core network. This small cell backhauling dilemma can be addressed with Integrated Access and Backhaul (IAB) using 5G NR, but densifying the network raises concerns about network reliability.
% Satellite-terrestrial edge computing networks (STECNs) have evolved as a global approach to sustain various IoT activities in the forthcoming 6G systems.
% Key enabling technologies, e.g., SDN, satellite edge computing (EC), and NFV, play a vital role in realizing the potential of STECNs.
A comprehensive analysis of network slicing (NS) for satellite-terrestrial MEC networks was presented in \cite{Esmat202314621}, including slice management and orchestration for hybrid architectures, satellite MEC, mmWave/THz, and AI schemes. Through MEC, modern 5G services can meet strict reliability and latency. However, the authors of \cite{Haber20216838} revealed that the inflexibility of ground-based MEC and its vulnerability to network infrastructure failures may hinder meeting these services' resiliency and strict demands.
UAVs can potentially provide flexible MEC capabilities through UAV-mounted cloudlets~\cite{li2021continuous,9922666,9387137}, reliable communications~\cite{9765746,9733205,9975284}, data collection~\cite{9875063,10100674}, and radio and video surveillance~\cite{10129074,9525335}, capitalizing on their mobility, cost-effectiveness, and LoS.
% \begin{figure}[htbp]
% \centering
% \includegraphics[width=8cm]  {59.png}
% \caption{\small The system model studied in \cite{Haber20216838} depicts IoT devices recurrently generating task offloading requests to UAV-mounted cloudlets. Each cloudlet serves as a server to deliver reliability in the face of potential failures. Tasks, characterized by input size, computation density, latency, and required reliability, follow Poisson processes. A UE is allowed to offload its tasks to multiple UAVs in parallel.} 

% \label{F:taxonomy.iot}
% \end{figure}
 %emami2021joint,li2022deep,li2015energy

Optimizing resources in 5G RAN is increasingly challenging in dynamic systems with many nodes and virtual network functions~\cite{8314676,8501940}. It was noted in \cite{DiCicco2022} that jointly optimizing multiple objectives while enforcing crucial application requirements, such as low latency, is essential. Furthermore, virtual network functions responsible for baseband processing are susceptible to cloud infrastructure failures, adding another layer of complexity.

Low latency and considerable computational resources are required for complex vehicular applications.  Vehicular MEC systems aim to address these challenges by enabling nearby vehicles and edge servers linked to BSs to share their computing and storage resources~\cite{8063331,8678697}. In practice, this technique is challenging due to the dynamic nature of network nodes, varying computing and energy loads, and rapid movement, resulting in frequent network failures.
With MEC~\cite{8274943} embedded in cellular-V2X (C-V2X), delay-sensitive services are offered to overcome vehicle resource limitations. As MEC servers are constrained in computing, storage, and communication resources, multi-domain resources must be orchestrated together\cite{Wang2023ssssss}.

\begin{figure}[t]
\centering
\includegraphics[scale=0.4]  {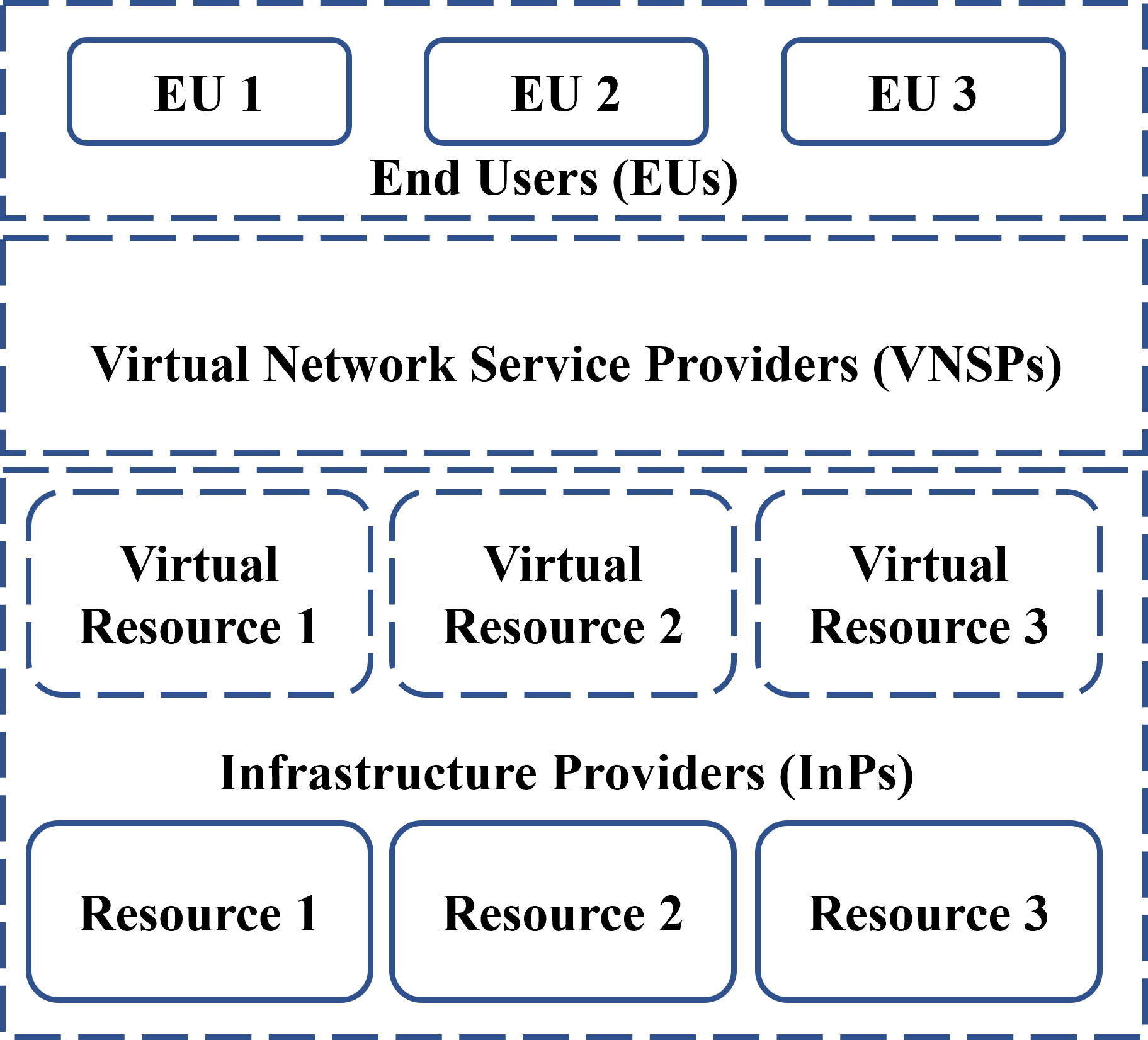}
\caption{\small The NFV-enabled network model defined in ~\cite{Cao202222492} with three major roles: Infrastructure Providers (InPs), Virtual Network Service Providers (VNSPs), and End Users (EUs). InPs handle physical network infrastructure and NFV technology. Network services are provided by VNSPs. EUs contract for these services. The model highlights the roles and responsibilities of each to prevent or mitigate failures.} 
\label{F:taxonomy.nfv}
\end{figure}

Vehicular Ad-hoc Networks (VANETs) are critical to 5G vertical applications. With NFV-enabled vehicular and 5G networks, all nodes, including vehicular, edge, and core components, can be fully virtualized. With NFV-enabled networks, virtual network services (VNS) can be offered with arbitrary topologies and customized resource demands~\cite{8570806}. The authors of \cite{Cao202222492} reveal that accidental failures of network elements, for example, nodes and links, can degrade the performance of VNSs; see Fig.~\ref{F:taxonomy.nfv}.
% Ensuring survivable services despite accidental network element failures is crucial for maintaining system reliability.

$\bullet$ \textbf{Power Failures}

The reliability of power is a fundamental concern essential for the smooth functioning of critical infrastructure. Despite the intelligent application of smart grid technologies to operate and control power delivery~\cite{9585464}, persistent challenges threaten the reliability of power communication networks. A particularly pressing issue is the recurring cascading failure in power grid networks, demanding specific attention \cite{cas_fail_power_3,cas_fail_power_5,cas_fail_power_6,cas_fail_power_7,Component_power_1,cps_compon_fail_1}. As previously discussed, component failures in one or several branches lead to power delivery redistribution due to the physical laws of circuit theory \cite{cas_fail_power_review}. The compensatory power flow can overload other branches attempting to fulfill the function of the failed branch, potentially triggering a complete power grid outage.

In future 6G networks, the integration of smart grids is expected to introduce novel solutions for Protection, Automation, and Control (PAC) in smart grids.  Innovative Fault Location, Isolation, and Service Recovery (FLISR) functions aim to enhance the responsiveness and coordination of grid defense mechanisms. Challenges, highlighted in \cite{NaitBelaid20225874}, suggest potential risks of power failures related to connectivity and energy consumption. Effective protection assets require crucial communication with FLISR functions, catering to both best-effort and URLLC services. 
Beyond connectivity  challenges, the energy consumption of contemporary telecommunication networks remains a concern.
% In addition to connectivity challenges, the energy consumption of modern telecommunication networks remains a critical factor in this context.

IoT sensors are crucial in reporting power equipment conditions, enabling preventive maintenance and repairs before failures occur. 
It was found in \cite{Ilahi2021} that monitoring Partial Discharge (PD) in a power system continuously and non-intrusively is vital for improving quality of service and preventing equipment damage. PD exhibits various measurable phenomena, with RF radiation being one of them. 
% Some PDs are characterized as Ultra-Wide Band (UWB), and their frequency spectrum may extend up to several GHz. 
Meanwhile, the authors of \cite{Vargas2023} investigated the outage failure of 5G arising in smart critical infrastructures. 

\subsubsection{Application Failures}

\begin{table*}[htbp]
\centering 
\caption{Application failures caused by resource depletion in 5G systems.} 
\begin{tabular}{p{1.5cm}<{\centering} | m{2.5cm} | m{3cm} | m{4cm} | m{1.5cm}<{\centering}} 
\hline 
\textbf{Failure type} & \textbf{5G Technology} & \textbf{Cause for Failure}  & \textbf{Countermeasure for failure} & \textbf{Perpetuate in 6G} \\ \hline 
 & MEC-based smart factory \cite{Qu2023} & Incidental disruption \cite{Qu2023} & Emergency offloading strategy scheme\cite{Qu2023}  & \checkmark \\ \cline{2-5} 
Service failure                &  Handover
management \cite{Prado20231845,Tayyab202299}  & Insufficient BS management \cite{Prado20231845}&  DQN BSs allocation mechanism \cite{Prado20231845}  & \checkmark \\ \cline{3-5}
 &     & Insufficient Radio Resource management\cite{Tayyab202299} & Discontinuous
Reception mechanism \cite{Tayyab202299}& \checkmark \\ \cline{1-5}
\end{tabular} 
\label{t.app.res}
\end{table*}

The practical deployment of 5G has revealed vulnerabilities to potential risks arising from unpredictable natural events or human-induced factors. These include environmental fluctuations, accidental disruptions or damages, and inefficient power management, all of which can significantly compromise 5G applications.
Table~\ref{t.app.res} summarizes the application failures caused by resource shortages in 5G systems.

$\bullet$ \textbf{Service Failures}

It was pointed out in \cite{Qu2023} that 5G has made it easier for smart factories to be realized. These factories use intelligent devices to monitor the environment, schedule production, and move autonomously. It might take longer to complete certain tasks if relying solely on their computational capabilities. These functions are delegated to edge servers and cloud servers. There are a number of risks that could lead to server resource failures, including natural disasters, network attacks, and hardware failures.
5G mobile networks for railways (5G-R) enhance reliability with overlapping coverage along railway. As a tradeoff, this improvement increases inter-cell interference.
The authors of \cite{Zhang202217936} explored how inter-cell interference affects the capacity of users situated at the edge of 5G-R systems with linear redundant coverage.  
% Furthermore, the study investigates coverage reliability, considering potential BS failures.

The authors of \cite{Haghrah2023} unveiled that the 3rd Generation Partnership Project (3GPP) standard for initiating handovers relies on comparing the quality of the received signal between the serving cell and its neighbors. Handover failures and inaccurate threshold values can compromise this process. Meanwhile, it was found in \cite{Gundogan2023} that handover failures in 5G-Advanced networks are caused by the implementation of handovers based on Layer1 measurements. Moreover, the authors of \cite{Prado20231845} pointed out that BSs are deployed at significantly higher density, resulting in more frequent handovers for users. 
% The seamless operation of 5G networks calls for advanced handover techniques to mitigate the risks of handover failures.

\subsection{Prevention of 5G Resource Depletion Failures}

Countermeasures have been proposed to address potential failures in 5G, focusing on mitigation or prevention in transmission, service, and network aspects. Their core mechanisms revolve around resilient resource allocation, incorporating strategies, such as backup transferring, autonomous reconfiguration, and self-healing.

$\bullet$ \textbf{Prevention of  Transmission Failures}

In \cite{Islam20231140}, a spatial-temporal correlation between radio communication and weather forecasts was considered to propose an LSTM-autoencoder-based communication link failure prediction scheme.
The authors of \cite{Li20232176} described an energy-efficient mechanism for delivering URLLC packets within a finite transmit power.
Using frequency-hopping and proactive dropping, this mechanism reduces the probability of uplink outages.
Addressing concurrent heterogeneous failures, the authors of \cite{Zheng20221} explored effective SFC delivery in edge networks.
The concept of $k$-heterogeneous fault tolerance was introduced, along with an enhanced protection graph called a $k$-connected service function slices layered graph (KC-SLG). 
% To address the problem of $k$-heterogeneous-faults-tolerant SFC embedding, an algorithm named fault-tolerant service function graph embedding (FT-SFGE) was introduced.
% FT-SFGE incorporates two innovative techniques: $k$-connected network slicing (KC-NS) and $k$-connected function slicing (KC-FS).

In the context of NFV, safeguarding against SFF failures is complex due to potential simultaneous failures of multiple SF instances resulting from a single SFF failure. According to \cite{Peng2023}, backup cost-effectiveness selection, backup auxiliary transferring, and adaptive fit backups are combined into a heuristic algorithm.
The authors of \cite{Zhang20235490} optimally combined the advantages of grant-based and grant-free random access, presenting a Hybrid Grant NOMA random access scheme. 
% This balances signaling overhead, access failure probability, and access delay.
The authors of \cite{Hasan20228895} introduced a resource-efficient Flow-Enabled Distributed Mobility Anchoring (FDMA) framework. As a result of varying parameters, such as the number of cells residence times and mobile routers, the performance of FDMA was assessed and compared with that of NEMO-BS and Proxy NEMO.

$\bullet$ \textbf{Prevention of Network Failures}

For the sake of mitigating and preventing the risks for cascading failures, a myriad of cascading failure analysis solutions have been proposed, e.g., model-based re-enactments solutions~\cite{zhang2020analysis,9342800,cas_fail_network_1,cas_fail_network_2},  isolation-based solutions~\cite{kaiser2021network}, and resource scheduling-based solutions~\cite{cas_fail_power_3}. In \cite{cas_fail_network_1}, the authors focused on cascading failures and proposed an invulnerability communication model to realize an optimal overall metric concerning performance, cost, and reliability. In \cite{kaiser2021network}, the authors proposed a general model framework to identify certain subgraphs to isolate the spread of cascading failure. In  \cite{cas_fail_power_3}, the authors proposed a novel communication network model to conduct congestion control for mitigating potential cascading failures.

% In \cite{Selim2022}, a self-healing strategy was employed to reduce or minimize the impact of backhaul failure. The utilization of IAB, in collaboration with neighboring gNBs, is proposed to counteract the effects of failed backhaul link(s). The objective is to establish a network prepared in advance to meet the minimum rate requirements for users, even during backhaul failures. A joint resource allocation/backhaul outage compensation optimization problem is formulated as a non-convex mixed-integer non-linear program. Given the complexity of this problem, it is bifurcated into two sub-problems, accompanied by applying various approximation and relaxation techniques to find a near-optimal solution.
% In \cite{Esmat202314621}, a robust NS design is developed to effectively manage failures and ensure uninterrupted service, irrespective of the integration architecture and across multiple domains. Strategies for achieving resilient networking and slicing in short-term evolution communication networks are outlined. These include preparing and allocating additional network resources, establishing guidelines for decomposing service level agreements, and employing cross-domain measures for detecting and addressing failures.
In \cite{Selim2022}, a self-healing strategy utilizing Integrated Access and Backhaul (IAB) and neighboring gNBs was proposed to mitigate backhaul failures, aiming to maintain minimum user rate requirements. This involves a complex optimization problem, divided into sub-problems and solved using approximation techniques. Esmat \textit{et al.} \cite{Esmat202314621} developed a robust Network Slicing (NS) design for resilient networking in short-term evolution communication networks, focusing on resource allocation, service level agreement decomposition, and cross-domain failure management.

The study in \cite{Haber20216838} revolved around offloading ultra-reliable low-latency computations with UAVs to facilitate future IoT services, mitigating potential failures with stringent requirements. UAV positions, offloading decisions, and resource allocations were optimized for serving requests while adhering to reliability and latency specifications. This problem was broken down into planning and operational stages. The planning stage involves optimizing UAV placement, while the operational stage involves optimizing offloading and resource allocation. Both stages are formulated as non-convex mixed-integer programs. A two-stage approximate algorithm is proposed to convert these into approximate convex programs.

The authors of \cite{DiCicco2022} presented the DUOpt algorithm for placing virtual functions in 5G-RAN. This algorithm solves multi-objective problems efficiently in medium to large networks, including static and dynamic traffic scenarios.
In \cite{Dandachi2022}, DRL and Monte Carlo methods were combined to embed virtual networks in mobile networks. In addition to providing solutions to network failures, it offers control-theory-based adjustments for exploration.
In \cite{deSouza20231}, bee colony-based task offloading was designed for task offloading in vehicular MEC systems. By scheduling tasks across servers, this algorithm reduces execution times.

% The authors of \cite{Cao202222492} proposed a dynamic virtual resource allocation method for NFV-enabled networks. Initially, they outline the business model and establish a framework for dynamic virtual resource allocation in the NFV-enabled networks. Secondly, they provide a detailed overview of all modules within the mechanism, emphasizing the initial resource allocation and re-allocation modules to ensure network services that can withstand disruptions, thereby mitigating the potential risk for network failures.

Redundancy is crucial in vehicle communication to prevent failures. Extensive redundancies, however, can increase costs. To balance reliability and cost efficiency, the authors of \cite{network_vehicle_1} explored the impact of network failure rates on overall performance. They aimed to assess and achieve an optimal configuration for redundancy. 
In scenarios involving C-V2X applications with dual dependencies on time and data, the authors of \cite{Wang2023ssssss} proposed an MEC hierarchical resource management framework.
By optimizing offloading, scheduling, and caching, this framework reduces system delays and prevents network failures. 
The approach has two parts: Resource management for a single MEC server using a scheduling algorithm and load balancing across multiple MEC servers.

$\bullet$ \textbf{Prevention of Power Failures}

To mitigate the recurrent cascading failure
in power grids mirrors that in communication networks, a myriad of model-based failure analysis solutions have been proposed\cite{cas_fail_power_5,cas_fail_power_6,cas_fail_power_7,cas_fail_power_8,cas_fail_power_9,cas_fail_power_10}. These works focus on the interdependence characteristic in smart grids to unveil the potential risks for cascading failure. In \cite{cas_fail_power_9}, the authors proposed a packet traffic model, comprehensively considering data packet network failures and power flow failures to investigate the interconnection between the two types of failures. Based on their proposed model, the authors of \cite{cas_fail_power_9} utilized a routing strategy to optimize the dispatching procedure for mitigating power flow failures.

 In \cite{NaitBelaid20225874}, the authors presented an energy-efficient route scheduler and link scheduler for 5G mobile network traffic. The problem is formulated as an ILP, and an optimal solution is provided. To ensure FLISR traffic adheres to the latency constraint and mitigate the risks of severe power failures, the objective is to determine the optimal trade-off between network throughput and energy consumption. 
 % Using a control flow application on top of an SD-RAN controller, the proposed approach aligns with the Software-Defined RAN (SD-RAN) paradigm.

The authors of \cite{Ilahi2021} concentrated on implementing a cost-effective and minimally intrusive method as a diagnostic tool for detecting Partial Discharge. Their innovative design solution introduces a UWB antenna designed specifically for 6G-IoT-based smart grid monitoring, operating within the frequency range of 3.02 GHz to 11.17 GHz. The antenna, featuring a cavity with five rectangular slots, demonstrates remarkable performance metrics, including a fractional bandwidth of 112.97\% and a maximum gain of 1.994 dB. The paper meticulously outlines the design parameters and presents simulation results, fostering a comprehensive discussion of its implications.

$\bullet$ \textbf{Prevention of  Service Failures}

In \cite{Gundogan2023}, a comprehensive system model was evaluated against baseline and conditional handover mobility procedures that are established for the higher layers. System-level simulations were employed to evaluate the handover failure risk of the lower-layer mobility procedure, and key performance indicators were used for comparisons with higher-layer handover mechanisms.
In \cite{Haghrah2023}, an approach based on fuzzy logic was presented to mitigate handover failures based on both serving and neighboring cells' estimated radio link quality (RLQ). For predicting RLQ for serving and neighbor cells, the system uses a second-order regressor and a simple fuzzy logic system. The final decision to trigger handover is based on a cascade fuzzy logic system, which addresses premature, delayed, and ping-pong handovers.

The concept of Radio Resource Management (RRM) relaxation was introduced in \cite{Tayyab202299}, focusing on optimizing UE power saving, particularly for UEs with ``low mobility'' and ``cell edge'' criteria. The study explored the benefits of RRM relaxation in conjunction with discontinuous reception (DRX) for reduced capability (RedCap) and new radio (NR) UEs. The impact on handover failures and packet delay was assessed.
In \cite{Prado20231845}, an optimization problem was formulated to achieve fairness in user data rates and minimize handovers. The study considered decisions on when to initiate a handover and which BS to assign to a user simultaneously. The proposed algorithm includes both a centralized and a multi-agent DQN-based approach. Comparative analysis against baselines demonstrated significant outperformance regarding handover failures, with performance consistently within 95\%.

The authors of \cite{Qu2023} introduced an emergency offloading strategy grounded in cloud-edge-end collaboration for smart factories. This strategy aimed to minimize both the total task execution delay and the critical task execution delay, forming an objective function. The resolution of this objective function was facilitated by a Fast Chemical Reaction Optimization (Fast-CRO) algorithm. Guided by the principle of prioritizing the offloading of crucial tasks during emergency scenarios, the algorithm swiftly made decisions for emergency offloading within the system.

\subsection{Potential Failures in 6G Systems}

\begin{table*}
\centering 
\caption{Potential operational failures caused by resource depletion in 6G.} 
\begin{tabular}{m{1.75cm}<{\centering} m{3.25cm} m{4.5cm} m{4.5cm}} 

\hline 
\textbf{Failure type} & \textbf{6G Technology} & \textbf{Cause for Failure}  & \textbf{Countermeasure for failure}  \\ \hline

& Visible light communication\cite{Chandran2022561}  & Variations in receiver performance under multi-
color channels\cite{Chandran2022561} &  Compensation  for distortions mechanism\cite{Chandran2022561}    \\ \cline{2-4}

&  Subnetworks \cite{Adeogun2021959} & Interference across shared frequency channels \cite{Adeogun2021959}&  Deep neural network based interference
mitigation mechanism\cite{Adeogun2021959}    \\ \cline{2-4}

& CRNs \cite{Khan20222726} & Incidental disruption for channels \cite{Khan20222726}&  Channel reservation
algorithm\cite{Khan20222726}   \\ \cline{2-4}

Transmission failure& CR-mIoT\cite{Abbas20227151}  & Limited network resources\cite{Abbas20227151} &  Idle channel prediction and
ranking algorithm\cite{Abbas20227151}   \\ \cline{2-4}

&  Air interface\cite{Lv20222831}   & Idle mode mobility \cite{Lv20222831}  &   double-layered flexible  architecture for mobility management\cite{Lv20222831}   \\ \cline{2-4} 

&  mmWave\cite{rca_fail_network_2}   & Inefficient management on extreme weather \cite{rca_fail_network_2}  &   LSTM-embedded RCA\cite{rca_fail_network_2}   \\ \cline{2-4} 
 
& URLLC and millimeter-wave \cite{Nishio202176}  & Channel blockage \cite{Nishio202176} &  Resilience of computer vision mechanism\cite{Nishio202176}    \\ \cline{1-4} 

& SDN \cite{Basu20216885} & Incidental disruption \cite{Basu20216885}&   Reverse path-flow mechanism\cite{Basu20216885}    \\ \cline{2-4}

&  Reconfigurable
wireless network slicing \cite{Khan2022} & Inefficient resource allocation \cite{Khan2022}&  Deep learning based resource allocation\cite{Khan2022}   \\ \cline{2-4}

Network failure&  Millimeter-wave
LAN \cite{Khan202231} & Overloading \cite{Khan202231} &  Hybrid deep learning-
enabled congestion control mechanism\cite{Khan202231}   \\ \cline{2-4}

&  Distributed intelligence \cite{Majumdar20222321} & Resource conflict \cite{Majumdar20222321}&  The cooperation based on of distributed intelligence\cite{Majumdar20222321}   \\ \cline{1-4}

\end{tabular} 
\label{t.po.op}

\end{table*}

 The coordination and integration of  resources to accomplish more sophisticated tasks pose a significant challenge in 6G. It is foreseeable that resource depletion failures will become prominent across diverse 6G scenarios.

\subsubsection{Operational Failures}
 It is crucial to highlight the unique challenges introduced by imbalanced requests in 6G, particularly concerning transmission, service, and network requirements. Given the envisaged significantly larger traffic in 6G, inefficient resource management could lead to severe blockages and resource wastage, causing failures.
Table~\ref{t.po.op} summarizes the potential operational failures caused by resource shortage in 6G.

$\bullet$ \textbf{Transmission Failures}

In visible light (VL) communication for 6G services, based on the multi-color channel between color LEDs and a photodiode, a desire for uniform communication performance across color channels exists. Typically, VL communication services under multiple color channels are utilized by a single VL receiver. It was found in \cite{Chandran2022561}  that the received signal experiences severe color distortion, potentially leading to channel transmission failure due to variations in receiver performance under multi-color channels. This distortion arose from photodiodes generating more electrical current in the red channel than in the green or blue channels.

% The mmWave technology is anticipated to be a pivotal component of future 6G networks, promising ultra-high throughput and ultra-low latency. In temporal microwave communication, ensuring reliability and identifying transmission failures is crucial. The authors of \cite{rca_fail_network_2} highlighted that failures in microwave communication, stemming from inefficient management during extreme weather conditions, pose a significant challenge in terms of discrimination. This difficulty arises from the automatic recovery mechanisms embedded in the devices themselves.

The authors of \cite{Adeogun2021959} pointed out that efficient deep learning-based methods can be used for interference mitigation, thereby mitigating failures
in independent wireless subnetworks. The focus was on dynamically allocating radio resources, treating resource allocation as a mapping from interference power measurements at each subnetwork to a class of shared frequency channels. 
Cognitive radio networks (CRNs) can enhance channel availability (CA) for primary and secondary users.  The authors of \cite{Abbas20227151} highlighted that the large-scale deployment of resource-constrained heterogeneous devices in CR-mIoT poses a challenge to the efficient utilization of limited device and network resources during the sensing process. However, the authors of \cite{Khan20222726} advocated that successful connection establishment is not guaranteed by CA alone; it also requires the assurance of receiver accessibility (RA) for mitigating potential failures.

\begin{figure}[t]
\centering
\includegraphics[width=6cm]  {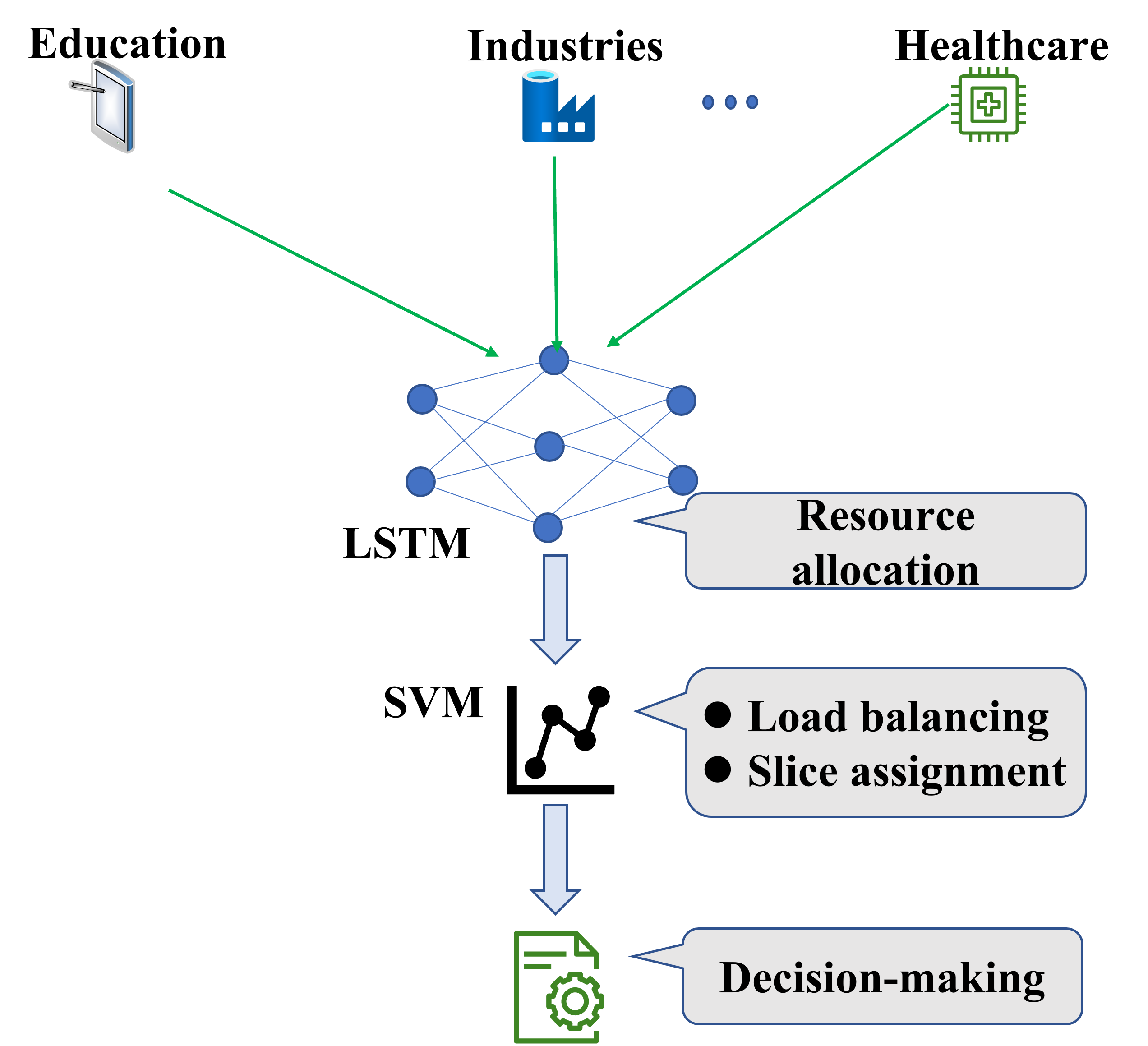}
\caption{\small The reconfigurable network slicing model developed in~\cite{Khan202231} for 5G/6G networks, which combines LSTM and Support Vector Machine (SVM) for intelligent decision-making. The LSTM handles resource allocation, while the SVM manages load balancing and alternate slice assignments in the event of failures caused by resource shortages. This model demonstrates high accuracy in various scenarios, validating its effectiveness in managing network traffic and slice allocation.}
\label{F:slicing}
\end{figure}

User-centric cell-free networking is a promising technology that ensures a ubiquitous user experience by dynamically grouping transmission points to create a user-specific cell. It is believed to be an effective remedy for single-point failures. However, it was found in \cite{Lv20222831} that practical deployment faces obstacles, such as computational complexity, signaling overhead, and challenges in idle mode mobility management. Additionally, considering the likely persistence of the 5G air interface in 6G, careful consideration must be given to backward compatibility.

In 6G systems, the authors of \cite{Nishio202176} introduced a groundbreaking intersection between computer vision and wireless communication. This fusion was crafted to { mitigate potential link failures, thereby empowering} mission-critical applications, such as autonomous and remote-controlled vehicles and visual-haptic virtual reality experiences. The collaboration between computer vision and wireless communication, fueled by recent advancements in machine learning (ML) and the accessibility of non-radio-frequency (NRF) data, was highlighted as a  catalyst for applications in B5G/6G.

The study \cite{Nishio202176} illustrated a significant improvement in wireless communication reliability while maintaining spectral efficiency. Particularly noteworthy is the role of computer vision as a vital tool for prediction in scenarios involving millimeter-wave channel blockages. This capability allows for the anticipation of blockages before they actually occur, contributing to a proactive approach to managing communication challenges.

\begin{table*}
\centering 
\caption{Potential application failures caused by resource depletion in 6G systems.} 
\begin{tabular}{m{1.75cm}<{\centering} m{3.25cm} m{4.5cm} m{4.5cm}} 

\hline 
\textbf{Failure type} & \textbf{6G Technology} & \textbf{Cause for Failure}  & \textbf{Countermeasure for failure}  \\ \hline 

&  Cloud-edge networks \cite{Shakeel2021969} & Inefficient management on dynamic environments \cite{Shakeel2021969}&  Multiple
Machine Access Learning with Collision Carrier Avoidance\cite{Shakeel2021969}    \\ \cline{2-4}

&  Terahertz communication \cite{Manogaran202214644} & Imbalanced service allocations \cite{Manogaran202214644} &   Service virtualization and flow management framework\cite{Manogaran202214644}   \\ \cline{2-4}

&  SDN \cite{Ortin20221287} & Imbalanced traffic demands \cite{Ortin20221287} &   Generic auto-scaling mechanism framework\cite{Ortin20221287}   \\ \cline{2-4}

Service failure &  Resilient LB \cite{Sarkar2022367} & Incidental  dynamic condition \cite{Sarkar2022367} &    Efficient user request handling mechanism\cite{Sarkar2022367}   \\ \cline{2-4}

&  MEC \cite{Sun202012240} & User mobility and the
volatile MEC environment \cite{Sun202012240} &    Digital Twin Edge Network based mobile offloading scheme\cite{Sun202012240}   \\ \cline{2-4}

&  Virtualization \cite{Mogyorosi20222453} & Sudden traffic
fluctuations \cite{Mogyorosi20222453} &    Latency-aware dual hypervisor placement and control
path design method\cite{Mogyorosi20222453}   \\ \cline{2-4}

&  MEC\cite{Saravanan2022851}  & User
mobility and the volatile MEC environment\cite{Saravanan2022851}  &  Lyapunov approach for optimization and
enhanced Actor-Critic learning combined with Digital Twin\cite{Saravanan2022851}   \\ \cline{2-4}

& URLLC \cite{Ganjalizadeh20224208}   & Random channel fluctuations \cite{Ganjalizadeh20224208}&  Deployment of distributed artificial intelligence\cite{Ganjalizadeh20224208}  \\   \hline

\end{tabular} 
\label{t.po.app}

\end{table*}

$\bullet$ \textbf{Network Failures}

The authors of \cite{Basu20216885} indicated that long-distance communication links pose a significant challenge to the reliability and resilience of cyber-physical systems (CPSs) in 5G and anticipated 6G networks. The quality index of network latency is at risk of disruption, leading to potential failures. Moreover, centralized network architectures prevalent in these systems exhibit low fault tolerance and susceptibility to security threats. Recognizing these vulnerabilities, virtualized software-defined network (vSDN)-enabled 5G networks address these issues. The authors of \cite{Basu20216885} redefined the existing network topology, strategically deploying controller and hypervisor instances to enhance overall reliability and security. The authors of \cite{Khan2022} suggested that implementing ML-enabled reconfigurable wireless network solutions becomes imperative for establishing a smart decision-making mechanism in network management and mitigating network slice failures.

% Future-generation networks, including millimeter-wave LAN, broadband wireless access systems, and 5G/6G networks, underscore the need for heightened security, low latency, and increased reliability and communication capacity. Within this context, efficient congestion control emerges as a key element enabling operators to run multiple network instances using a single infrastructure to enhance the overall QoS. AI/MLs play crucial roles in reconfiguring and optimizing the performance of 5G/6G wireless networks.
It was found in \cite{Khan202231} and illustrated in Fig.~\ref{F:slicing} that integrating a smart decision-making mechanism is essential for managing incoming network traffic, ensuring load balancing, restricting network slice failures, and providing alternative slices in case of failure or overloading.
The advent of 6G networks foresees a significant role for distributed automation in network management. This approach circumvents the drawbacks of a single point of failure and the signaling overhead inherent in a centralized paradigm. The authors of \cite{Majumdar20222321} found that conflicts arise in a distributed architecture, potentially impairing system Key Performance Indicators (KPIs). Considering the conflict, questions remain regarding the scalability of distributed automation to fully realize the potential of 6G networks.

\subsubsection{Application Failures}

 Vulnerability to factors like atmospheric attenuation poses risks that may hinder the widespread adoption of 6G. Proactively addressing these concerns is crucial.
Table~\ref{t.po.app} summarizes the potential application failures caused by resource shortage in 6G.

$\bullet$ \textbf{Service Failures}

Cloud computing serves as a critical technology, providing a broad pool of elastic resources to consumer appliances\cite{phy_vir_mixed_failure_1}. The authors of \cite{Shakeel2021969} pointed out that the heterogeneous network encounters communication collisions, which detrimentally impacts overall network performance. In anticipation of addressing these challenges, future cloud-edge networks are envisioned to accommodate a diverse array of clients and servers, including those in the IoT and 6G networks. Flexibility in solutions becomes paramount for the effective management of such dynamic environments.

Moving into the 6G communication leads to high interoperability through terahertz data transfer and latency-less service sharing. The interoperable nature of 6G allows for the seamless integration of heterogeneous networks, such as the IoT and cloud RANs (CRANs). This integration is aptly managed by deploying SDNs to mitigate potential risks for failures and ensuring a consistent QoS experience for users, regardless of the specific application in use\cite{Manogaran202214644}.

As mobile networks undergo softwarization, optimizing resource utilization becomes paramount. This involves dynamically scaling and re-assigning resources in response to variations in demand. The authors of \cite{Ortin20221287} pointed out that striking a right balance between efficiently anticipating traffic demands, preventing service disruptions, and avoiding the wasteful activation of surplus servers becomes crucial, particularly in the context of the stringent reliability requirements of 5G applications and the inherent fallibility of servers. 
Within the context of scalable 5G Core (5GC), the significance of efficient Load Balancers (LBs) cannot be overstated. It was pointed out in  \cite{Sarkar2022367} that inefficiencies in LBs at any Network Function (NF) can lead to a catastrophic failure of the entire system, resulting in a complete disruption of High Availability (HA) services.

Envisioning the 6G landscape, wireless communication and computation take center stage through the digitalization and connectivity of everything, e.g., MEC is a key enabling factor. The authors of \cite{Saravanan2022851} found that in MEC, a key enabler for mobile downloads, considerable challenges emerge due to the dynamic and unpredictable 6G network environment. Existing literature has overlooked the implications of user mobility and the volatile MEC environment. The authors of \cite{Sun202012240} came to the conclusion that a notable gap in existing literature lies in the oversight of the impacts of user mobility and the volatile MEC environment. Mission-critical and tactile Internet applications run on the same physical infrastructure. It was found in \cite{Mogyorosi20222453} that network hypervisors, enabling such virtualization, must exhibit resilience to failures and adapt to sudden traffic fluctuations instantaneously. This preparedness becomes crucial in the face of unpredictable environmental changes. Remarkable advancements have been achieved in communication services through the application of distributed AI, spanning fault-tolerant factory automation to smart cities. As pointed out in \cite{Ganjalizadeh20224208}, the execution of distributed learning across a network of connected wireless devices encounters challenges stemming from random channel fluctuations and simultaneous operations of incumbent services on the same network, impacting the efficacy of distributed learning.

\subsection{Prevention of 6G Resource Depletion Failures}

Various countermeasures have been proposed in response to potential failures in 6G systems, particularly focusing on transmission and service aspects. The basic mechanisms involve designing resilient resource allocation and scheduling strategies to address practical incidental disruptions effectively. Key strategies include dynamic offloading, hierarchical resource management, joint routing and link scheduling, and dynamic virtual resource allocation.

$\bullet$ \textbf{Prevention of  Transmission Failures}

In addressing distortions in multi-color channel-based VL 6G communication service,  the authors of \cite{Chandran2022561} proposed an ML-based solution. This approach effectively estimated and compensated for distortions across different color channels. According to \cite{Chandran2022561}, the solution demonstrated its capability to overcome communication failures within the entire range of communication distances through compensation for various color channels. In \cite{Adeogun2021959}, a DNN was trained to approximate a mapping obtained through the centralized graph coloring (CGC). This trained network was subsequently deployed at each subnetwork for distributed channel selection.

Addressing channel failures in ultra-reliable communication within 6G IoT, the authors of \cite{Khan20222726} emphasized the necessity of the receiver's accessibility for successful connection establishment. The authors proposed a channel reservation algorithm that optimizes spectrum resource utilization efficiency while considering receiver accessibility. They further predicted and ranked idle resources, offering dynamic mitigation against the detrimental consequences of channel failures.

The work in \cite{Abbas20227151} introduced a novel multiparameter-based flexible scheme for idle channel prediction and ranking, accounting for user priorities and heterogeneity. Using a probabilistic approach and simultaneous consideration of multiple parameters, the scheme evaluated channel suitability before selection for transmission. It addressed the challenge of channel obsolescence inherent in channel prediction and ranking. In \cite{rca_fail_network_2}, an LSTM-embedded RCA approach was developed to discriminate transmission failures in microwave communication. The approach can leverage environmental information and network data to comprehensively make judgement on failure causes. It was reported in \cite{rca_fail_network_2} that the proposed LSTM-embedded RCA can achieve shows 95\% accuracy.

In the domain of RAN architecture, the authors of \cite{Lv20222831} proposed a double-layered flexible architecture to mitigate risks for transmission failures. This architecture provided multiple transmission points with joint processing gains similar to user-centric cell solutions but with lower complexity and backward compatibility of the 5G air interface. The architecture decoupled access and service network functions, with an access layer for traditional cellular network functions and a service layer for serving users using a virtual, flexible cell dynamically formed by multiple transmission points. A mobility management algorithm based on trajectory prediction, leveraged user similarity within clusters.

The authors of \cite{Nishio202176} underscored the significance of RF-based sensing and imaging in fortifying the resilience of computer vision applications against occlusion and failure. 
To exemplify these concepts, they presented a case study involving RF-based image reconstruction. This use case highlights the correction of image failures on the receiver side, resulting in reduced retransmission needs and lower latency. By emphasizing the convergence of RF and non-RF modalities, the authors of \cite{Nishio202176} advocated for a transformative approach to enable ultra-reliable communication and the realization of truly intelligent 6G networks.

$\bullet$ \textbf{Prevention of  Network Failures}

Root cause analysis (RCA) has been used as a powerful auxiliary solution for identifying the cause inducing the network failures\cite{rca_fail_network_1,rca_fail_network_2,rca_fail_network_3}. The traditional RCA generally relies on the formulated rules based on expert knowledge to understand the cause of failure. However, the solution relying on expertise is comparatively inefficient to employ experts with different discipline backgrounds for formulating identification rules, not to mention the coverage of rules reliant on manpower. The more complicated failure types in future 6G scenarios could further exacerbate the weakness of traditional RCA techniques. 

In recent years, a myriad of AI-based RCA solutions, e.g., LSTM~\cite{rca_fail_network_2} and CNN~\cite{rca_fail_network_1}, relying on the statistical historical information have emerged. These AI-based approaches could comprehensively utilize the historical data collected in the operation process to extract useful information, and  make judgments on the cause of failures. 

In \cite{Basu20216885}, an approach was proposed to dynamically deploy controller-hypervisor (C-H) pair(s) for various network functions, such as differentiating between control and data signals, implementing various translation functions, etc., with ultra-low latency. The system model employed four well-defined network latency matrices. A mixed-integer linear programming model was utilized to optimize latency objectives to mitigate failure risk. The resulting reverse path-flow mechanism (RPFM) ensured feasible solutions by maintaining network load and controller capacity within tolerance limits. 

To address network management challenges in 5G and 6G networks, a hybrid deep learning model was proposed in \cite{Khan2022}, combining CNN and LSTM. The CNN handled resource allocation, network reconfiguration, and slice selection, while the LSTM managed statistical information related to network slices, such as load balancing and error rates. The model's applicability was validated under various conditions, including unknown devices, slice failures, and overloading. The authors of \cite{Khan202231} introduced a hybrid deep learning-enabled congestion control mechanism, incorporating LSTM and SVM. Its effectiveness was demonstrated through simulations over a one-week period, considering scenarios with multiple unknown devices, slice failures, and overloads.
The scalability of distributed intelligence, specifically based on Q-learning, was validated in \cite{Majumdar20222321} through Q-learning for Cooperation (QLC) framework. QLC comprised intelligent agents cooperating on a discrete state space. The results indicated that QLC  scales well compared to the optimal solution computed by a centralized approach. These findings suggest the promising applicability of QLC to other use cases in 6G, especially when convergence speed is not a significant concern.

$\bullet$ \textbf{Prevention of  Service Failures}

In the 6G IoT, the authors of \cite{Shakeel2021969}  developed a Multiple Machine Access Learning with Collision Carrier Avoidance (MMALCCA) protocol to enhance communication service effectiveness. Utilizing the THz band, this protocol employed a Media Access Control (MAC) protocol for synchronization in high-speed wireless communication networks. The protocol leveraged a classification and regression learning method to make decisions, enhancing the efficiency of MAC synchronization. In \cite{Manogaran202214644}, a Service Virtualization and Flow Management Framework (SVFMF) was optimized for resource utilization in a 6G-cloud environment. Addressing imbalances in service requests and responses due to overloaded and idle virtual resources, SVFMF introduced service virtualization,  user allocation modules, and a linear decision-making process to identify overloaded services for reallocation.

An analysis of a generic auto-scaling mechanism for communication services was presented in  \cite{Ortin20221287}, focusing on server activation and deactivation based on occupation thresholds. The impact of activation delay and finite server lifetimes on power consumption and failure probability was modeled. An algorithm for optimal threshold configuration was derived from this model.
The LOCOMOTIVE 5GC~\cite{Sarkar2022367} was introduced as a resilient alternative to traditional hot standby configurations. Outperforming hot standby in HA and resilience under dynamic conditions, LOCOMOTIVE demonstrated superior user request handling during LB failures. Feasibility was validated in a 3GPP-compliant 5G testbed, showcasing its availability and resilience.

A vision of Digital Twin Edge Networks (DITEN)  was presented in \cite{Sun202012240}, where digital twins of edge servers estimate states. The digital twins of the entire MEC system provided training data for offloading decisions. The mobile offloading scheme in DITEN minimized latency to mitigate the risk of failures while considering accumulated service migration costs during user mobility. Leveraging Lyapunov optimization and Actor-Critic DRL, the scheme circumvented  long-term migration cost constraints. The studies in \cite{Saravanan2022851} and \cite{10234427} contributed to training data for offload decisions in digital edge servers and evaluating the edge servers' status and Digital Twin for MEC environment. The systems reduced download delay while considering the cumulative expense of service relocation for user mobility. Leveraging Lyapunov optimization and enhanced Actor-Critic learning, the systems can reduce average offload delay, failure rate, and operation migration rate.

A latency-aware dual hypervisor placement and control path design method was proposed to protect against single-link and hypervisor failures in \cite{Mogyorosi20222453}. The methodology, ready for unknown future changes, addressed NP-hard challenges with optimal and heuristic algorithms. Simulations demonstrated the efficiency of the method in real-world optical topologies.
In \cite{Ganjalizadeh20224208}, the interaction between a concurrently operating distributed AI workflow and URLLC services was investigated over a network. Through 3GPP-compliant simulations within a factory automation use case, the impact of different distributed AI settings, e.g., model size and the number of participating devices,  was investigated on the convergence time of distributed AI and the application layer performance of URLLC. Simulations indicated a substantial impact of distributed AI on the availability of URLLC unless 5G-NR QoS handling mechanisms were utilized to segregate traffic from the two services.

\section{Security-Related Failures}
\label{sec_V}

\begin{table*}
\centering 
\caption{Security-related operational failures inherited from 5G systems.} 

\begin{tabular}{m{1.5cm}<{\centering} m{3.5cm} m{3cm} m{4cm} m{1.5cm}<{\centering}} 
\hline 
\textbf{Failure Type} & \textbf{5G Technology} & \textbf{Cause of Failure}  & \textbf{Countermeasure for Failure} & \textbf{Perpetuate in 6G} \\ \hline

& 5G authentication protocol \cite{Munilla2021} &  location confidentiality attacks\cite{Munilla2021} & Enhancement to the symmetric-key authentication and key agreement protocol\cite{Munilla2021}  & \checkmark \\ \cline{2-5}

&  MTC \cite{Yan2022} & Lack of mutual authentication \cite{Yan2022} & Key forward
secrecy authentication protocol \cite{Yan2022} & \checkmark \\ \cline{2-5}
&  Zero Touch Management \cite{BenSaad20231612} & Poisoning attacks \cite{BenSaad20231612} & Deep reinforcement based dynamical trustworthiness mechanism \cite{BenSaad20231612} & \checkmark \\ \cline{2-5}
Authentication failure &  Precision time protocol \cite{Shi202361} & Malicious insiders \cite{Shi202361} & Time crowdsourcing based Byzantine-resilient network  \cite{Shi202361} & \checkmark \\ \cline{2-5}
& Vehicular Networks \cite{Yan20221678} & Lack of mutual authentication \cite{Yan20221678} & Specific authentication protocol \cite{Yan20221678} & \checkmark \\ \cline{2-5}
& Industrial Automation \cite{Xu20226368} & Complicated cross-layer devices \cite{Xu20226368} & Quantum encryption \cite{Xu20226368} & \checkmark \\ \cline{2-5}
& Internet of Drones \cite{Feng20226224} & Incapability of cross-domain authentication \cite{Feng20226224} & Blockchain-based authentication mechanism \cite{Feng20226224} & \checkmark \\ \hline
Transmission failure &  Handover
management \cite{Yang202313959}  & Inefficient protocol in satellite and terrestrial networks \cite{Yang202313959} & Handover authentication protocol \cite{Yang202313959}& \checkmark \\ \cline{2-5}
 &  Small BSs \cite{Ming2022} & Security threats \cite{Ming2022} &  Semi-supervised learning-based framework \cite{Ming2022}  & \checkmark \\ \cline{1-5}
&  Multidimensional resources coding \cite{Xu20212429} & Uncertainty of attacks \cite{Xu20212429} &   Quantum learning-based nonrandom superimposed coding method\cite{Xu20212429}  & \checkmark \\ \cline{2-5}
Network failure&   Slice \cite{Wijethilaka2022915} &  Deficiency of FL \cite{Wijethilaka2022915} &   Coordinated security
orchestration architecture \cite{Wijethilaka2022915}  & \checkmark \\ \cline{2-5}
& IoT-Enabled Cloud Manufacturing \cite{Hewa20227174}& Incidental disruption for device \cite{Hewa20227174} & Blockchain-based and fog-computing-enabled security service mechanism \cite{Hewa20227174} & \checkmark \\ \hline 
\end{tabular} 
\label{t.secur.op}
\end{table*}

In communication systems, ``security" generally refers to the overall state of protection against unauthorized access, attacks, and potential breaches. It encompasses measures and practices designed to safeguard information, systems, and communication channels from threats. Security measures include encryption, authentication, access control, firewalls, and other mechanisms to ensure the confidentiality, integrity, and availability of data and services.

``Security-related failures" specifically denote instances where the security mechanisms or protocols in a communication system fall short, resulting in vulnerabilities, breaches, or unauthorized access. These failures can manifest as weaknesses in encryption algorithms, flaws in authentication processes, susceptibility to specific types of attacks, or other shortcomings that compromise the intended security posture.

\subsection{Failures Inherited from 5G}
In anticipation of stringent security requirements in 6G, enhancing authentication mechanisms is crucial, and many security-related authentication failures observed in 5G are anticipated to persist in diverse 6G scenarios~\cite{hu20}.

\subsubsection{Operational Failures}
% {\color{blue}
 In 5G, authentication failures have revealed vulnerabilities in location confidentiality. 
Table~\ref{t.secur.op} summarizes the security-related operational failures inherited from 5G systems.

$\bullet$ \textbf{Authentication Failures}

Symmetric-key authentication and key agreement (AKA) protocols, such as those developed in~\cite{7931680} and \cite{braeken2020symmetric}, underpin security architectures, relying on the exchange of failure messages for mutual authentication in 5G. However, vulnerabilities, particularly in location confidentiality, have been identified~\cite{Munilla2021}. 
Recent research, exemplified by~\cite{Taskou20222328,8989788,10.1016/j.ipm.2021.102492,DUAN2023102897,MAKHDOOM2019251}, harnesses Blockchain technology to decentralize authentication in 5G NFV, IoT, and general cloud platforms. In 5G and B5G, FL within the ZSM concept is vital. Nevertheless, FL faces construction failures due to poisoning attacks, posing a significant threat to slice management~\cite{zheng2021federated,BenSaad20231612,zheng2022exploring}. 
Despite the  PTP's role in achieving time synchronization in 5G, it remains susceptible to Byzantine failures from malicious insiders~\cite{Shi202361}. Anticipating stringent security requirements in 6G and enhancing existing authentication mechanisms to ensure reliable communication will be crucial. 

\begin{figure}[htbp]
\centering
\includegraphics[width=8cm]  {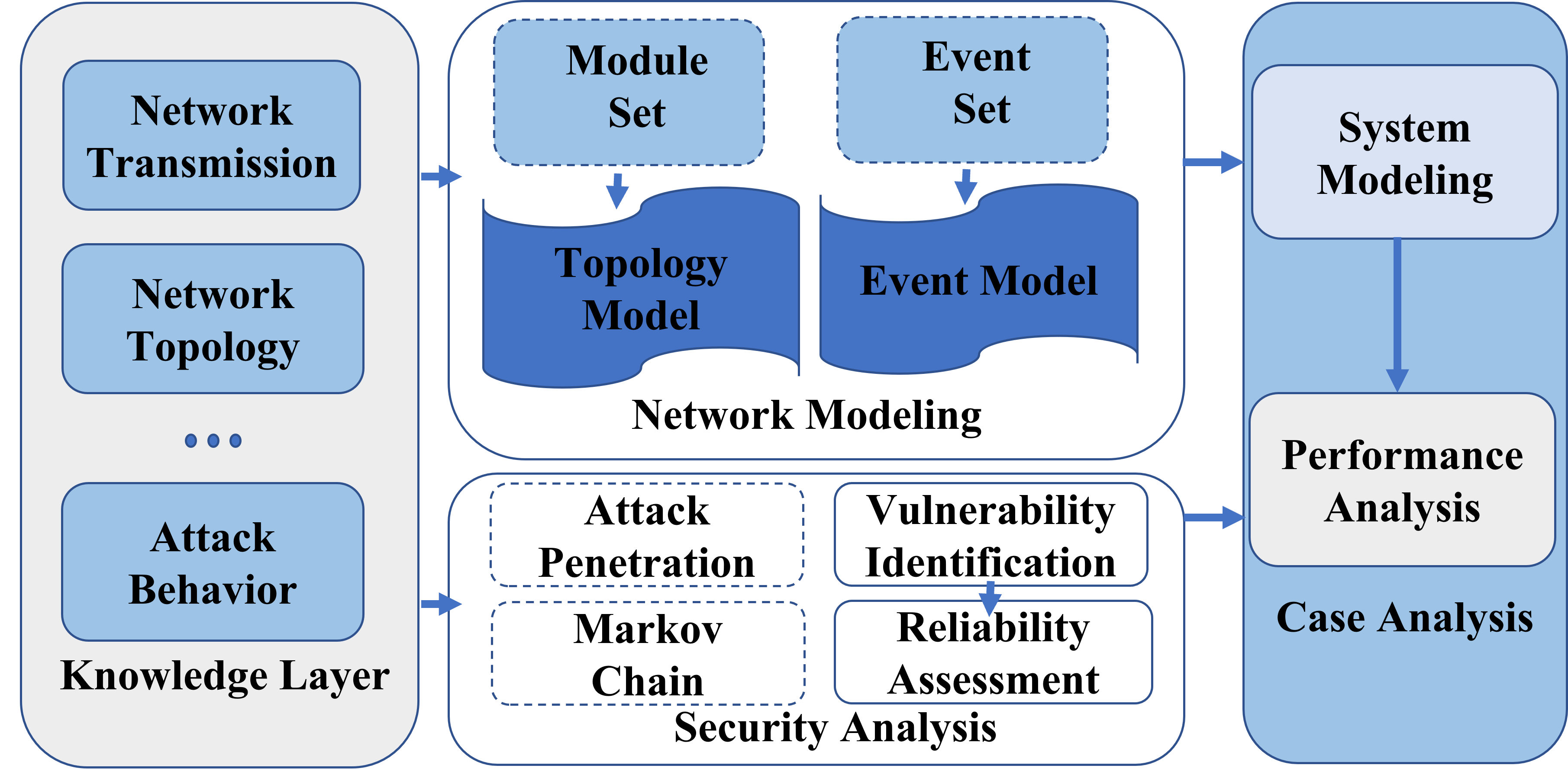}
\caption{\small Model-driven security analysis architecture developed in~\cite{Li2023889} for 5G networks. This architecture offers a comprehensive security analysis in 5G networks, incorporating network modeling and case studies on network impacts. It details the structure of the 5G network, including access, bearer, and core networks, emphasizing the need for topological modeling to understand and mitigate the risks of transmission failures due to potential attacks in various network domains.}
\label{F:Li2023889}
\end{figure}

The authentication failure in 5G VANETs, resulting from a lack of mutual authentication, has been thoroughly examined in \cite{Yan20221678,Yan20233104}. Similarly, the authentication failure in 5G-enabled Internet of Drones, stemming from the incapability of cross-domain authentication, is investigated in \cite{Feng20226224}. 
Within the 5G-enabled industry, the complexities introduced by cross-layer devices leading to authentication failures were explored in \cite{Xu20226368}. The authors of \cite{Li2023889} specifically evaluated authentication failures by considering both attack and defensive models, as depicted in Fig.~\ref{F:Li2023889}.

$\bullet$ \textbf{Transmission Failures}

Recent research~\cite{shahriar2014phy} has shown that pilot-based mechanisms in 5G can suffer from a high risk of pilot-aware attack, a physical-layer threat that can acquire, jam, spoof and null pilot sequences of interest. Paralyzing uplink access through tampering with pilots, preferred by attack, is easier and more efficient than directly disturbing data transmission. 
The requirements for ensuring physical layer security will be foreseeable in 6G. The security-related transmission failure issue arising from 5G will continuously perpetuate in all kinds of 6G technical scenarios. 

$\bullet$ \textbf{Network Failures}

Network slicing, a foundational feature of 5G, allows multiple virtual networks \cite{Thiruvasagam20212502,Thiruvasagam20211491} to operate on a single physical infrastructure tailored to different services or customer needs. However, this granularity introduces security concerns. The existing works on FL-empowered 5G network issues have investigated privacy leakage failures~\cite{Liu202024}. The joint research on both FL and network slicing~\cite{Wijethilaka2022915} has also investigated privacy leakage failures. 

UAV networks have played an important role in 5G systems and are envisioned to be continuously a crucial technology in 6G~\cite{cps_secu_fail_3,bi_tifs_2024,kurunathan2023machine,li2020joint}. However, the communications among UAVs and between a UAV and ground equipment are vulnerable to eavesdropping, jamming, and blockage, thereby leading to  security failures~\cite{Bastami20215018,li2019energy, yuan18capacity}.
 UAV-enabled target tracking can suffer from loss of Global Positioning System (GPS) signals and visual blockage,
leading to network failures~\cite{hu21}.

The requirements for establishing a space-air-ground integrated network and realizing global coverage and full application bring new challenges in 6G~\cite{Qiu2020Multiple, Chen2020Performance}.
How to effectively exploit  UAV networks for secure relay communication, safeguard edge networks and network endogenous security in diversified 6G scenarios will become more prominent.
The security-related network failures arising from 5G will continuously perpetuate in all kinds of 6G technical scenarios~\cite{li2021bloothair}.
The single point failure in 5G Industrial IoT-Enabled Cloud Manufacturing has been investigated in \cite{Hewa20227174}.

\subsubsection{Application Failures}

Many critical applications empowered by 6G systems are also vulnerable to security breaches and failures, including vehicular networks or IoT, due to their distributed network architecture and subsequently enlarged attack surfaces. Following are some of the latest discussions on application failures resulting from security breaches in 5G systems.

\subsection{Prevention and Defense of 5G Security Failures}

This section delves into prevention and mitigation techniques for authentication, transmission, and network failures in security-related failures associated with 5G systems. Techniques include symmetric-key protection, handover authentication protocols, frameworks for detecting sleeping cell failures, and quantum learning-based coding methods.

$\bullet$ \textbf{Prevention of  Authentication Failures}

Achieving complete unlinkability and mitigating susceptibility to failure message attacks can be accomplished through symmetric-key protection. However, this method introduces a trade-off between privacy and availability, potentially rendering the protocol vulnerable to DoS attacks. In a recent study \cite{Munilla2021}, an enhancement to the symmetric-key authentication and key agreement protocol was proposed,  involving updating the shared key after each successful authentication, providing a potential solution to the identified challenges, and offering an improved balance between privacy and availability.

To address authentication failures in FL within B5G networks, a novel approach was presented in \cite{BenSaad20231612}. The authors proposed a deep reinforcement learning framework that dynamically selects a trusted participant and employs unsupervised learning to identify malicious participants, contributing to enhanced security in FL within B5G networks.
Responding to Byzantine failures within 5G, Shi \textit{et al.}~\cite{Shi202361} strategically adopted time crowdsourcing to design a Byzantine-resilient network. This approach aims to bolster the network's resilience in the face of Byzantine failures. 
To combat key forward secrecy failure in MTC resulting from a lack of mutual authentication, Yan \textit{et al.}\cite{Yan2022} introduced an authentication protocol applicable to all handover scenarios of MTC.

The existing solutions for the authentication failure in 5G-enabled VANETs or  Internet of Drones mainly consist of designing specific authentication protocol~\cite{Yan20221678}, and blockchain-based authentication mechanism~\cite{Feng20226224,Gupta20211712,Xue20225284}. Quantum encryption~\cite{Xu20226368} has also been proposed to mitigate authentication failures in the 5G-enabled industry.

$\bullet$ \textbf{Prevention of  Transmission Failures}

\begin{figure}[htbp]
\centering
\includegraphics[width=8cm]  {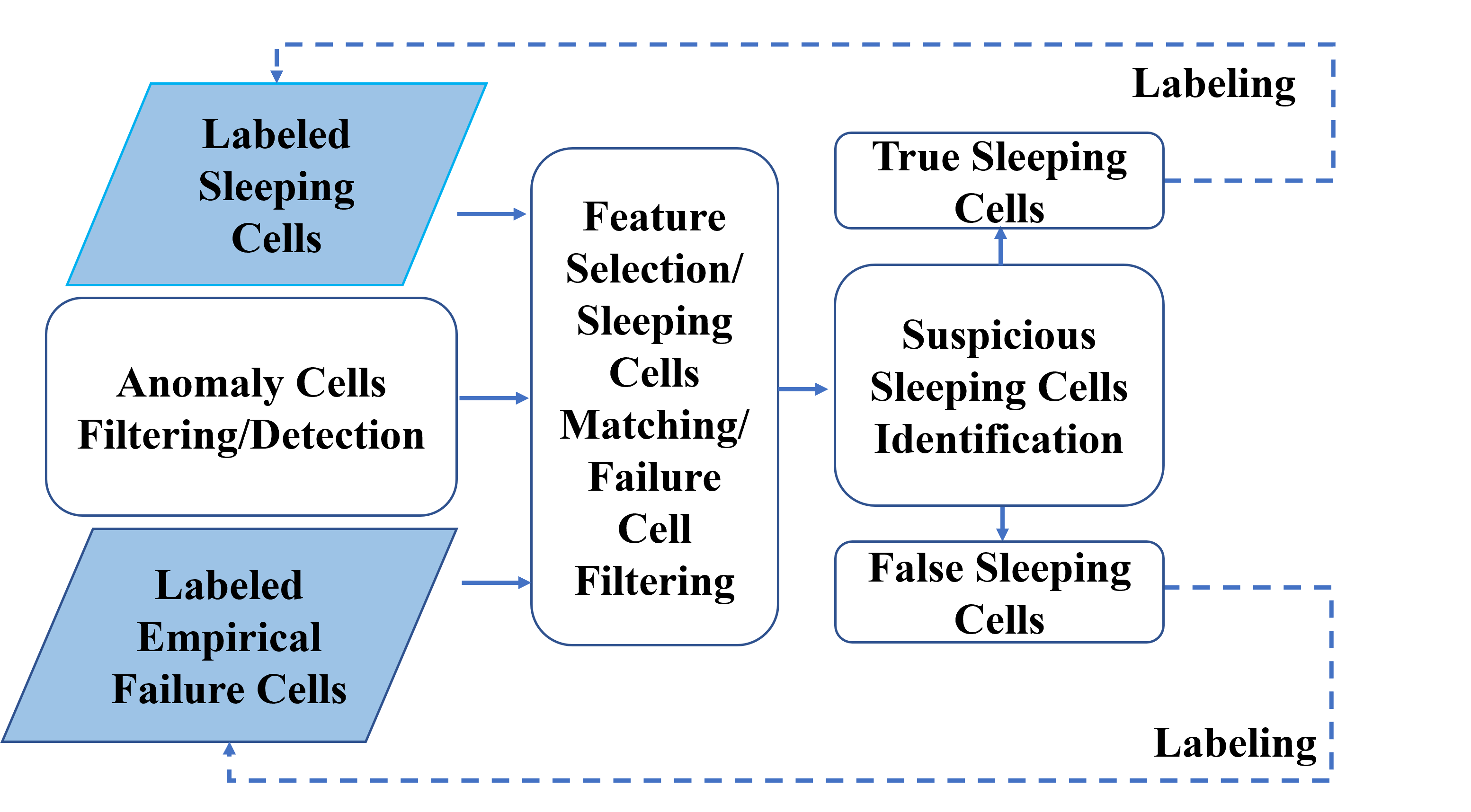}
\caption{\small The semi-supervised learning framework developed in~\cite{Ming2022} for sleeping cell detection. This frame is a multi-step framework combining anomaly detection, feature selection, classification, and clustering for sleeping cell detection. The framework uniquely utilizes unlabeled KPI data for clustering after each detection period, enriching training data for failure detection.}
\label{Ming2022}
\end{figure}

Handover failures in satellite and terrestrial networks were addressed in \cite{Yang202313959}. A handover authentication protocol was proposed to facilitate high-speed rail (HSR) connectivity. In the context of 5G small BSs (SBSs), the authors of \cite{Ming2022} directed attention to sleeping cell failures triggered by security threats. They put forth a semi-supervised learning-based framework for the timely detection of sleeping cells. As shown in Fig.~\ref{Ming2022}, this framework relies on the measurement data of the resiliency of the SBSs to enhance security.

The authors of \cite{Xu20212429} presented a coding method based on quantum learning to enable the encoding and decoding of pilots on multidimensional resources in  5G networks. This method was geared towards swiftly learning and accurately eliminating uncertainties arising from potential attacks to mitigate potential failures. Encoding involves using distinguishable subcarrier activation patterns to encode multiuser pilots during uplink access. The gNB decodes the pilots based on observed subcarrier activation patterns.

$\bullet$ \textbf{Prevention of  Network Failures}

In~\cite{Wijethilaka2022915}, a coordinated security orchestration architecture was developed based on FL to manage security operations within a slicing ecosystem centrally. This architecture preserves data privacy while enabling proactive security measures. It enhances maintaining a steady security level independent of the slicing strategy.
Addressing network failures induced by Distributed Denial of Service (DDoS) attacks in 5G SDN, the authors of \cite{Abdulqadder2021866} delivered a Hybrid Fuzzy with Artificial Neural Network (HF-ANN) classifier. This classifier effectively discriminates between malicious packets and normal operational packets, providing an advanced defense mechanism against DDoS attacks.

RISs can potentially augment UAV communication networks by enhancing desired signals, and suppressing interference or jamming signals, thereby mitigating risk for potential network failures~\cite{yuan23tcom}. The studies in~\cite{hu23ris} and~\cite{yuan23tifs} delved into RIS-assisted anti-jamming UAV communication and legitimate UAV eavesdropping systems to mitigate potential failure risks. DRL is employed to train a UAV for trajectory design and RIS configuration. 
The blockchain and fog-computing-enabled security service architecture has been proposed by \cite{Hewa20227174} to mitigate single-point failures in 5G-enabled manufacturing.

\subsection{Security Failures in 6G Systems}

Some known failures in 6G authentication stem from the vulnerability of relying on single-point mechanisms.  Task integrity failures resulting from software bugs, hardware malfunctions, or network issues highlight the importance of robust integrity measures. The spectrum scarcity presents another known failure, requiring effective strategies for sharing spectrum and managing interference. All this emphasizes the ongoing need for robust security protocols and real-time monitoring in the 6G development.

 \subsubsection{Operational Failures}

% Authentication concerns revolve around the vulnerability of relying on single-point mechanisms.  Traditional centralized approaches are challenged, prompting joint authentication methods for increased resilience.
% Distributed identity management may help decentralize control and functionality, mitigating the risk of complete system breaches.
% Task integrity failure may face challenges with beam disruptions and recovery delays. 
% Physical security failures, encompassing eavesdropping and physical attacks, pose severe threats in 6G.
Table~\ref{t.po.secur} summarizes the potential security-related operational failures in 6G systems.

\begin{table*}[h]
\centering 
\caption{Potential security-related operational failures in upcoming 6G systems. } 
\begin{tabular}{m{1.75cm}<{\centering} | m{4cm} | m{4.5cm} | m{5cm}} 
\hline 
\textbf{Failure Type} & \textbf{6G Technology} & \textbf{Cause of Failure}  & \textbf{Countermeasure for failure}  \\ \hline 

 &  Distributes identity management \cite{Garzon2022}&  Fortified trust connections between verified domains \cite{Garzon2022}& Control and functionality are
evenly spread among various trust areas\cite{Garzon2022}  \\ \cline{2-4} 

Authentication Failure    &  MEC \cite{Fang20232091} & Increased delays and the vulnerability & Collaborative authentication scheme \cite{Fang20232091} \\ \cline{2-4}

&  Broadband radio services \cite{Zhang2020SDSFSDFSD}  & Cyber attacks \cite{Zhang2020SDSFSDFSD} & Blockchain-centric framework\cite{Zhang2020SDSFSDFSD} \\ \cline{2-4}

&  V2X \cite{Chai20224620} &  DDoS \cite{Chai20224620} &  Diffused practical Byzantine fault tolerance mechanism\cite{Chai20224620}   \\ \cline{1,2,3-4}

& Cognitive radio networks \cite{Ghourab2022} & Malicious attacks \cite{Ghourab2022} & Resilient cognitive radio
framework\cite{Ghourab2022} \\ \cline{2-4}

&  Secure computation \cite{Lin20217204} & Privacy data leakage \cite{Lin20217204} & Network-in-box-based blockchain framework\cite{Lin20217204}  \\ \cline{2-4}

Physical Security Failure &  IIoT \cite{Tanwar2022} &  Data modification and sniffing \cite{Tanwar2022} &  Blockchain-based distributed architecture\cite{Tanwar2022}   \\ \cline{2,3-4}

&  Power electronic hardware \cite{Chandwani2020} &  Data integrity attacks \cite{Chandwani2020} &  Attack-induced failure analysis \cite{Chandwani2020}   \\ \cline{1,2,3-4}

\end{tabular} 
\label{t.po.secur}

\end{table*}

$\bullet$ \textbf{Authentication Failures}

Since multiple 6G network operators can provide services to users, trusted third parties could be the potential targets for authentication failures\cite{Garzon2022}. It is critical to implement a system that distributes identity management across 6G networks while permitting secure authentication between various network units without a trusted intermediary.  An initial framework for such distributed identity management in 6G was provided in \cite{Garzon2022}, where control and functionality are evenly spread among various trust areas within interlinked and diversified 6G environments.

Security and access control in 6G depends on a single authentication mechanism or process in single-point authentication failures. It is impossible to prevent an attacker from gaining full access if this single authenticating node is compromised, fails, or experiences a fault. Vehicle-to-everything (V2X) applications can be developed using the Internet of Vehicles (IoVs), with authentication ensuring a reliable vehicular environment. The authors of \cite{Chai20224620} found that prevalent schemes predominantly rely on centralized systems involving third parties like certificate authorities (CAs) or key generation centers (KGCs). 
This centralized model is susceptible to threats like DDoS and single point of failure attacks. Hence, vehicle owners hesitate to store personal information on servers due to privacy concerns.

To address this, a joint authentication approach is studied in~\cite{Fang20232091}, which involves edge devices functioning as collaborative partners. Edge devices can assist the service provider in authentication by analyzing users' received signal strength indicators (RSSI) and movement patterns. In \cite{Zhang2020SDSFSDFSD}, a distributed citizens broadband radio service-blockchain model with a unique consensus technique establishes a sound consensus mechanism and safeguards the spectrum allocation from single-point failures; see Fig.~\ref{Lin20217204}. This approach can potentially decrease the administrative costs associated with  dynamic access systems.

$\bullet$ \textbf{Physical Security Failures}

Physical security failure encompasses various threats, including eavesdropping attacks, where malicious actors intercept data transmission, leading to severe privacy breaches and loss of sensitive information\cite{Lin20217204}.
This is particularly concerning in 6G due to its anticipated high-speed, high-volume data transfer and pervasive connectivity. Additionally, physical attacks\cite{Ghourab2022}, for instance, tampering or destruction of infrastructure, such as BSs and sensors, pose a significant threat,  and cause system-wide failures. These vulnerabilities highlight the urgent need for advanced security protocols, real-time monitoring, and robust physical defenses to safeguard the integrity and trustworthiness of 6G networks.

 The industrial Internet-of-Things (IIoT) connects numerous devices and machines for real-time data transfer. According to \cite{Tanwar2022}, and \cite{wlan_network_fail_1}, large numbers of connected devices and machines bring about security concerns, such as data modification and sniffing. Blockchain, e.g., with adaptive sharding~\cite{8954616,10201805}, effectively addresses these problems with lower costs, while the 6G network enhances communication speed and reliability.

Cyber-physical security in power systems underpins critical infrastructure construction. 6G will need reliable power to realize the objective of endogenous intelligence. Large quantities of power devices have been deeply coupled to underpin the needs of complex scenarios \cite{security_power_1}. The authors of \cite{cas_fail_secu_power_4} pointed out the vulnerability nodes responsible for the greatest impact on the whole power network failure should be specially considered. In \cite{Chandwani2020}, researchers examined the impact of different data integrity attacks on power electronic hardware in EV chargers to protect electric vehicles and their onboard charging systems (OBCs). Cyberattacks could manipulate the logic and data of the main charger controller, a Field Programmable Gate Array (FPGA) in the study; create false communication between the charging controller and other electronic control units connected through the same controller area network (CAN) bus; and disrupt battery functionality.

\subsubsection{Application Failures}
A secure and reliable network architecture would be needed for new 6G applications, such as IoV, IIoT, V2X, and DT. In order to deliver these applications, security breaches may cause some failures.

\begin{table*}[h]
\centering 
\caption{Potential security-related failures in upcoming 6G systems.} 
\begin{tabular}{m{2cm}<{\centering} m{3.25cm} m{3.5cm} m{6cm}} 
\hline 
\textbf{Failure Type} & \textbf{6G Technology} & \textbf{Cause for Failure}  & \textbf{Countermeasure for failure}  \\ \hline 

 &  MEC-powered V2X \cite{Cai20231}  &  Safety fault \cite{Cai20231}  &  Digital twin-based network architecture\cite{Cai20231}   \\ \cline{2,3-4}

 &  M2M \cite{Mezair2022164} &  Unable to handle heterogeneous data \cite{Mezair2022164} &   Integrated Deep Learning framework consisting of LSTM, CNN, and GNN for failure analysis \cite{Mezair2022164}   \\ \cline{2-4}

 Task Integrity Failure&  Millimeter-wave \cite{Jayaweera2022} & Disruption by obstacles \cite{Jayaweera2022} & Multiple light detection and ranging-based
approach\cite{Jayaweera2022}  \\ \cline{2-4}

 &  Air interface \cite{Rathinavel2022594}  & spectrum anomalies \cite{Rathinavel2022594} & Analyzing the metadata derived
from the monitoring air interface anomaly signal
approach\cite{Rathinavel2022594}  \\ \cline{2-4}

 & Trustworthy network \cite{Anand2022283} & Differential
attack \cite{Anand2022283} & Trustworthy and dedicated cipher approach\cite{Anand2022283}  \\ \cline{2-4}

 &  Core network \cite{Fernandez-Fernandez202328,Faisal202380} & Inefficient coordination \cite{Fernandez-Fernandez202328}, Risky transactions \cite{Faisal202380} & Blockchain-centric framework\cite{Fernandez-Fernandez202328,Faisal202380}  \\ \cline{1-4}
 % &  & Risky transactions \cite{Faisal202380}&  \\ \cline{1-4}

\end{tabular} 
\label{table:res}
\end{table*}

$\bullet$ \textbf{Task Integrity Failures}

\begin{figure}[t]
\centering
\includegraphics[width=8cm]  {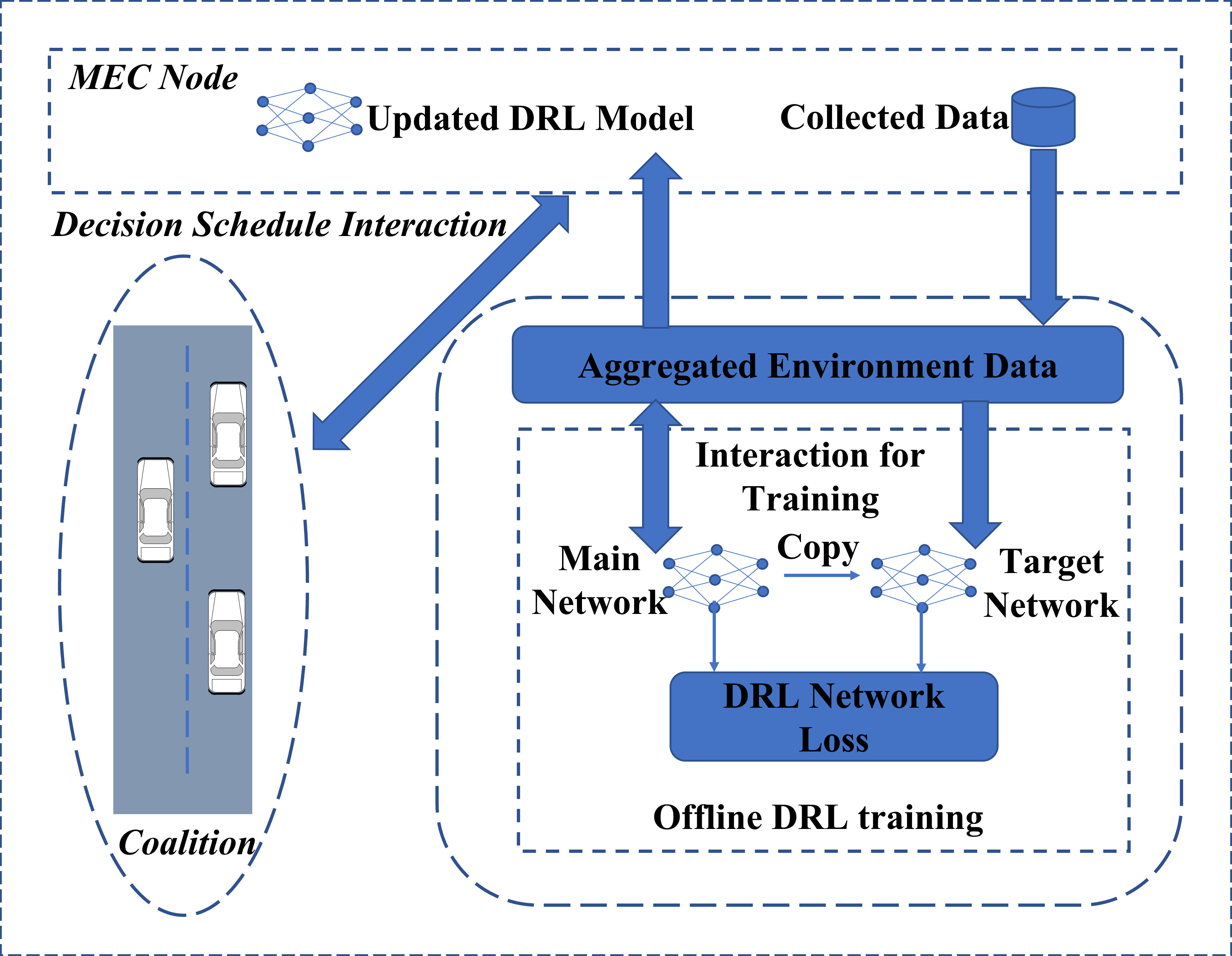}
\caption{\small Illustration of V2X channel scheduling developed in~\cite{Cai20231}. A branch-lane vehicle is modeled as a DRL agent, which learns from environmental data obtained via a simulator. The DRL agent's state represents traffic flow in the ramp area, action dictates vehicle merging decisions, and the reward is a weighted sum of average traffic speeds. The training process involves an experience replay memory, a target network, and an evaluation network. This coalition-based V2X channel scheduling for vehicle merging decisions, facilitated by the MEC node, processes the DRL network's output to guide vehicles in the ramp area for efficient merging.}
\label{Cai20231}
\end{figure}

It was pointed out in~\cite{Cai20231} that the application of DT in 6G V2X communications faces two challenges. The first is identifying which DT capabilities can seamlessly integrate with 6G V2X networks. The second challenge is centered on translating these DT capabilities into tangible enhancements in V2X network performance. To tackle these challenges probably leading to potential failures, the authors of~\cite{Cai20231}  explored the incorporation of DT capabilities within a network architecture that synergizes DT and MEC in 6G V2X communications, as depicted in Fig.~\ref{Cai20231}. The approach introduces three specific DT capabilities: 
Enhancing human-machine interaction through the analysis of driving behaviors; 
boosting traffic safety by applying knowledge-based methods for vehicle failure detection; and
examining spatio-temporal traffic patterns through data aggregation.    

The normal operation of vehicle-to-ground communication plays a vital role in subway vehicles. In~\cite{cps_fail_2}, the authors investigated the impact of interference on vehicle-to-ground communication and conducted failure analysis to identify the key communication equipment that is faulty in a failure.
In Machine-to-Machine (M2M) communications, the integration of 5G and B5G/6G contributes to the intelligence of Industry 4.0. However, the aspiration for a sustainable, self-monitored industry remains unfulfilled~\cite{Mezair2022164}. Heterogeneous data challenges the current state-of-the-art failure detection algorithms based on deep learning. Despite employing multiple failure detection computational devices, they did not effectively leverage the combination of information available in diverse formats. Often, these algorithms rely on inefficient hyperparameter tuning.

High-frequency communication in 6G faces vulnerabilities due to obstacles that obstruct signals. To mitigate the risk of failure in a millimeter-wave (mmWave) 6G system with stationary obstacles, the authors of \cite{Jayaweera2022} proposed using access points paired with light detection and ranging (LiDAR) sensors. Using fixed LiDAR maps, a strategy was designed to detect LoS shifts 400 ms in advance.
The authors of \cite{Rathinavel2022594} outlined the challenge of monitoring spectrum anomalies using metadata sourced from high-frequency radio signals. Their approach emphasizes scalable resolution for anomaly detection without requiring supervision and is bandwidth efficient.  Models were trained with non-malicious data to identify anomalies using unsupervised anomaly detection.

The authors of \cite{Anand2022283} described a fault attack on the security of ciphers in 6G systems, and demonstrated the possibility of retrieving the entire internal state by exploiting faults. During this attack, the adversary assumed knowledge of the fault location.
To address single-point failures, an architectural component called smart resource and service discovery was introduced in \cite{Fernandez-Fernandez202328}. Through decentralized telecommunication marketplaces with data-informed discovery features, this component aimed to improve service distribution in 6G networks. Additionally, the study in \cite{Faisal202380} proposed utilizing service-level agreements as contractual tools to optimize network usage and apply penalties related to service failures. As part of these agreements, permission-based distributed ledgers were developed to reduce the risk of single-point failure.

\subsection{Prevention and Defense of 6G Security Failures}
To prevent a cascade of errors and failures, it is essential to prevent and mitigate failures, once identified, promptly.
% Due to software bugs, hardware malfunctions, incorrect input data, or network congestion and overload, a task, process, or operation of the 6G device may not be completed correctly or as intended, leading to errors and data corruption. Thus, task integrity is crucial in 6G to ensure that all processes are executed correctly and that data integrity is maintained throughout the operation. When task integrity fails, it can result in incomplete or incorrect task execution, potentially causing a cascade of errors or data integrity issues within the system.

$\bullet$ \textbf{Prevention of Authentication Failures}

In addressing the single-point authentication failure potentially associated within 6G trustworthy distributed networks, the authors of \cite{Garzon2022} advocated for a decentralized identity management. This authentication framework actualized a distributed identity management system within 6G. Instead of relying on a trusted third party, this framework relied on individual entities to validate credentials. 6G network management can be collaboratively countered with this decentralized methodology. 

Meanwhile, the authors of \cite{Zhang2020SDSFSDFSD} addressed authentication failures in 6G broadband radio services by introducing a blockchain-centric framework. By incorporating a proof-of-strategy into the spectrum allocation procedure, failure risks were effectively mitigated. To explore single-point attack-induced failures within  6G edge networks, the authors of \cite{Fang20232091} delineated a collaborative authentication scheme to detect malevolent attackers in  6G distributed networks promptly.
For the attack-induced failure analysis in 6G-enabled IoVs, the authors of \cite{Chai20224620} proposed a diffused practical Byzantine fault tolerance mechanism to accelerate the authentication process while reducing consensus latency.

$\bullet$ \textbf{Prevention of  Physical Security Failures}

The authors of \cite{Ghourab2022} introduced an elastic vehicle-based cognitive radio architecture overseen by a blockchain framework that is responsive to varying circumstances. In real-time, the architecture rearranged network topology and data pathways. Moving-target defense technology enhanced protection against cyber-attacks and system failures. As part of a decentralized trust management system, blockchain technology was also integrated.
\begin{figure}[t]
\centering
\includegraphics[width=8cm]  {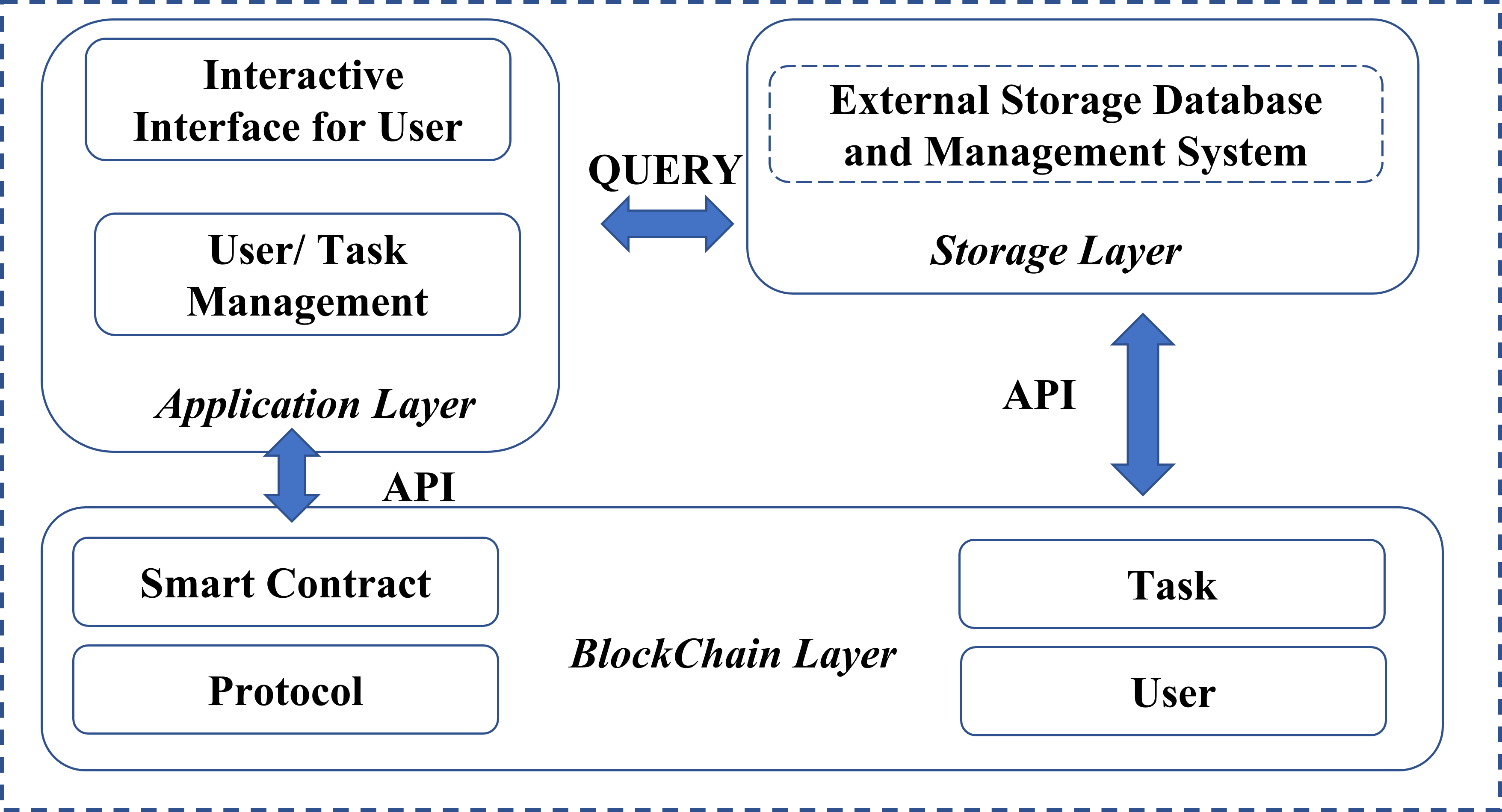}
\caption{\small The architecture of Blockchain-based parallel distributed computing~\cite{Lin20217204}, which has three layers: application layer, blockchain layer, and storage layer. Using smart contracts, the blockchain layer records tasks to prevent single-point failures. Users and tasks are managed through the application layer, which provides an API.}
\label{Lin20217204}
\end{figure}

In \cite{Lin20217204}, a Blockchain-based Privacy-aware Distributed Collection (BPDC) algorithm was developed for distributed data aggregation, aimed at safeguarding 6G-enabled Non-IP Based (NIB) industrial applications from internal collusion attacks while bolstering privacy security. Various attributes specify the security level of various tasks and the requisite credentials for task recipients. BPDC involved decomposing sensitive tasks and categorizing task receivers according to their security level requirements, as depicted in Fig.~\ref{Lin20217204}.

Addressing security-induced single-point failures in 6G-enabled industrial automation, the authors of \cite{Tanwar2022} proposed a Blockchain-based distributed architecture to ensure secure and reliable communication between pairwise industrial IoT devices. In a more detailed exploration of attack-induced failure analysis on the IoV, the authors of \cite{Chandwani2020} provided insights into potential vulnerabilities and failure scenarios.
The study in \cite{Prathiba202119} delved into attack-induced failure analysis in 6G-enabled IoVs. As a result of attack-induced failures, rerouting was implemented, significantly improving response latency. 
In \cite{cas_fail_secu_power_4}, a model capturing the interdependence between a power network and a communication network, and a failure analysis pinpointed the critical nodes that contributed to power communication network reliability, thereby mitigating cascading failures.

$\bullet$ \textbf{Prevention of  Task Integrity Failures}

In addressing attack-induced failure analysis in 6G-enabled V2X communication, the authors of \cite{Cai20231} introduced a digital twin-based network architecture. This approach enhances attack-induced failure analysis through knowledge-based data analysis. 
For security-induced failure analysis in 6G-enabled heterogeneous data surveillance within Industrial 4.0, the authors of \cite{Mezair2022164} integrated LSTM, CNN, and Graph Neural Network (GNN) components for comprehensive failure analysis. This framework accommodates various types of heterogeneous data.

% \subsubsection{Prevention of  Task Integrity Failure}
The task integrity failure on a trustworthy 6G core network has been investigated in \cite{Anand2022283}, where the authors presented an efficient and trusted cipher that was applied to 6G systems to deal with the failures caused by differential attacks. The critical failure issue on a trustworthy 6G core network deeply driven by blockchain technology can also be found in \cite{Faisal202380,Fernandez-Fernandez202328}. 
For task integrity failure detection in 6G air-interface, the authors of \cite{Rathinavel2022594} proposed to identify anomalous activities by analyzing metadata derived from monitoring air interface anomaly signal. Analyzing receiver operating characteristics data performed reasonably well on anomaly-induced failure detection.
For failure prevention in 6G sub-THz communication, the authors of \cite{Jayaweera2022} suggested a multiple light detection and ranging-based approach to detect non-line-of-sight (NLoS) link failures and further ensure secure communications. This approach can predict link failures caused by LoS-NLoS transitions.

\begin{table*}
\centering 
\caption{Incidental operational failures inherited from 5G systems.} 
\begin{tabular}{p{1.5cm}<{\centering} m{3cm} m{4cm} m{4.5cm} m{1.5cm}<{\centering}} 
\hline 
\textbf{Failure Type} & \textbf{5G Technology} & \textbf{Cause for Failure}  & \textbf{Countermeasure for Failure} & \textbf{Perpetuate in 6G} \\ \hline 
 &  Massive MIMO \cite{Fernandes20221580}  & Defect of compute-and-forward serial connection \cite{Fernandes20221580} & Markov chain
Monte Carlo simulations \cite{Fernandes20221580} & \checkmark \\ \cline{2-5}
 &  PONs \cite{Khare2023}  & Early-stage risk \cite{Khare2023} & Hybrid single-mode
FSO wavelength division multiplexed gigabit PON architecture\cite{Khare2023} & \checkmark \\ \cline{2-5}
Transmission
Failure  &  mmWave \cite{Ferreira20235178}  & Susceptibility
to high propagation loss \cite{Ferreira20235178} & Extensive measurement and evaluation\cite{Ferreira20235178} & \checkmark \\ \cline{2-5}
 &  backhaul \cite{Yaghoubi2022616}   & Vulnerability to channel fluctuations \cite{Yaghoubi2022616} & Design of cost-efficient
and reliable wireless backhaul networks under correlated
failures\cite{Yaghoubi2022616} & \checkmark \\ \cline{2-5}

 &  SDN \cite{Yasin20225}  & Signal homogeneity
and environmental impact \cite{Yasin20225} & Adaptive Multipath number  mechanism\cite{Yasin20225} & \checkmark \\ \cline{2-5}

 &  Active
Antenna Unit BS \cite{Shi2022}   & Defect and crack of electric device \cite{Shi2022} & The alternative
training framework for generative adversarial networks for analysing failure sample\cite{Shi2022} & \checkmark \\ \cline{1-5}

Power Failure & D2D \cite{Xu202277} & Incidental disruption to cross-layer connection \cite{Xu202277} & The dual-plane redundancy in substation and heterogeneous hand in hand connection\cite{Xu202277} & \checkmark \\ \cline{1-5}

Component Failure & IoV \cite{Wang2023118} & Incidental engine disruption \cite{Wang2023118} & Engine test system \cite{Wang2023118} & \checkmark \\ \cline{2-5}

& Antenna Unit \cite{Latry2022} & Incidental aging disruption to GaN transistors \cite{Latry2022} & Degradation mechanism of surface pitting \cite{Latry2022} & \checkmark \\ \hline

\end{tabular} 
\label{t.incident.op}

\end{table*}

\section{Incidental Failures}\label{sec_VI}
 The failure rate of products and systems increases with age, including deterioration and unavailability. The purpose of this section is to emphasize the importance of resilient or self-healing frameworks. Reliable and long-lasting systems are enhanced by these frameworks. 

\subsection{Failures Inherited From 5G}

During practical deployments of 5G systems, disruptions, damages, and inefficient power management have been observed. In order to gain insights into 5G operations and applications, these failures are extensively examined.

\subsubsection{Operational Failures}
Incidental operational failures could occur to transmissions and services,
as summarized in Table~\ref{t.incident.op} and delineated in the following.

$\bullet$ \textbf{Transmission Failures}

Cell-free (CF) massive Multiple-Input Multiple-Output (mMIMO) networks are integral for 5G and beyond~\cite{Fernandes20221580}. A compute-and-forward architecture with serially interconnected access points (APs) may pose reliability issues. 
Survivability, reliability, and transmission latency awareness are early-stage design challenges for  5G systems \cite{Khare2023}. Passive Optical Networks (PONs) emerge as a key solution to address future network requirements.

For 5G wireless backhaul, WiGig protocols, e.g., IEEE 802.11ad and 802.11ay, are considered. The susceptibility of the mmWave band to high propagation loss \cite{Ferreira20235178}, necessitates the use of directional antennas.
% Designing reliable wireless backhaul networks faces challenges due to vulnerability to channel fluctuations. 
The authors of \cite{Yaghoubi2022616} emphasized the importance of considering the correlation among link failures. Overlooking this aspect may lead to failed topology designs under correlated scenarios.

Despite the high speed of 5G, link failures can impact service quality \cite{Yasin20225}. Signal homogeneity and environmental impact are addressed by strategic BS placement, particularly on highways. SDN aids in efficient management, offering link robustness to prevent service unavailability. 
A specific case study in \cite{Shi2022} investigated the field failure of a radio frequency differential amplifier within the 5G Active Antenna Unit BS. Contributes to predicting reliability risks by providing insights into die crack failure analysis.

A highly integrated space-air-ground-sea communication network poses complex challenges in 6G. Leveraging  RISs in this context requires cooperation between intelligent Electromagnetic (EM) devices to facilitate intelligent transmission, addressing channel fluctuations and  latency\cite{6G_wang_2023}. 
While addressing the incidental transmission failure issue in 5G, this challenge is anticipated to persist in 6G.

$\bullet$ \textbf{Power Failures}

A power optical communication network is a specialized network that caters to power grids, and its survival is crucial for their secure and steady functioning. The analysis in \cite{Wu2022304} focused on the dependability of communication networks and calculated a collection of risk links that can identify the likelihood of power failures. The set of probabilistic risk management links resulted in three distinct shared development path algorithms. Ad-hoc systems operate without infrastructure. User equipment (UE)  establishes quick networking but cannot ensure the reliability of D2D connections.  5G offers a reliable network infrastructure, with which D2D enables visualized connections, such as power grids \cite{Xu202277}.

$\bullet$ \textbf{Component Failures}

In the automobile industry, improvements in reliability testing and intelligent fault diagnosis have increased the importance of engines. As the primary component and the most prone to failures in a vehicle, the engine's significance cannot be understated \cite{Wang2023118}. IoT-based automotive engine inspection systems are the future trend. The goal of \cite{Wang2023118} was to gain a deeper understanding of smart car engine systems in the era of 5G IoT. This study optimizes the engine's dynamic performance through dynamic testing and characterization.
In \cite{Latry2022}, an assessment of the time to failure (TTF) of GaN transistors in applications  to 5G and RADAR was assessed. Based on RF pulsed life tests involving various input powers and duty cycles, TTF values were estimated using Arrhenius curves.  The study \cite{Latry2022} described the method  to estimate temperatures during the aging tests under different operating conditions.

\subsubsection{Application Failures}

Table~\ref{t.incident.app} summarizes the incidental application failures inherited from 5G systems.

\begin{table*}[h]
\centering 
\caption{Incidental application failures inherited from 5G systems.} 
\begin{tabular}{P{1.5cm} m{2cm} m{4cm} m{5cm} m{1.5cm}<{\centering} } 
\hline 
\textbf{Failure Type} & \textbf{5G Technology} & \textbf{Cause for Failure} & \textbf{Countermeasure for Failure} & \textbf{Perpetuate in 6G} \\ \hline 
&  SFC \cite{Cherrared20212515,Bai20232811} &  Network visibility and dynamic topologies \cite{Cherrared20212515} &Self-modeling approach and an active diagnosis process\cite{Cherrared20212515} & \checkmark  \\ \cline{3-5} 

 Service Failure &    & Device aging  \cite{Bai20232811}& Semi-Markov model exploring transient availability and steady-state dependability \cite{Bai20232811} & \checkmark \\ \cline{2-5}

&  MEC \cite{Chantre20222478}  & Stringent quality requirement \cite{Chantre20222478} & 1: N: K protection scheme\cite{Chantre20222478} & \checkmark \\ \cline{2-5}

  &  Core network \cite{Zhu202248,Ahmad2023754}  & The redundancy of the 5G core network \cite{Zhu202248} & Leveraging the Network Data Analytics Function  for intelligent analysis\cite{Zhu202248} & \checkmark \\ \cline{3-5}

 &    & The fault of control plane \cite{Ahmad2023754}& Abstraction of reliable access to cellular services while ensuring lower latency\cite{Ahmad2023754} & \checkmark \\ \cline{1-5}

\end{tabular} 
\label{t.incident.app}
\end{table*}

$\bullet$ \textbf{Service Failures}

The adaptability of  5G systems is crucial for  accommodating diverse functions and infrastructure, supporting specific service requirements through SFC~\cite{10146517}. However, as highlighted in \cite{Cherrared20212515}, effective failure management is indispensable for meeting SFC requirements and ensuring the reliability of 5G. Through a dependency model, model-based approaches (MBs) explicitly represent system structure and behavior. Despite current methodologies within network virtualization, challenges persist, such as lack of visibility and dynamic topologies.

The integrated framework of MEC and NFV enables the execution of customized services structured as SFCs. The study  in \cite{Bai20232811} highlighted memory-related software aging in SFs as a new threat exacerbated risk for failure, severely threatening MEC-SFC reliability. The issue of SF aging must be countered by proactive rejuvenation techniques.

With MEC and slicing techniques, mobile networks can meet stringent QoS requirements and mitigate potential service failures \cite{Chantre20222478}. To prevent service failures in 5G core networks, operators must balance cost and reliability \cite{Zhu202248}. Networks lacking redundant deployment may face difficulties in achieving high reliability.

The demands of Industry 4.0 place stringent criteria on 5G  systems, necessitating high reliability, availability, and low latency. However, it was noted in \cite{Baldvinsson2022621}  that training reinforcement learning algorithms and simulating rare events require access to diverse failure data.
 End-user applications are directly affected by the time required for cellular control plane operations. A key component of cellular core networks is the control plane \cite{Ahmad2023754}.

\vspace{-3mm}
\subsection{Prevention of 5G Incidental Failures}

For 5G networks, technologies have been developed to address incidental or accidental failures. We review the technologies and highlight their potential to improve network performance and dependability.

$\bullet$ \textbf{Prevention of Transmission Failures}

The authors of \cite{Fernandes20221580} used Markov chain Monte Carlo simulations to study the effects of failures in APs and fronthaul segments in Cloud-Fog mMIMO systems.
In \cite{Khare2023}, the authors proposed a hybrid gigabit PON architecture combining single-mode fiber (SMF) and free-space optics (FSO), integrated with wavelength division multiplexing. This system, offering failure mitigation, supported direct internetworking data transmission, inter- and intra-optical distribution network (ODN) data flows, and broadcasting, reducing inter-ODN transmission latency and failure risks by 55\%.
% introduced a novel hybrid architecture employing single-mode fiber and free-space optics (FSO) in a wavelength division multiplexed gigabit PON. 
% This architecture featured failure mitigation capabilities and facilitated the direct transmission of internetworking data. It enabled the transmission of inter-optical distribution network (ODN) and intra-ODN data, along with broadcasting. This architecture was demonstrated to reduce transmission latency and potential transmission failures for inter-ODN transmissions by 55\%.
The authors of \cite{Mandal2022} presented a hybrid Wavelength division multiplexing-free space optics-passive optical network capable of 4×10 Gbps downlink and 2×10 Gbps uplink, serving both wired and wireless users. Spanning 60 km of SMF and 650 m of FSO or a 62 km SMF link, it demonstrated enhanced fault tolerance and uninterrupted data transfer between SMF and FSO links.

The authors of \cite{Ferreira20235178} conducted an extensive measurement and cross-layer analysis on physical (PHY), medium access control (MAC), and transport layers metrics under short-term and long-term blockages. It was discovered that high modulation and coding schemes (MCSs) in long-term blocked channels can cause packet errors up to 100\%, round-trip-times (RTTs) of several seconds, and packet losses up to 90\%. This degradation, more severe in short-term links, is aggravated by rapid MCS changes during sudden obstructions.
% maintaining consistent and higher modulation and coding schemes (MCSs) in long-term blocked channels could result in packet errors reaching 100\%, round-trip-time (RTTs) in the order of a few seconds, and packet losses as high as 90\%. This degradation potentially leading to failures is more exacerbated in short-term links, experiencing more extreme MCS changes during sudden obstructions.
The authors of \cite{Yaghoubi2022616} designed cost-efficient, reliable wireless backhaul networks resistant to correlated failures, particularly rain disturbances. They included a penalty cost to model path correlations, formulating the network topology design as a quadratic integer program to find optimal solutions under these correlations.

The authors of \cite{Yasin20225} introduced an approach where the multipath count is contingent on the reliability of the primary path. 
As link reliability increases, fewer alternative paths are needed, streamlining the calculation process. They combined shortest distance and reliability factors to improve service availability in 5G networks, reducing latency and traffic overhead during link failure recovery.
% The number of alternative paths decreases as the reliability of the link increases, resulting in reduced time required for calculating alternative paths. The authors of \cite{Yasin20225} integrated the shortest distance factor with the reliability factor. This approach enhanced service availability in 5G networks by minimizing latency rates and traffic overhead during link failure recovery.
The authors of \cite{Baldvinsson2022621} introduced IL-GAN, a training model for generative adversarial networks (GANs) harnessing incremental learning (IL). This approach allowed GANs to understand the tail behavior of distributions with few samples, demonstrating its effectiveness in a 5G factory automation scenario simulation.

$\bullet$ \textbf{Prevention of  Power Failures}

In \cite{Wu2022}, simulation calculations and comparative analyses of three algorithms were conducted on the CERNET network topology aimed at business applications. This analysis held immense importance in ensuring the secure and dependable operation of grid systems while minimizing large-scale grid accidents.
The authors of \cite{Xu202277} proposed employing dual-plane redundancy in substations and a heterogeneous hand-in-hand connection for distribution power lines to integrate 5G D2D communication with data terminal equipment in power grid automation and protection. This approach enabled cross-layer connection failure detection and autonomous maintenance.

$\bullet$ \textbf{Prevention of  Component Failures}

{\blue
The study in \cite{Wang2023118} focused on the PUMAOPEN test system by AVL and its application in the intelligent transformation of key components to mitigate potential component failures. The study involved managing test processes and assessing dynamic conditions in intelligent vehicles, comparing these to steady states using both qualitative and quantitative methods.
In another study \cite{Latry2022}, the authors investigated surface pitting degradation in HEMT AlGaN/GaN transistors, extending DC to RF life test scaling to include duty cycles. This study underlined the transistors' high reliability, especially in RF pulsed conditions, thereby reducing component failures in power bars.

% In a comprehensive test program presented in \cite{Marozau2022}, the reliability of miniaturized electromechanical relays (MEMR) used in RF applications was evaluated, with a focus on meeting requirements for space applications. The test program, aligned with ESA standards and specifically tailored for RF applications, successfully evaluated the reliability of MEMR devices, mitigating potential component failures for space and terrestrial applications, like communication satellites or 5G  equipment.

The authors of \cite{Marozau2022} conducted a comprehensive test program assessing the reliability of miniaturized electromechanical relays (MEMR) for RF applications, emphasizing space application standards. Tailored to ESA standards and RF-specific requirements, the program verified MEMR reliability for both space and terrestrial applications, including satellites and 5G equipment.
}

$\bullet$ \textbf{Prevention of  Service Failures}

% A self-modeling approach and an active diagnosis process for virtual networks were proposed in \cite{Cherrared20212515}, combining learned and acquired knowledge through fault injection, to address identified limitations. This approach integrated acquired and learned knowledge through fault injection, expanding and validating the model. Experimental results applied to a real-world virtual IP Multimedia Subsystem (vIMS) use case to demonstrate the effectiveness of self-modeling and active diagnosis procedures in determining root causes of failures and explaining fault propagation.

% The study in \cite{Bai20232811} developed a semi-Markov model to explore transient availability and steady-state dependability of MEC-SFC services. The model accommodated any SFs and captures complex time-dependent aging, failure, and recovery behaviors. Through simulation, the study identified potential bottlenecks in a MEC-SFC system through sensitivity analysis and explored the impact of event-time interval distributions on steady-state dependability. 

% Investigating the challenge of locating MECs and slices within a 5G infrastructure protected by a 1: N: K protection scheme, the authors of \cite{Chantre20222478} aimed to meet high reliability and low latency demands at an economical cost. They introduced a bi-objective non-linear formulation and utilized the non-dominated sorting genetic algorithm (NSGA)-II to derive the solution. 

A self-modeling approach and an active diagnosis process for virtual networks were proposed in \cite{Cherrared20212515}, combining learned and acquired knowledge through fault injection, to address identified limitations. This method was validated in a real-world virtual IP Multimedia Subsystem (vIMS) case, proving effective in identifying failure root causes and explaining fault propagation.

The study in \cite{Bai20232811} created a semi-Markov model to analyze the transient availability and steady-state dependability of MEC-SFC services, accounting for complex aging, failure, and recovery patterns. The model, validated through simulations, identified MEC-SFC system bottlenecks via sensitivity analysis and examined the effects of event-time intervals on dependability.
The authors of \cite{Chantre20222478} investigated efficient MEC and slice placement in 5G networks under a 1: N: K protection scheme, aiming to balance high reliability, low latency, and cost. They formulated a bi-objective non-linear problem and applied the non-dominated sorting genetic algorithm (NSGA)-II to find solutions.

\begin{table*}
\centering 
\caption{Potential incidental operational failures in the upcoming 6G systems.} 
\begin{tabular}{m{1.75cm}<{\centering} | m{3.25cm} | m{5cm} | m{5cm}} 
\hline 
\textbf{Failure Type}  & \textbf{6G Technology} & \textbf{Cause for Failure}  & \textbf{Countermeasure for Failure}  \\ \hline 
 &  Intelligent active phased array \cite{Nielsen20225044}  & Electromagnetic
Device
deterioration \cite{Nielsen20225044} & DNN-based base-
band signal analysis \cite{Nielsen20225044} \\ \cline{2-4}

    &  Non-terrestrial networks \cite{Demir2022}  & Non-geostationary
satellite moving
coverage \cite{Demir2022} & Measurement-based mechanism \cite{Demir2022} \\ \cline{2-4}

 Transmission Failure  &  Femtocells \cite{Maiwada2022486}   & Femtocell mobility \cite{Maiwada2022486} & Mobility state detection \cite{Maiwada2022486} \\ \cline{2-4}

  & Counseling AI \cite{Taniguchi20223899}& Occasional connection malfunction\cite{Taniguchi20223899}s & Slice-based mechanism \cite{Taniguchi20223899} \\ \cline{2-4}

   & V2X \cite{Linsalata20221,Morandi2021} & Blocked by obstacles \cite{Linsalata20221} & Proactive relaying \cite{Linsalata20221} \\ \cline{3-4}

&  & Occasional connection malfunction\cite{Morandi2021} & Beam selection \cite{Morandi2021} \\ \hline

Network Failure &  Core network \cite{Li20221292,Wang2021sasasas}   & Insufficient high-quality labeled
data \cite{Li20221292} & Robust belief weighting \cite{Li20221292}  \\ \cline{3-4}

  &    & Network
operation
malfunction \cite{Wang2021sasasas} & AI-based failure prediction \cite{Wang2021sasasas} \\ \cline{1-4}

 & Industrial IoT \cite{Zheng2021sasasasas,Mohamed2020}&  & FPGA-based intelligent analysis \cite{Zheng2021sasasasas} \\ \cline{4}

Component Failure&   & Electromagnetic device deterioration \cite{Zheng2021sasasasas,Lvssssssss20217,Mohamed2020,Zhong2022} & Inter-disciplinary approach integrating wireless sensor networks with ML \cite{Mohamed2020} \\ \cline{2,4}

& Aircraft security communication service \cite{Lvssssssss20217}  &  & Flexible surveillance \cite{Lvssssssss20217} \\ \cline{2,4}

& Aircraft positioning service\cite{Zhong2022} &  & Deep Belief Network-based failure prediction \cite{Zhong2022} \\ \hline

\end{tabular} 
\label{t.po.incident.op}

\end{table*}

A method leveraging 5G-advanced and 6G concepts was proposed in \cite{Zhu202248} to enhance Network Data Analytics Function (NWDAF) for intelligent control and scheduling. With this approach, the Home Subscriber Server (HSS) backs up Unified Data Management (UDM), and NWDAF triggers automated transfers when UDM fails. Fault Tree Analysis (FTA) and probability theory were used to assess the method's effectiveness.
The authors of \cite{Ahmad2023754} introduced Neutrino, a cellular control plane designed to offer users a reliable abstraction of access to cellular services with a focus on latency. Neutrino increases control procedure completion times by up to 3.1 times without control plane failures and 5.6 times with them.

\vspace{-3mm}
\subsection{Incidental Failures in 6G}

% We now elaborate on the incidental failures that potentially arise in 6G system architecture and technologies, as well as its applications. 

\subsubsection{Incidental Operational Failures}
The operations of services, transmissions, and networks can be susceptible to incidents and accidents and result in hardware, software, and system failures.
Table~\ref{t.po.incident.op} summarizes the potential incidental operational failures in  6G systems.

{\color{black}

$\bullet$ \textbf{Transmission Failures}

Nielsen \textit{et al.}~\cite{Nielsen20225044}  focused on examining the components' accidental failures within a 6G intelligent active phased array communication block, as illustrated in Fig.~\ref{Nielsen20225044}. These failures extend beyond antenna elements, and can manifest in front-end circuits such as power amplifiers (PAs) or phase shifters, thereby presenting a complex multidimensional challenge for fault diagnosis.
Demir \textit{et al.} \cite{Demir2022} investigated link failures occurring in 6G non-terrestrial networks.  Maiwada \textit{et al.} \cite{Maiwada2022486} explored accidental radio link failures in deploying 6G femtocells.

\begin{figure}[htbp]
\centering
\includegraphics[width=8.5cm]  {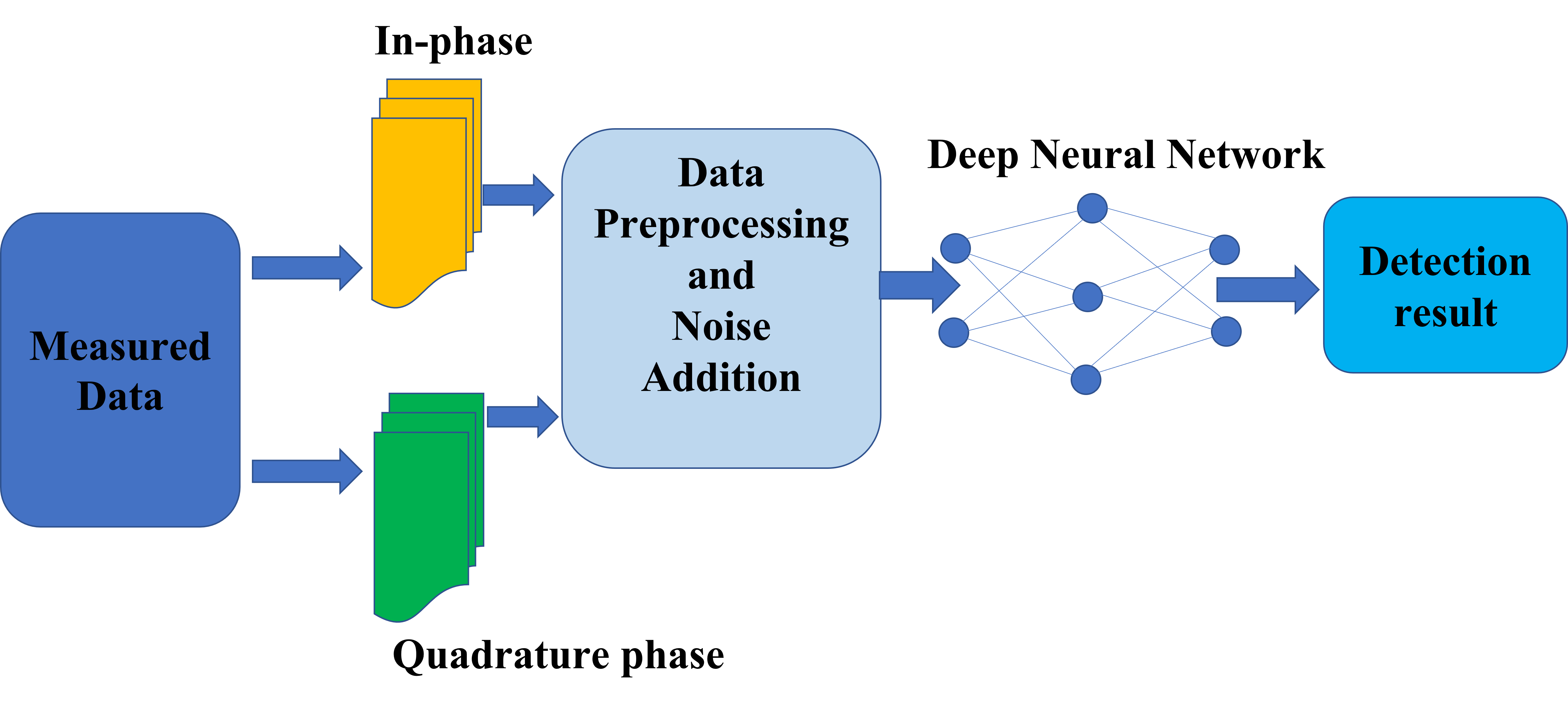}
\caption{\small Illustration of the APA Diagnosis Concept developed in~\cite{Nielsen20225044}, where mathematical estimators use measured electric fields and antenna patterns, and a DNN-based approach uses in-phase and quadrature (IQ) baseband signals. DNN training includes IQ signal acquisition, preprocessing, noise addition for robustness testing, and detection outcomes like the confusion matrix.}
\label{Nielsen20225044}
\end{figure}

A counseling robot was investigated in~\cite{Taniguchi20223899} to assist people with mental distress through spoken interaction.
The system, however, has issues such as word omissions, lags, and intermittent connectivity issues. Its ability to monitor and support users is compromised.
CAVs operate at mmWave frequencies, offering high bandwidth (over 1GHz) and data rates (10Gbit/s). Physical barriers can easily obstruct LoS transmission at such high frequencies. This issue can be mitigated through the use of relays. In dynamic environments, conventional relays, which respond to link failures and use current data, are hindered~\cite{Linsalata20221}.

$\bullet$ \textbf{Network Failures}

6G core networks are more vulnerable to failure due to the exponential growth in their size and complexity. This poses a substantial challenge to the QoS and overall reliability, as emphasized in studies \cite{Li20221292} and \cite{Wang2021sasasas}. By leveraging limited labeled data and a simplified knowledge base consisting of elementary belief rules, the studies address these challenges.

$\bullet$ \textbf{Component Failures}

A collaborative research effort, substantiated by \cite{Zheng2021sasasasas,Mohamed2020,Malakoutian2022,Huang2023}, focused on addressing accidental failures in 6G-enabled industrial applications.  A range of service demands can be met with 6G cellular networks, including autonomous failure detection and prediction, optimization of operations, and proactive control. These advancements equip industrial facilities with an advanced, ``sixth sense'' reasoning capability, optimizing operations and preventing failures.

The rising power density is a significant trend in electronics applications. Due to the increased power density, the device channel experiences Joule heating and elevated temperatures, resulting in performance degradation. Diamond integration close to the hot spot helps dissipate heat by enhancing the heat transfer coefficient~\cite{Malakoutian2022}.
Micro-hole drilling faces new difficulties with the high-frequency, high-speed PCB required for 5G/6G  due to increased board thickness and reduced hole diameter (aspect ratio exceeding 20), and further ignites failure risks from micro-drill fracture\cite{Huang2023}.
Furthermore, an engine failure could take place in a 6G-based aircraft system~\cite{Lvssssssss20217,Zhong2022}.
To {\color{blue}prevent} the accidents of commercial passenger aircraft {\color{blue}resulting from} an engine fire, cloud sea computing (CSC) and an overlapping fault-tolerant large passenger aircraft (OFTLPA) architecture were presented in \cite{Lvssssssss20217}.}

\subsubsection{Incidental Application Failures}
\begin{table*}[h]
\centering 
\caption{Incidental application Failure Analysis in 6G} 
\begin{tabular}{P{2cm} | m{3.5cm}<{\centering}| P{3cm} |P{5cm}} 
\hline 
\textbf{Failure Type} & \textbf{6G Technology} & \textbf{Cause for Failure} & \textbf{Countermeasure for Failure} \\ \hline 

Service Failure &  NFV \cite{Shaghaghi2022437}   &  Edge node deterioration \cite{Shaghaghi2022437,Mismar20213330}  & DRL-based
proactive failure recovery mechanism \cite{Shaghaghi2022437}   \\ \cline{2,4} 
& MEC \cite{Mismar20213330}& 
 & Unsupervised learning self-diagnosis mechanism \cite{Mismar20213330}  \\ \cline{1-4}

\end{tabular} 
\label{t.po.incident.app}
\end{table*}

The new applications powered by 6G technology, such as counseling robots for medical care, V2X, and Industrial 4.0, can be vulnerable to accidental failures, including connection, link, and engine failures; see Table~\ref{t.po.incident.app}. A collective effort \cite{Shaghaghi2022437,Reddy202243,Babou2022295,Shayesteh2021217,Sethi2020} has been directed towards analyzing service failures within 6G NFV. 
According to \cite{Mismar20213330}, VNF software is susceptible to physical node failures and software malfunctions, when operating on physical nodes.

\subsection{Prevention and Defense of Incidental Failures}
Some initial investigations have been carried out to prevent, mitigate, or eliminate the potential operational and application failures for 6G systems.

$\bullet$ \textbf{Prevention of  Transmission Failures}

In addressing accidental component failure analysis within 6G  active phased array blocks, the authors of \cite{Nielsen20225044} proposed a DNN-based approach. In 6G active phased arrays, this method identified, classified, and analyzed hidden features in baseband signals. An attractive candidate for on-site component failure analysis is the approach that efficiently locates and diagnoses failed components.

In investigating accidental link failure within 6G non-terrestrial networks, the authors of \cite{Demir2022} observed that handovers could be induced by the moving coverage area of non-geostationary satellites, resulting in link failures. According to the authors, a measurement-based approach outperforms alternative approaches in mitigating handover-induced link failures. Similarly, the authors of \cite{Maiwada2022486} explored accidental radio link failures in 6G femtocell deployment. The authors focused on femtocell mobility states to enhance the QoS in the 6G core network. 6G femtocell deployment can be mildly improved by the improved mobility state detection mechanism proposed by the authors.

For accidental connection failures in 6G-enabled counseling robot services, the authors of \cite{Taniguchi20223899} proposed a 6G slice solution to handle occasional connection failures. The approach can efficiently enhance counseling quality by leveraging  6G slices.  
For accidental link failures arising in  6G-enabled V2X, the authors of~\cite{Linsalata20221} proposed a proactive relaying strategy to dynamically select relays based on the LoS-map generated by autonomous vehicles and the environment.  The strategy can generate a dynamic LoS map and predict link failure quickly.

$\bullet$ \textbf{Prevention of  Network Failures}

For accidental failures arising from insufficient high-quality labeled data for 6G core networks, the authors of \cite{Li20221292} proposed a robust belief weighting framework for few-shot failure prevention. The framework uses abductive learning and belief rule structures. The framework can be re-trained iteratively to improve the coarse data set. With the framework, communication services in 6G networks became more reliable and fault-tolerant. A similar failure data analysis for 6G core networks can be found in \cite{Wang2021sasasas}.

$\bullet$ \textbf{Prevention of  Component Failures}

In 6G-enabled industrial applications, an FPGA-based on-site approach mimics natural immunity to extract the underlying data information in real-time~\cite{Zheng2021sasasasas}. Troubleshooting in industrial IoT could benefit from the approach. 
In \cite{Mohamed2020}, a cross-disciplinary method was introduced that combines wireless sensor networks with ML-enhanced industrial facilities. An example is a failure detection and prediction system in a wireless network equipped with sensors and actuators. A chemical plant applied these strategies to detect and predict failures accurately.
In \cite{Huang2023}, experiments indicated micro-drill fracture during micro-hole drilling is primarily due to excessive torque, not thrust force. 
% Upon analyzing the thrust force and torque during drilling, it was found that the fracture occurred not because of excessive thrust force in traditional PCB drilling but due to excessive torque caused by friction and resistance in chip removal. A micro-drill fracture solely happens during the speed conversion stage, where no material is removed and the drilling torque is the largest.
Friction and chip removal resistance cause this fracture during the speed conversion stage, when drilling torque peaks and no material is removed.

For accidental engine failures arising in  6G-enabled aircraft service, the authors of \cite{Lvssssssss20217} proposed an architecture comprising a coupling robot (CR), a flying wing load-carrier,
and a commercial passenger aircraft with two semi-embedded rear propulsion engines.
{\blue 
In the event of engine failure, the CR facilitates the safe detachment and departure of the aircraft from the carrier. Quick failure detection and enhanced flight safety are achieved using 6G technology.
}

% The operation of CR and the disassembly of an OFTLPA are governed by a CSC model. 
% If the engine failure occurs in the load-carrying aircraft, the passenger aircraft is rapidly deployed and released from the OFTLPA through the grasping mechanism of the CR.
% The passenger aircraft flies away from the carrier, ensuring a safe flight. The effectiveness of this approach lies in its ability to leverage the advanced capabilities of 6G technology, enabling prompt detection of engine failures and ensuring the safety of the flight.

\begin{figure}[t]
\centering
\includegraphics[width=8.5cm]  {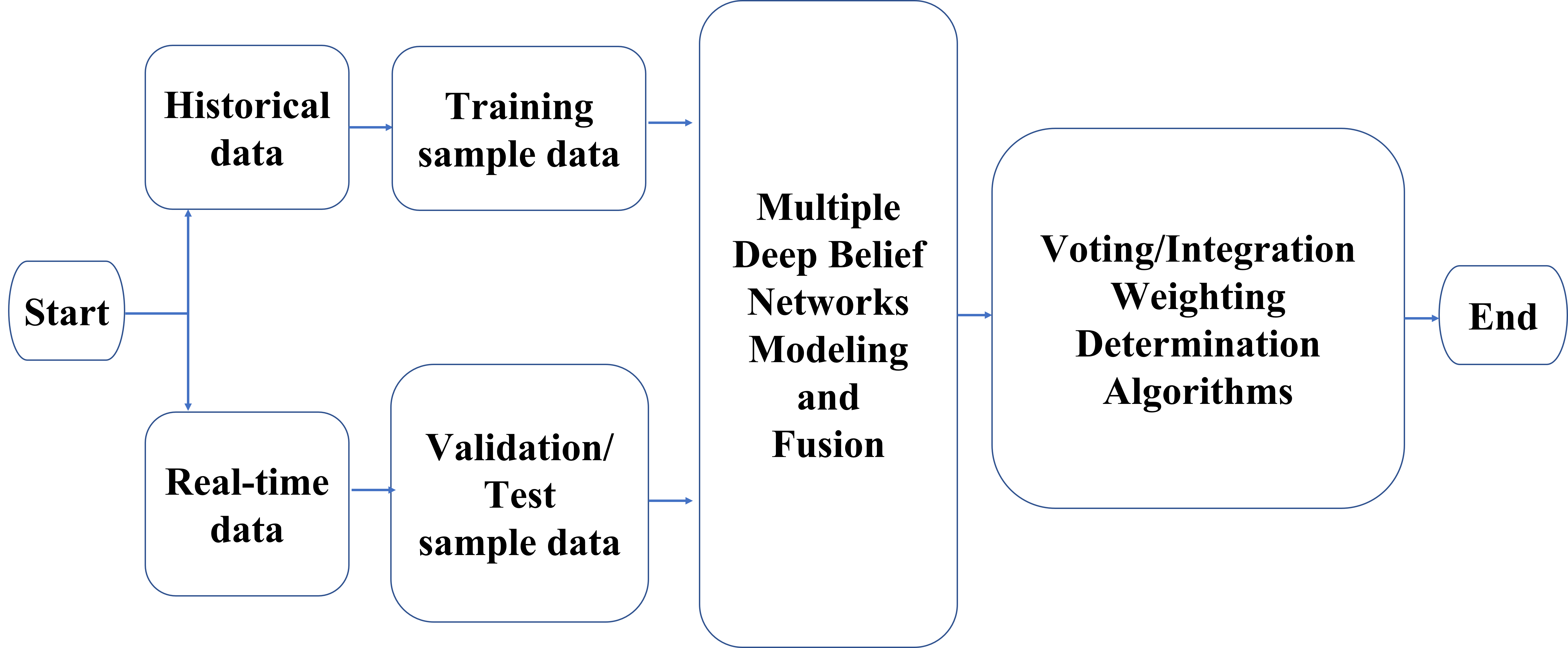}
\caption{\small Multi-model fusion failure prognostic framework developed in \cite{Zhong2022} for airborne equipment. Under complex stress conditions, this multi-model fusion framework integrates various predictive models to enhance failure prediction accuracy. An improved weighted voting algorithm considers model-specific performance across degradation stages based on a quantitative health assessment technique for real-time and historical flight data. Correcting and predicting equipment health indices, the framework overcomes the limitations of single DBN models.}
\label{F:multimodal}
\end{figure}

Distance Measuring Equipment (DME) has been used for aircraft positioning, typically relying on multiple ground beacon stations (GBSs). Zhong \textit{et al.} \cite{Zhong2022} introduced a method based on ML and signal processing techniques to predict and assess the health status and degradation trend of air-borne DME receivers; see Fig.~\ref{F:multimodal}.

$\bullet$ \textbf{Prevention of  Service Failures}
 
For accidental failures arising in 6G NFV, the authors of \cite{Shaghaghi2022437} proposed a proactive failure recovery framework based on {\color{blue}DRL} to mitigate ramifications caused by impending failures. This involved implementing a {\color{blue}DRL}  by integrating soft-actor-critic, proximal-policy-optimization, and {\color{blue}LSTM}. The approach also utilized the age of information (AoI) to evaluate the trade-off between real-time and scheduling-based monitoring. 
In contrast, the authors of \cite{Mismar20213330} proposed a series of strategies for resilient recovery from failures in  6G edge networks. The authors detected anomalous performance and discovered the root causes of failures, configuration issues, or network procedure failures.

\section{Standardization Efforts}

% \begin{figure*}[htbp]
% \centering
% \includegraphics[width=0.8\textwidth]  {fail_hist.jpg}
%  \caption{\label{1} The timeline for the standardization of failures in communication. }
%  \label{F:timeline2}
%  \end{figure*}

\begin{table*}[h]
\centering
\caption{Comparison of Standardization Activities on Failure Analysis and Prevention}
\label{tab:comparison_standards}
\renewcommand{\arraystretch}{1}
\begin{tabular}{l p{8.5cm} p{6cm} }
\hline
\textbf{Org.} & \textbf{Key Contributions} & \textbf{Noteworthy Standards} \\
\hline
\textbf{ITU} &
\begin{itemize}
    \item Standards on intelligent network analytics and diagnostics.
    \item Mobile network energy efficiency assessment.
    \item Common equipment management function requirements.
\end{itemize} &
ITU-T standards on network failure analysis, mobile network energy efficiency, and equipment management. \\
\hline
\textbf{ATIS} &
\begin{itemize}
    \item Examination of resilience against equipment failures.
    \item Report on ``Trust, Security, and Resilience for 6G Systems.''
\end{itemize} &
Resilience against equipment failures on trust, security, and resilience for 6G. \\
\hline
\textbf{3GPP} &
\begin{itemize}
    \item Standardized Network Function (NF) Service and User Plane Path Failures.
    \item Release-17 with failure prediction and data analytics.
    \item Standards on digital cellular telecommunications, NG-RAN, UE conformance, and radio transmission.
\end{itemize} &
3GPP standards addressing various aspects of network failures, including Release-15, Release-16, and ongoing efforts in Release-17. \\
\hline
\textbf{IEEE} &
\begin{itemize}
    \item Standards on failure analysis for maintenance, reliability, predictions, and assessment.
    \item Organized Asian Test Symposium and the International Symposium on Physical and Failure Analysis of Integrated Circuits.
\end{itemize} &
IEEE standards covering failure analysis and significant conferences on related topics. \\
\hline
\textbf{NGMN} &
\begin{itemize}
    \item Testing framework for 5G pre-commercial networks.
    \item Standard on 5G trust, incl. service trustworthiness failure.
\end{itemize} &
NGMN contributions include a testing framework for 5G networks and a standard on 5G trust. \\
\hline
\textbf{WWRF} &
\begin{itemize}
    \item Published an outlook report on a simulation tool for mobile communication system reliability.
\end{itemize} &
WWRF's outlook report on a simulation tool for reliability in mobile communication systems. \\
\hline
\textbf{One6G} &
\begin{itemize}
    \item Publications discussing mechanisms against server failures, drawbacks of single-point failures, and the role of failure tolerance.
\end{itemize} &
One6G publications on mechanisms against server failures and the role of failure tolerance in 6G. \\
\hline
\end{tabular}
\end{table*}

This section encompasses the global initiatives to standardize and mitigate failures in 6G systems. Key international organizations like the International Telecommunications Union~(ITU), the Alliance for Telecommunications Industry Solutions (ATIS), and the 3GPP have been developing standards tailored to address various challenges in 6G, ranging from network diagnostics and energy efficiency to equipment failures and network service reliability. 
The IEEE and the Next Generation Mobile Networks (NGMN) Alliance have contributed significantly to communication layer reliability and radio link failure, respectively. 
% Fig.~\ref{F:timeline2} depicts the timeline of the standardization, and 
Table~\ref{tab:comparison_standards} compares the standardization activities on failure analysis and prevention.

Concurrently, the integration of AI into 6G systems introduces a new spectrum of challenges, particularly due to AI's vulnerability to typical software faults and data-related issues. Recognizing the critical role of AI in 6G infrastructure, leading global entities have been proactively developing AI-specific standards and regulatory frameworks. Notable advancements include the AI Risk Management framework by the U.S. National Institute of Standards and Technology (NIST), the European Union's white paper and subsequent regulations for AI in critical infrastructures, and similar initiatives by the UK, Canada, and Australia. Collectively, these efforts aim to safeguard against AI-induced failures in 6G systems.

\subsection{Earlier Standardization Effort }

\subsubsection{ITU}

The ITU has issued a set of communication standards, particularly addressing failure analysis and prevention to enhance standardized procedures.

In January 2020, ITU-T released standards on intelligent network analytics and diagnostics, delving into network failure analysis. By September 2020, ITU-T standardized mobile network energy efficiency assessment, discussing failures like handover and coverage issues and computational solution failure rates. October 2020 saw ITU-T standardize equipment management function requirements, focusing on hardware failures, e.g., transmission and link issues, and failure localization and detection strategies.

Transitioning from ITU's efforts, the Alliance for Telecommunications Industry Solutions (ATIS) also contributed significantly. On March 22, 2022, ATIS examined resilience against equipment failures due to unintentional disruptions in critical infrastructure like power grids, telecommunications, and finance systems. On August 18, 2022, ATIS released a report on ``Trust, Security, and Resilience for 6G Systems,'' targeting the mitigation of single-point failures in 6G systems.

\subsubsection{3GPP}

Furthering the development in this field, the 3GPP introduced a series of releases focusing on various aspects of network failures. The 3GPP's Release-15 standardized NF Service and User Plane Path Failures for enhanced detection and prevention \cite{3gpp2018}. A work item for Release-16, issued in 2018, agreed radio link failure issues to reduce connection interruption delays \cite{3gpp2020}. Release-17, published in 2021, introduced failure prediction as a key feature of Management Data Analytics (MDA). 3GPP TS 28.532 version 17.5.2 Release 17 also standardized failure reasons, statuses, and types for generic management services configuration \cite{3gpp2023}.

In 2023, 3GPP continued its efforts in standardization to address a spectrum of network failures. July 2023 saw 3GPP release standards on digital cellular telecommunications and mobile station conformance, detailing failures like initialization, tunnel, handover, synchronization, and TCP issues \cite{3gpp20231}. 
In January 2023, 3GPP standardized NG-RAN, focusing on positioning measurement and activation failures \cite{3gpp20232}. In the same month, standards on UE conformance were released, addressing link, handover, and connection establishment failures \cite{3gpp20233}. Another standardization in July 2023 covered radio transmission, reception, and resource management test cases, focusing on beam failure and recovery \cite{3gpp20234}.
Building upon these developments, 3GPP's Release-17 work plan included Policy and Charging Rules Function (PCRF) failure and restoration. The recovery of beam failure for supporting NR sidelink CA operation is reported to be included in 3GPP's upcoming Release-18~\cite{3gpp20236}.

\subsubsection{IEEE}
The IEEE has made significant contributions to this field. The IEEE has published various standards on failure analysis for maintenance, reliability, predictions, and assessment in communication support layers \cite{ieee1}. They have also organized significant conferences like the Asian Test Symposium (ATS) \cite{ieee2} and the International Symposium on Physical and Failure Analysis of Integrated Circuits (IPFA) \cite{ieee3} to discuss these issues.

\subsubsection{NGMN}
In parallel with the above advancements, the NGMN Alliance has also been actively involved in addressing network failures. In July 2019, NGMN released a testing framework for 5G pre-commercial networks, discussing radio link failure in NR service connectivity. On July 26, 2021, NGMN released a standard on 5G trust, focusing on service trustworthiness failure for performance evaluation \cite{ngmn2}.

\subsubsection{Other Standardization Activities}
In February 2022, the Wireless World Research Forum (WWRF) published an outlook report, mentioning their independent evaluation group (IEG)'s design of a simulation tool for mobile communication system reliability, addressing link-level and system-level failures \cite{wwrf1}.
In addition, One6G's publications in June and November 2022 and June 2023 discussed mechanisms against server failures in 6G Vertical Use Cases, the drawbacks of single-point failures in centralized architectures, and the critical role of failure tolerance in communication systems for 6G and robotics \cite{one6g1,one6g2,one6g3}.

\vspace{-3mm}
\subsection{Recent Effort for AI Failures}

\begin{table*}[h]
\centering
\caption{Comparison of Global Initiatives on AI Risk Management}
\label{tab:comparison_ai_risk}
\renewcommand{\arraystretch}{1.5}
\begin{tabular}{l p{7cm} p{6cm}}
\hline
\textbf{Country/Region} & \textbf{Initiatives on AI Risk Management} & \textbf{Key Regulations and Actions} \\
\hline
\textbf{USA} &
NIST released a framework in January 2023 for AI Risk Management \cite{nist}. &
Significant step by the U.S. Department of Commerce in addressing AI risks. \\
\hline
\textbf{EU} &
Released a white paper in February 2020 on AI to standardize AI requirements and address its risks \cite{eu1}. Proposed regulations in April 2021 focusing on privacy, security, and safety \cite{eu2}. Adopted the AI Act in June 2023 \cite{eu3}. &
Proactive initiatives to harmonize AI development and regulate its risks, especially in managing critical infrastructure. \\
\hline
\textbf{UK} &
Presented a pro-innovation approach to AI regulation to Parliament in March 2023 \cite{uk1}. &
% Balances AI's risks and opportunities, emphasizing 
safety, security, robustness, accountability, governance, fairness, contestability, and redress. \\
\hline
\textbf{Canada} &
Released several proposals and directives, including the Directive on Automated Decision-Making  \cite{cd1} and the Guide on the use of Generative Artificial Intelligence \cite{cd2}. &
Actively involved in ensuring the ethical and responsible use of AI, managing AI-related risks. \\
\hline
\textbf{Australia} &
Published a paper in June 2023 outlining strategies to mitigate safety risks associated with AI \cite{au1}. &
Recognizing the importance of AI's safe and responsible use, actively addressing safety risks. \\
\hline
\textbf{Other} &
Global efforts with a growing awareness and proactive stance in managing AI's unique risks. &
Acknowledges the importance of responsible AI development and application globally. \\
\hline
\end{tabular}
\end{table*}

Software systems often fail because of undetected bugs or defects during preliminary design, detailed design, and coding phases. The self-adaptation and self-learning abilities of AI, which are heavily influenced by external data, set AI apart from traditional software. Data quality, especially biased, imbalanced, or malicious data, can significantly impact AI's decision-making, resulting in deviations from its intended functionality. To address AI failure modes, it is crucial to understand that it is susceptible to inherent software risks and external data influences.

Concerning that the promising endogenous intelligent architecture~\cite{9810299,10123399} and derivative AI-empowered, advanced solutions, e.g., semantic communication (SemCom)~\cite{yang_tut}, are envisioned to be pervasively deployed in 6G.  AI failure must be addressed with specific standards, particularly in critical communication systems or infrastructure that rely on continuous high-frequency data. Various countries and regions have initiated regulatory frameworks to oversee AI development and application responsibly,
as summarized in Table~\ref{tab:comparison_ai_risk}.

\subsubsection{USA}
In the United States, the NIST released a framework in January 2023 for Artificial Intelligence Risk Management to prevent various types of AI failures \cite{nist}. This initiative represents a significant step by the U.S. Department of Commerce in addressing AI risks.

\subsubsection{EU}
The European Union (EU) has been proactive in this regard. In February 2020, the EU released a white paper on AI to standardize AI requirements and address its risks, promoting harmonized AI development \cite{eu1}. By April 2021, the EU proposed regulations to harmonize AI development, focusing on privacy, security, and safety in managing critical infrastructure \cite{eu2}. On June 14, 2023, the European Parliament adopted its position on the AI Act, moving towards a consensus among EU countries \cite{eu3}.

\subsubsection{UK}
In the United Kingdom, a pro-innovation approach to AI regulation was presented to Parliament in March 2023~\cite{uk1}. The UK's policy balances AI's risks and opportunities, emphasizing safety, security, robustness, accountability, governance, fairness, and contestability.

\subsubsection{Canada}
Canada has also been actively involved in ensuring the ethical and responsible use of AI. Since March 2019, the Canadian government has released several proposals and directives, like the Directive on Automated Decision-Making in March 2019 \cite{cd1} and the Guide on the use of Generative Artificial Intelligence in September 2023 \cite{cd2}, to manage AI-related risks.

\subsubsection{Australia}
Australia's government, recognizing the importance of AI's safe and responsible use, published a paper in June 2023 outlining strategies to mitigate safety risks associated with AI \cite{au1}.

% \subsubsection{Other Activities}
% These global efforts highlight a growing awareness and proactive stance in managing AI's unique risks, ensuring its development and application are safe and responsible.

\section{Lessons Learned and Open Challenges}

\subsection{Lessons Learned}
This section summarizes the major failure analysis and mitigation lessons learned with a focus on situational awareness, AI/ML for incident prediction, and decentralized identity management.

\subsubsection{Sensing for Situational Awareness}

Incidental failures in 6G applications manifest in diverse domains, including smart healthcare\cite{Taniguchi20223899}, connected vehicles\cite{Linsalata20221}, industrial plants\cite{Zheng2021sasasasas}, electronics\cite{Huang2023}, power sectors\cite{Ilahi2021}, and aircraft transportation\cite{Lvssssssss20217}. The focus is on addressing these failures through innovative sensing solutions and technologies tailored to each specific domain, highlighting the crucial role of proactive measures in averting disruptions and ensuring the robustness of critical systems.

\begin{table*}
\centering
\caption{Concise Overview of Challenges in 6G Failure Analysis.}
\label{table:challenges}
\begin{tabular}{p{2.5cm}p{4.5cm}p{5cm}p{3cm}}
\hline
\textbf{Challenge} & \textbf{Current Limitations} & \textbf{Potential Solutions} & \textbf{Impacted Areas} \\
\hline
Standardized Procedure & Existing solutions lack a comprehensive guide for 6G failure analysis. & Develop a step-by-step standardized procedure. & Implementation, Procedure Design \\
\hline
Failure Datasets & Limited data for specific failures hampers proposed solution evaluation. & Create standardized data sets for each 6G failure scenario. & Data Collection, Evaluation, AI Performance \\
\hline
Heterogeneous Data & Lack of lightweight models for diverse data sources. & Investigate and develop lightweight AI models for diverse data handling. & Data Handling, AI Model Design \\
\hline
Imbalanced Data & Imbalanced data affects AI performance. & Explore few-shot learning for addressing imbalanced data. & AI Performance, Data Balance \\
\hline
Complicated Multiple Failures & Coordinated failures in 6G require efficient identification. & Develop methods to identify and distinguish complex failures. & System Coordination, Failure Discrimination \\
\hline
% LLM for Efficiency & Building traditional knowledge graphs is costly. & Incorporate LLMs to enhance domain knowledge graph efficiency. & Knowledge Graph, AI Performance, Data Balance \\
% \hline
AI Trust and Interpretability & Trust and interpretability of AI in 6G applications are scrutinized. & Research techniques for enhancing trust, interpretability, and transparency. & AI Trust, Interpretability, Transparency, Security \\
\hline
\end{tabular}
\end{table*}

\subsubsection{AI/ML for Incident Prediction and Prevention}

The recurrence of familiar failures, including single-point failures \cite{Mogyorosi20222453}, channel failures\cite{Nishio202176}, and service failures\cite{Saravanan2022851}, persists in 6G applications due to the critical role of reliable communication. Countermeasures, such as data-reliant AI approaches\cite{yuan23tcom}, remain relevant by leveraging traffic data for surveillance.
Furthermore, critical failures in 6G-enabled industrial automation \cite{Ganjalizadeh20224208}, smart grid\cite{NaitBelaid20225874}, and smart healthcare\cite{Pradhan2023807} have been identified. Existing countermeasures aim to ensure low latency and high infrastructure availability for failure prevention and recovery\cite{Ganjalizadeh20224208,NaitBelaid20225874}. Real-time operational data collected from sensors near the infrastructure is anticipated to fuel AI models for timely risk evaluation of potential failures\cite{Pradhan2023807,Ganjalizadeh20224208}.
The deployment of AI is discussed as a crucial element in addressing failures, emphasizing tailoring AI to specific scenarios. Distributed AI deployments are highlighted as particularly promising, handling the growing influx of failure data efficiently \cite{Ganjalizadeh20224208,het4}. This approach alleviates computation pressure on central processors and mitigates communication overhead by integrating MEC to offload computation loads \cite{modal1,het4}.

\subsubsection{Decentralized Identity Management and Collaborative Assessment}

The decentralized architecture is emphasized as offering significant improvements in computation efficiency and reduced communication overhead~\cite{Garzon2022}. However, implementing a decentralized system brings challenges regarding security.
Decentralized trust management is implemented through blockchain-based techniques to address security concerns \cite{Garzon2022}. With Proof of Work (PoW), Proof of Stake (PoS), and Practical Byzantine Fault Tolerance (PBFT), blockchain is a robust solution to avert single-point failures \cite{Zhang2020SDSFSDFSD}.
% The application of blockchain in higher layers, e.g., in data security, is promising, addressing issues like data transmission and privacy-aware data protection of 6G systems.
When relying on centralized authorization servers, edge devices are vulnerable to malicious attacks because of lower security levels. New solutions involve decentralizing authorization processes through collaborative assessments by edge devices\cite{Fang20232091}. As part of 6G security, AI modules are introduced to identify malicious attacks or anomalies based on diverse surveillance data.

\subsection{Challenges and Open Issues}
\label{sec_future}

% While previous sections have delved into potential 6G failures and their respective countermeasures proposed by existing works, critical challenges and gaps within current research demand heightened attention. 
Building upon this foundation, we underscore the insufficiency present in 6G failure analysis; see Table~\ref{table:challenges}.

\subsubsection{Establishing a Standardized Procedure for 6G Failure Analysis and Prevention}
Current works often tailor solutions to specific failures within distinct application scenarios~\cite{BI2023107172,cas_fail_power_5,Mogyorosi20222453,Nishio202176}. 
% 6G implementations are hampered by the absence of a standardized, step-by-step procedure. 
There is a critical need for a comprehensive failure analysis and prevention procedure. Tailoring these procedures to different application scenarios is a crucial open issue.

\subsubsection{Establishing 6G Failure Datasets and Specifications}
Effective failure analysis hinges on comprehensive background data collection~\cite{BI2023107172}. The quality and performance of failure analysis are heavily dependent on the data AI algorithms are fed. 6G robustness requires standardized failure data sets and compatible data specifications. A standard data set, guiding collection, and compatible data specifications are needed for each 6G failure. RF data sampling frequencies, image resolution specifications, and AI algorithm metrics should be included in transmission failure data sets. 6G standards must cater to diverse scenarios.

\subsubsection{Developing Effective Approaches for Handling Heterogeneous Data}
The widespread interconnection of IoT devices in 6G results in vast amounts of multi-modal and heterogeneous data~\cite{Bi_iot0222}. In addition to this diversity, 6G's versatile roles complicate its architecture. A key challenge lies in designing an effective approach to handle the immense volumes of multi-modal and heterogeneous data for failure analysis~\cite{modal1,modal2,modal3,modal4,het1,het2,het3,het4,het5,het6}. Lightweight AI models for effective failure analysis in 6G have yet to be explored, magnifying the challenge.

\subsubsection{Coping with the Imbalanced Data Issue}
With the endorsement and application of AI algorithms in failure analysis, the imbalance in available data for training, validating, and testing becomes a critical factor influencing performance~\cite{hu21dml}. 
Negative-label data (failure data) is substantially insufficient compared to positive-label data (normal operation data)~\cite{Zhangsasasasas2020,Jia2020120974,Xudsfsdfds202177,Ansdffsdfs2021, Liqweqweqweq2019,Sunewqewqewqewq2021,Haovxdsfsdfs2020}. Despite recent AI approaches addressing minor uni-modal samples, such as few-shot learning\cite{fewshot1,fewshot2,fewshot3}, zero-shot learning~\cite{zeroshot1,zeroshot2,zeroshot3}, or meta-learning~\cite{meta1,meta2,meta3,meta4,meta5,meta6,meta7,meta8}, practical deployment in the complex and pluralistic 6G landscape remains a formidable challenge, emphasizing the pressing need for innovative solutions.

\subsubsection{Coping with More Complicated Multiple Failures}
The multi-functional nature of 6G necessitates coordination among different components and modules, leading to complex failures~\cite{subservice1,subservice2,subservice3}. In 6G, where failures in one service may be caused by deeper failures in transmission and network components, it is crucial to identify, discriminate, and position deep failures efficiently. Identifying primary failure modes from secondary failure modes is a challenging task that requires comprehensive solutions.

% \subsubsection{Leveraging Large Language Models for Enhanced Failure Analysis Efficiency}
% Failure analysis, reliant on domain expert experience\cite{expert_fail_1}, can benefit from innovative solutions, such as knowledge graphs~\cite{KG1,KG2,KG3,KG4,KG5,KG6}. However, building these graphs is often costly and manual. Leveraging large language models (LLMs) introduces a potential solution with their powerful generalization ability and vast knowledge~\cite{LLM1,Sch_tse_2023,yan_tps_2023,LLM5}. Incorporating LLM to enhance domain knowledge graph efficiency~\cite{DKG1,DKG2,DKG3} in failure analysis, along with utilizing LLM for data augmentation to address imbalanced challenges, represents a formidable challenge that requires careful consideration and innovative approaches.

\subsubsection{Addressing Failures in AI for 6G}
AI, deeply embedded in 6G development, has been employed for failure analysis~\cite{cas_fail_power__ml_review_1,BI2023107172}. Trustworthiness, interpretability, and transparency of AI remain controversial in 6G. Vulnerabilities, such as adversarial attacks on neural networks~\cite{10263803,10129254}, necessitate comprehensive analysis and identification of underlying AI failures. It requires extensive research efforts to overcome inherent limitations and enhance AI's reliability in failure analysis scenarios in order to ensure robustness of AI-intensive applications in 6G.

\section{Conclusions}
\label{sec_conclusion}

This survey aspires to propel discussions on failures, failure analysis, and countermeasures in the context of 6G and critical communication infrastructures. To ensure continuity, security, and availability of robust and resilient critical infrastructures, we systematically identified and classified existing research endeavors. 
In-depth exploration of these crucial research areas led to a thorough review of representative works. It shed light on typical failures in 6G systems and applications, and also contributed constructive countermeasures to address them. 

We also delivered a comprehensive and standardized procedure for meticulously addressing critical and typical failures in 6G. 
Due to the envisioned intelligence embedded in 6G networks, we highlighted key agendas for failure standardization, including those in AI. 
The survey also identified future challenges and research directions for 6G failures and critical infrastructure development.
This survey could serve as a practical guide for researchers and practitioners involved in failure analysis and prevention.

\bibliographystyle{IEEEtran}
\bibliography{bibF6G}

\begin{thebibliography}{100}
\providecommand{\url}[1]{#1}
\csname url@samestyle\endcsname
\providecommand{\newblock}{\relax}
\providecommand{\bibinfo}[2]{#2}
\providecommand{\BIBentrySTDinterwordspacing}{\spaceskip=0pt\relax}
\providecommand{\BIBentryALTinterwordstretchfactor}{4}
\providecommand{\BIBentryALTinterwordspacing}{\spaceskip=\fontdimen2\font plus
\BIBentryALTinterwordstretchfactor\fontdimen3\font minus \fontdimen4\font\relax}
\providecommand{\BIBforeignlanguage}[2]{{%
\expandafter\ifx\csname l@#1\endcsname\relax
\typeout{** WARNING: IEEEtran.bst: No hyphenation pattern has been}%
\typeout{** loaded for the language `#1'. Using the pattern for}%
\typeout{** the default language instead.}%
\else
\language=\csname l@#1\endcsname
\fi
#2}}
\providecommand{\BIBdecl}{\relax}
\BIBdecl

\bibitem{8412482}
K.~David and H.~Berndt, ``{6G} vision and requirements: {I}s there any need for beyond {5G}?'' \emph{IEEE Veh. Technol. Mag.}, vol.~13, no.~3, pp. 72--80, Sep. 2018.

\bibitem{wang2015}
X.~Wang, Y.~Zhang, G.~B. Giannakis \emph{et~al.}, ``Robust smart-grid-powered cooperative multipoint systems,'' \emph{IEEE Trans. Wireless Commun.}, vol.~14, no.~11, pp. 6188--6199, 2015.

\bibitem{hu20tgcn}
S.~Hu, X.~Chen, W.~Ni \emph{et~al.}, ``Modeling and analysis of energy harvesting and smart grid-powered wireless communication networks: {A} contemporary survey,'' \emph{IEEE Trans. Green Commun. Netw.}, vol.~4, no.~2, pp. 461--496, 2020.

\bibitem{8808168}
K.~B. Letaief, W.~Chen, Y.~Shi \emph{et~al.}, ``The roadmap to {6G:} {AI} empowered wireless networks,'' \emph{IEEE Commun. Mag.}, vol.~57, no.~8, pp. 84--90, August 2019.

\bibitem{hu22tvt}
S.~Hu, W.~Ni, X.~Wang \emph{et~al.}, ``Disguised tailing and video surveillance with solar-powered fixed-wing unmanned aerial vehicle,'' \emph{IEEE Trans. Veh. Tech.}, vol.~71, no.~5, pp. 5507--5518, May 2022.

\bibitem{8766143}
Z.~Zhang, Y.~Xiao, Z.~Ma \emph{et~al.}, ``{6G} wireless networks: {V}ision, requirements, architecture, and key technologies,'' \emph{IEEE Veh. Technol. Mag.}, vol.~14, no.~3, pp. 28--41, Sep. 2019.

\bibitem{8792135}
E.~Calvanese~Strinati, S.~Barbarossa, J.~L. Gonzalez-Jimenez \emph{et~al.}, ``{6G}: {T}he next frontier: {F}rom holographic messaging to artificial intelligence using subterahertz and visible light communication,'' \emph{IEEE Veh. Technol. Mag.}, vol.~14, no.~3, pp. 42--50, 2019.

\bibitem{8760275}
B.~Zong, C.~Fan, X.~Wang \emph{et~al.}, ``{6G} technologies: {K}ey drivers, core requirements, system architectures, and enabling technologies,'' \emph{IEEE Veh. Technol. Mag.}, vol.~14, no.~3, pp. 18--27, 2019.

\bibitem{hu23visual}
S.~Hu, X.~Yuan, W.~Ni \emph{et~al.}, ``Visual camouflage and online trajectory planning for unmanned aerial vehicle-based disguised video surveillance: {R}ecent advances and a case study,'' \emph{IEEE Veh. Technol. Mag.}, pp. 2--11, 2023.

\bibitem{notable_fail_2024_01_10}
\BIBentryALTinterwordspacing
``Katrina response: {A} failure to communicate,'' 2023, accessed on 2024-1-10. [Online]. Available: \url{https://veoci.com/blog/katrina-response-a-failure-to-communicate/}
\BIBentrySTDinterwordspacing

\bibitem{Bi_power_2019}
S.~Bi, Z.~Fang, X.~Yuan \emph{et~al.}, ``Joint base station activation and coordinated downlink beamforming for {HetNets}: {E}fficient optimal and suboptimal algorithms,'' \emph{IEEE Trans. Veh. Technol.}, vol.~68, no.~4, pp. 3702--3712, 2019.

\bibitem{5g_smart_grid}
A.~Kumari, S.~Tanwar, S.~Tyagi \emph{et~al.}, ``Fog computing for smart grid systems in the {5G} environment: {C}hallenges and solutions,'' \emph{IEEE Wireless Commun.}, vol.~26, no.~3, pp. 47--53, 2019.

\bibitem{5g_gas}
S.~Shi, J.~Wang, Q.~Yu \emph{et~al.}, ``Research of hybrid integrated energy station based on gas turbine,'' in \emph{Proc. ACPEE}, 2021.

\bibitem{hu16}
S.~Hu, Y.~Zhang, X.~Wang \emph{et~al.}, ``Weighted sum-rate maximization for {MIMO} downlink systems powered by renewables,'' \emph{IEEE Trans. Wireless Commun.}, vol.~15, no.~8, pp. 5615--5625, 2016.

\bibitem{hu18}
S.~Hu, C.~Xu, X.~Wang \emph{et~al.}, ``A stochastic {ADMM} approach to distributed coordinated multicell beamforming for renewables powered wireless cellular networks,'' \emph{IEEE Trans. Veh. Technol.}, vol.~67, no.~9, pp. 8595--8607, 2018.

\bibitem{5g_bank}
G.~O. Boateng, D.~Ayepah-Mensah, D.~M. Doe \emph{et~al.}, ``Blockchain-enabled resource trading and deep reinforcement learning-based autonomous {RAN} slicing in {5G},'' \emph{IEEE Trans. Netw. Serv. Manag.}, vol.~19, no.~1, pp. 216--227, 2022.

\bibitem{5g_gov}
X.~J. Xu, ``Intelligent government control system construction based on 5g {I}o{T},'' \emph{J. Sensors}, vol. 2021, pp. 1--12, 2021.

\bibitem{5g_transport}
X.~Wang, S.~Garg, H.~Lin \emph{et~al.}, ``Heterogeneous blockchain and {AI}-driven hierarchical trust evaluation for {5G}-enabled intelligent transportation systems,'' \emph{IEEE Trans. Intell. Transp. Syst.}, vol.~24, no.~2, pp. 2074--2083, 2023.

\bibitem{5g_emergency}
Z.~Yao, W.~Cheng, W.~Zhang \emph{et~al.}, ``Resource allocation for {5G-UAV}-based emergency wireless communications,'' \emph{IEEE J. Sel. Areas Commun.}, vol.~39, no.~11, pp. 3395--3410, 2021.

\bibitem{5g_water}
Z.~Kapelan, ``Digital water approach for smarter water management in cities with interconnected infrastructure,'' in \emph{Proc. EGU}, 2020.

\bibitem{8782879}
P.~Yang, Y.~Xiao, M.~Xiao \emph{et~al.}, ``{6G} wireless communications: {V}ision and potential techniques,'' \emph{IEEE Network}, vol.~33, no.~4, pp. 70--75, July 2019.

\bibitem{8869705}
W.~Saad, M.~Bennis, and M.~Chen, ``A vision of {6G} wireless systems: {A}pplications, trends, technologies, and open research problems,'' \emph{IEEE Network}, vol.~34, no.~3, pp. 134--142, May 2020.

\bibitem{notable1}
\BIBentryALTinterwordspacing
``Telecom and trust services incidents in 2020: {S}ystem failures on the rise,'' accessed on 2023-12-19. [Online]. Available: \url{https://www.enisa.europa.eu/news/enisa-news/telecom-trust-services-incidents-in-2020-system-failures-on-the-rise}
\BIBentrySTDinterwordspacing

\bibitem{flevent3}
\BIBentryALTinterwordspacing
``The power outages and telecommunication failure caused by the extreme snowstorm in the central {U}nited {S}tates,'' 2023, accessed on 2024-1-7. [Online]. Available: \url{https://www.americanoversight.org/investigation/the-february-2021-texas-power-outage}
\BIBentrySTDinterwordspacing

\bibitem{notable2}
\BIBentryALTinterwordspacing
``Massive internet outage: {W}ebsites and apps around the world go dark,'' accessed on 2023-12-19. [Online]. Available: \url{https://edition.cnn.com/2021/06/08/tech/internet-outage-fastly/index.html}
\BIBentrySTDinterwordspacing

\bibitem{notable3}
\BIBentryALTinterwordspacing
``The internet outage that took down several major websites seems to be fixed,'' accessed on 2023-12-19. [Online]. Available: \url{https://www.cnbc.com/2021/07/22/several-major-websites-go-down-in-widespread-internet-outage.html}
\BIBentrySTDinterwordspacing

\bibitem{notable5}
\BIBentryALTinterwordspacing
``The failure for data center in {L}ondon,'' accessed on 2023-12-19. [Online]. Available: \url{https://new.qq.com/rain/a/20221226A02X7200}
\BIBentrySTDinterwordspacing

\bibitem{flevent4}
\BIBentryALTinterwordspacing
``The communication outages caused by equipment failure in {J}apan,'' 2023, accessed on 2024-1-7. [Online]. Available: \url{https://www.theregister.com/2022/07/04/massive_telecom_outage_in_japan/}
\BIBentrySTDinterwordspacing

\bibitem{flevent5}
\BIBentryALTinterwordspacing
``The devastating flooding threatening the reliability of critical infrastructures,'' 2023, accessed on 2024-1-7. [Online]. Available: \url{https://edition.cnn.com/2021/10/11/weather/infrastructure-flood-risk-climate-first-street/index.html}
\BIBentrySTDinterwordspacing

\bibitem{flevent8}
\BIBentryALTinterwordspacing
``37 million {T}-mobile customers were hacked,'' 2023, accessed on 2024-1-7. [Online]. Available: \url{https://edition.cnn.com/2023/01/19/tech/tmobile-hack/index.html}
\BIBentrySTDinterwordspacing

\bibitem{notable4}
\BIBentryALTinterwordspacing
``The failure for server room in {H}ong {K}ong,'' accessed on 2023-12-19. [Online]. Available: \url{https://new.qq.com/rain/a/20221226A02X7200}
\BIBentrySTDinterwordspacing

\bibitem{flevent9}
\BIBentryALTinterwordspacing
``Why cell phone service is down in {M}aui,'' 2023, accessed on 2024-1-7. [Online]. Available: \url{https://edition.cnn.com/2023/08/09/tech/cell-service-outages-maui-fires/index.html}
\BIBentrySTDinterwordspacing

\bibitem{flevent7}
\BIBentryALTinterwordspacing
``Telecom companies in {H}aiti report severed fiber optic cables,'' 2023, accessed on 2024-1-7. [Online]. Available: \url{https://apnews.com/article/haiti-digicel-gangs-cables-cut-fcce6d69803d59d6b19ea02f98f5e296}
\BIBentrySTDinterwordspacing

\bibitem{notable6}
\BIBentryALTinterwordspacing
``The failure for nationwide bank data communication system in {J}apan,'' accessed on 2023-12-19. [Online]. Available: \url{https://new.qq.com/rain/a/20231010A07D8500.html}
\BIBentrySTDinterwordspacing

\bibitem{flevent6}
\BIBentryALTinterwordspacing
``Optus outage affects millions of {A}ustralians, phone and internet connections still down,'' 2023, accessed on 2024-1-7. [Online]. Available: \url{https://edition.cnn.com/2023/11/07/tech/australia-optus-network-outage-intl-hnk/index.html}
\BIBentrySTDinterwordspacing

\bibitem{9040264}
M.~Giordani, M.~Polese, M.~Mezzavilla \emph{et~al.}, ``Toward {6G} networks: {U}se cases and technologies,'' \emph{IEEE Commun. Mag.}, vol.~58, no.~3, pp. 55--61, March 2020.

\bibitem{9040431}
H.~Viswanathan and P.~E. Mogensen, ``Communications in the {6G} era,'' \emph{IEEE Access}, vol.~8, pp. 57\,063--57\,074, 2020.

\bibitem{9144301}
M.~Z. Chowdhury, M.~Shahjalal, S.~Ahmed \emph{et~al.}, ``{6G} wireless communication systems: {A}pplications, requirements, technologies, challenges, and research directions,'' \emph{IEEE Open J. Commun. Soc.}, vol.~1, pp. 957--975, 2020.

\bibitem{9145564}
I.~F. Akyildiz, A.~Kak, and S.~Nie, ``{6G} and beyond: {T}he future of wireless communications systems,'' \emph{IEEE Access}, vol.~8, pp. 133\,995--134\,030, 2020.

\bibitem{9023459}
G.~Gui, M.~Liu, F.~Tang \emph{et~al.}, ``{6G}: {O}pening new horizons for integration of comfort, security, and intelligence,'' \emph{IEEE Wireless Commun.}, vol.~27, no.~5, pp. 126--132, October 2020.

\bibitem{9349624}
W.~Jiang, B.~Han, M.~A. Habibi \emph{et~al.}, ``The road towards {6G}: {A} comprehensive survey,'' \emph{IEEE Open J. Commun. Soc.}, vol.~2, pp. 334--366, 2021.

\bibitem{9397776}
C.~D. Alwis, A.~Kalla, Q.-V. Pham \emph{et~al.}, ``Survey on {6G} frontiers: {T}rends, applications, requirements, technologies and future research,'' \emph{IEEE Open J. Commun. Soc.}, vol.~2, pp. 836--886, 2021.

\bibitem{8922617}
T.~Huang, W.~Yang, J.~Wu \emph{et~al.}, ``A survey on green {6G} network: {A}rchitecture and technologies,'' \emph{IEEE Access}, vol.~7, pp. 175\,758--175\,768, 2019.

\bibitem{scopus1}
\BIBentryALTinterwordspacing
``Scopus,'' accessed on 2023-12-19. [Online]. Available: \url{https://www.scopus.com/}
\BIBentrySTDinterwordspacing

\bibitem{fa_book_1}
B.~A. Miller, R.~J. Shipley, R.~J. Parrington \emph{et~al.}, \emph{{Failure Analysis and Prevention}}.\hskip 1em plus 0.5em minus 0.4em\relax ASM International, 01 2021.

\bibitem{ieee_home}
\BIBentryALTinterwordspacing
``{IEEE},'' accessed on 2023-12-19. [Online]. Available: \url{https://ieeexplore.ieee.org/Xplore/home.jsp}
\BIBentrySTDinterwordspacing

\bibitem{else1}
\BIBentryALTinterwordspacing
``Elsevier,'' accessed on 2023-12-19. [Online]. Available: \url{https://www.elsevier.com/}
\BIBentrySTDinterwordspacing

\bibitem{spring1}
\BIBentryALTinterwordspacing
``Springer,'' accessed on 2023-12-19. [Online]. Available: \url{https://link.springer.com/}
\BIBentrySTDinterwordspacing

\bibitem{failure_mode}
\BIBentryALTinterwordspacing
``What is failure mode? {D}efinition and examples,'' accessed on 2023-12-19. [Online]. Available: \url{https://marketbusinessnews.com/financial-glossary/failure-mode}
\BIBentrySTDinterwordspacing

\bibitem{aemo}
\BIBentryALTinterwordspacing
``Communication system failure guidelines,'' accessed on 2024-1-7. [Online]. Available: \url{https://aemo.com.au/-/media/files/electricity/nem/network_connections/stage-6/communication-system-failure-guidelines.pdf}
\BIBentrySTDinterwordspacing

\bibitem{zhang2020analysis}
C.~Zhang, X.~Xu, and H.~Dui, ``Analysis of network cascading failure based on the cluster aggregation in cyber-physical systems,'' \emph{Reliab. Eng. Syst. Saf.}, vol. 202, p. 106963, 2020.

\bibitem{kaiser2021network}
F.~Kaiser, V.~Latora, and D.~Witthaut, ``Network isolators inhibit failure spreading in complex networks,'' \emph{Nat. Commun.}, vol.~12, no.~1, p. 3143, 2021.

\bibitem{9342800}
A.~Prakash and S.~Kar, ``Network failure assessment based on graph signal processing,'' in \emph{Proc. ANTS}, 2020, pp. 1--6.

\bibitem{6G_wang_2023}
C.-X. Wang, X.~You, X.~Gao \emph{et~al.}, ``On the road to {6G}: {V}isions, requirements, key technologies, and testbeds,'' \emph{IEEE Commun. Surv. Tut.}, vol.~25, no.~2, pp. 905--974, 2023.

\bibitem{harsh_fail_1}
Z.~E. Khaled and H.~Mcheick, ``Case studies of communications systems during harsh environments: {A} review of approaches, weaknesses, and limitations to improve quality of service,'' \emph{Int. J. Distrib. Sens. Netw.}, vol.~15, no.~2, p. 1550147719829960, 2019.

\bibitem{9430901}
Q.~Cui, Z.~Zhu, W.~Ni \emph{et~al.}, ``Edge-intelligence-empowered, unified authentication and trust evaluation for heterogeneous beyond {5G} systems,'' \emph{IEEE Wirel. Commun.}, vol.~28, no.~2, pp. 78--85, 2021.

\bibitem{9838599}
K.~Li, Q.~Cui, Z.~Zhu \emph{et~al.}, ``Lightweight, privacy-preserving handover authentication for integrated terrestrial-satellite networks,'' in \emph{Proc. ICC}, 2022, pp. 25--31.

\bibitem{soft_fail_1}
S.~Gupta, A.~Mishra, and M.~Chawla, ``Analysis and recommendation of common fault and failure in software development systems,'' in \emph{Proc. SCOPES}, 2016, pp. 1730--1734.

\bibitem{zhou_dpa}
Z.~Shuai, W.~Zhangzhao, Q.~Baojun \emph{et~al.}, ``Destructive physical analysis methods of flip chip packaging devices for high reliability,'' in \emph{Proc. ICEPT}, 2021.

\bibitem{zhao_ndt_ai}
Y.~Zhao, M.~Xiao, H.~Lv \emph{et~al.}, ``Research on scanning acoustic image defects detection of integrated circuits based on {YOLOX},'' in \emph{Proc. ICEPT}, 2022.

\bibitem{Islam20231140}
M.~A. Islam, H.~Siddique, W.~Zhang \emph{et~al.}, ``A deep neural network-based communication failure prediction scheme in {5G RAN},'' \emph{IEEE Trans. Netw. Serv. Manag.}, vol.~20, no.~2, pp. 1140 -- 1152, 2023.

\bibitem{Li20232176}
K.~Li, P.~Zhu, Y.~Wang \emph{et~al.}, ``Joint uplink and downlink resource allocation toward energy-efficient transmission for urllc,'' \emph{IEEE J. Sel. Areas Commun.}, vol.~41, no.~7, pp. 2176 -- 2192, 2023.

\bibitem{Zheng20221}
D.~Zheng, G.~Shen, Y.~Li \emph{et~al.}, ``Service function chaining and embedding with heterogeneous faults tolerance in edge networks,'' \emph{IEEE Trans. Netw. Serv. Manag.}, pp. 1--1, 2022.

\bibitem{Peng2023}
C.~Peng, D.~Zheng, Y.~Zhong \emph{et~al.}, ``Off-site protection against service function forwarder failures in {NFV},'' \emph{Comput. Netw.}, vol. 221, 2023.

\bibitem{9239911}
X.~Lyu, C.~Ren, W.~Ni \emph{et~al.}, ``Online learning of optimal proactive schedule based on outdated knowledge for energy harvesting powered {I}nternet-of-{T}hings,'' \emph{IEEE Trans. Wirel. Commun.}, vol.~20, no.~2, pp. 1248--1262, 2021.

\bibitem{Zhang20235490}
N.~Zhang and X.~Zhu, ``A hybrid grant {NOMA} random access for massive {MTC} service,'' \emph{IEEE Internet Things J.}, vol.~10, no.~6, pp. 5490 -- 5505, 2023.

\bibitem{Hasan20228895}
M.~K. Hasan, S.~Islam, I.~Memon \emph{et~al.}, ``A novel resource oriented {DMA} framework for internet of medical things devices in {5G} network,'' \emph{IEEE Trans. Ind. Informat.}, vol.~18, no.~12, pp. 8895 -- 8904, 2022.

\bibitem{Selim2022}
M.~Y. Selim and A.~E. Kamal, ``Self-backhauling failure mitigation using {5G} new radio,'' \emph{Comput. Netw.}, vol. 217, 2022.

\bibitem{Esmat202314621}
H.~H. Esmat, B.~Lorenzo, and W.~Shi, ``Toward resilient network slicing for satellite-terrestrial edge computing {I}o{T},'' \emph{IEEE Internet Things J.}, vol.~10, no.~16, pp. 14\,621 -- 14\,645, 2023.

\bibitem{Haber20216838}
E.~E. Haber, H.~A. Alameddine, C.~Assi \emph{et~al.}, ``{UAV}-aided ultra-reliable low-latency computation offloading in future {I}o{T} networks,'' \emph{IEEE Trans. Commun.}, vol.~69, no.~10, pp. 6838 -- 6851, 2021.

\bibitem{DiCicco2022}
N.~Di~Cicco, F.~Tonini, V.~Cacchiani \emph{et~al.}, ``Optimization over time of reliable {5G-RAN} with network function migrations,'' \emph{Comput. Netw.}, vol. 215, 2022.

\bibitem{Dandachi2022}
G.~Dandachi, S.~Cerf, Y.~Hadjadj-Aoul \emph{et~al.}, ``A robust control-theory-based exploration strategy in deep reinforcement learning for virtual network embedding,'' \emph{Comput. Netw.}, vol. 218, 2022.

\bibitem{Wang2023ssssss}
Y.~Wang, L.~Zhang, C.~Wei \emph{et~al.}, ``Joint optimization of resource allocation and computation offloading based on game coalition in {C-V2X},'' \emph{Ad Hoc Netw.}, vol. 150, 2023.

\bibitem{Cao202222492}
H.~Cao, H.~Zhao, D.~X. Luo \emph{et~al.}, ``Dynamic virtual resource allocation mechanism for survivable services in emerging {NFV}-enabled vehicular networks,'' \emph{IEEE Trans. Intell. Transp. Syst.}, vol.~23, no.~11, pp. 22\,492 -- 22\,504, 2022.

\bibitem{NaitBelaid20225874}
M.~O. Nait~Belaid, V.~Audebert, B.~Deneuville \emph{et~al.}, ``{SD-RAN} based approach for smart grid critical traffic routing and scheduling in {5G} mobile networks,'' in \emph{Proc. IEEE GLOBECOM}, 2022.

\bibitem{Ilahi2021}
F.~Ilahi, S.~Dutta, M.~M. Hasan \emph{et~al.}, ``Development of a novel uwb antenna for {6G-I}o{T} based smart grid device monitoring system,'' in \emph{Proc. GECOST}, 2021.

\bibitem{li2022data}
K.~Li, W.~Ni, Y.~Emami \emph{et~al.}, ``Data-driven flight control of internet-of-drones for sensor data aggregation using multi-agent deep reinforcement learning,'' \emph{IEEE Wireless Commun.}, vol.~29, no.~4, pp. 18--23, 2022.

\bibitem{cas_fail_power_review}
M.~Z. Islam, Y.~Lin, V.~M. Vokkarane \emph{et~al.}, ``Cyber-physical cascading failure and resilience of power grid: {A} comprehensive review,'' \emph{Frontiers in Energy Research}, vol.~11, 2023.

\bibitem{cas_fail_network_1}
Y.~Qu, M.~Gao, Y.~Chen \emph{et~al.}, ``An analysis of the invulnerability for communication networks base on cascading failure model,'' in \emph{Proc. ICRIS}, 2020, pp. 154--157.

\bibitem{cas_fail_network_2}
S.~Gao, X.~Ma, F.~Ma \emph{et~al.}, ``Cascading failure analysis of uniform double-layer hyper-networks based on the couple map lattice model,'' in \emph{Proc. IEEE HPCC/DSS/SmartCity/DependSys}, 2022, pp. 1449--1456.

\bibitem{cas_fail_power_3}
X.~Li and T.~S. Cheng, ``Cascading failure analysis of cyber-physical power systems under communication congestion,'' in \emph{Proc. ICCSI}, 2021, pp. 1--6.

\bibitem{8994208}
X.~Lyu, C.~Ren, W.~Ni \emph{et~al.}, ``Cooperative computing anytime, anywhere: {U}biquitous fog services,'' \emph{IEEE Wirel. Commun.}, vol.~27, no.~1, pp. 162--169, 2020.

\bibitem{8793221}
------, ``Optimal online data partitioning for geo-distributed machine learning in edge of wireless networks,'' \emph{IEEE J. Sel. Areas Commun.}, vol.~37, no.~10, pp. 2393--2406, 2019.

\bibitem{9187796}
S.~He, X.~Lyu, W.~Ni \emph{et~al.}, ``Virtual service placement for edge computing under finite memory and bandwidth,'' \emph{IEEE Trans. Commun.}, vol.~68, no.~12, pp. 7702--7718, 2020.

\bibitem{li2021continuous}
K.~Li, W.~Ni, and F.~Dressler, ``Continuous maneuver control and data capture scheduling of autonomous drone in wireless sensor networks,'' \emph{IEEE Trans. Mobile Comput.}, vol.~21, no.~8, pp. 2732--2744, 2021.

\bibitem{9922666}
A.~V. Savkin, C.~Huang, and W.~Ni, ``Joint multi-{UAV} path planning and {LoS} communication for mobile-edge computing in {IoT} networks with {RISs},'' \emph{IEEE Internet Things J.}, vol.~10, no.~3, pp. 2720--2727, 2023.

\bibitem{9387137}
C.~Sun, W.~Ni, and X.~Wang, ``Joint computation offloading and trajectory planning for {UAV}-assisted edge computing,'' \emph{IEEE Trans. Wirel. Commun.}, vol.~20, no.~8, pp. 5343--5358, 2021.

\bibitem{9765746}
Z.~Wang, T.~Lv, J.~Zeng \emph{et~al.}, ``Placement and resource allocation of wireless-powered multiantenna {UAV} for energy-efficient multiuser {NOMA},'' \emph{IEEE Trans. Wirel. Commun.}, vol.~21, no.~10, pp. 8757--8771, 2022.

\bibitem{9733205}
T.~Mir, M.~Waqas, S.~Tu \emph{et~al.}, ``Relay hybrid precoding in {UAV}-assisted wideband millimeter-wave massive {MIMO} system,'' \emph{IEEE Trans. Wirel. Commun.}, vol.~21, no.~9, pp. 7040--7054, 2022.

\bibitem{9975284}
A.~V. Savkin, C.~Huang, and W.~Ni, ``On-demand deployment of aerial base stations for coverage enhancement in reconfigurable intelligent surface-assisted cellular networks on uneven terrains,'' \emph{IEEE Commun. Lett.}, vol.~27, no.~2, pp. 666--670, 2023.

\bibitem{9875063}
X.~Xiong, C.~Sun, W.~Ni \emph{et~al.}, ``Three-dimensional trajectory design for unmanned aerial vehicle-based secure and energy-efficient data collection,'' \emph{IEEE Trans. Veh. Technol.}, vol.~72, no.~1, pp. 664--678, 2023.

\bibitem{10100674}
C.~Sun, X.~Xiong, Z.~Zhai \emph{et~al.}, ``Max–min fair {3D} trajectory design and transmission scheduling for solar-powered fixed-wing {UAV}-assisted data collection,'' \emph{IEEE Trans. Wirel. Commun.}, vol.~22, no.~12, pp. 8650--8665, 2023.

\bibitem{10129074}
A.~V. Savkin, W.~Ni, and M.~Eskandari, ``Effective {UAV} navigation for cellular-assisted radio sensing, imaging, and tracking,'' \emph{IEEE Trans. Veh. Technol.}, vol.~72, no.~10, pp. 13\,729--13\,733, 2023.

\bibitem{9525335}
H.~Huang, A.~V. Savkin, and W.~Ni, ``Navigation of a {UAV} team for collaborative eavesdropping on multiple ground transmitters,'' \emph{IEEE Trans. Veh. Technol.}, vol.~70, no.~10, pp. 10\,450--10\,460, 2021.

\bibitem{8314676}
X.~Lyu, C.~Ren, W.~Ni \emph{et~al.}, ``Distributed optimization of collaborative regions in large-scale inhomogeneous fog computing,'' \emph{IEEE J. Sel. Areas Commun.}, vol.~36, no.~3, pp. 574--586, 2018.

\bibitem{8501940}
X.~Chen, W.~Ni, I.~B. Collings \emph{et~al.}, ``Automated function placement and online optimization of network functions virtualization,'' \emph{IEEE Trans. Commun.}, vol.~67, no.~2, pp. 1225--1237, 2019.

\bibitem{8063331}
X.~Lyu, W.~Ni, H.~Tian \emph{et~al.}, ``Optimal schedule of mobile edge computing for {I}nternet of {T}hings using partial information,'' \emph{IEEE J. Sel. Areas Commun.}, vol.~35, no.~11, pp. 2606--2615, 2017.

\bibitem{8678697}
C.~Ren, X.~Lyu, W.~Ni \emph{et~al.}, ``Profitable cooperative region for distributed online edge caching,'' \emph{IEEE Trans. Commun.}, vol.~67, no.~7, pp. 4696--4708, 2019.

\bibitem{8274943}
X.~Lyu, H.~Tian, W.~Ni \emph{et~al.}, ``Energy-efficient admission of delay-sensitive tasks for mobile edge computing,'' \emph{IEEE Trans. Commun.}, vol.~66, no.~6, pp. 2603--2616, 2018.

\bibitem{8570806}
X.~Chen, W.~Ni, T.~Chen \emph{et~al.}, ``Multi-timescale online optimization of network function virtualization for service chaining,'' \emph{IEEE Trans Mob. Comput.}, vol.~18, no.~12, pp. 2899--2912, 2019.

\bibitem{9585464}
X.~Chen, H.~Wen, W.~Ni \emph{et~al.}, ``Distributed online optimization of edge computing with mixed power supply of renewable energy and smart grid,'' \emph{IEEE Trans. Commun.}, vol.~70, no.~1, pp. 389--403, 2022.

\bibitem{cas_fail_power_5}
A.~Salehpour, I.~Al-Anbagi, K.-C. Yow \emph{et~al.}, ``A realistic failure propagation model for smart grid networks,'' in \emph{2022 10th International Conference on Smart Grid (icSmartGrid)}, 2022, pp. 66--71.

\bibitem{cas_fail_power_6}
Y.~Wang, G.~Bai, Y.-A. Zhang \emph{et~al.}, ``Cascading failure analysis and robustness assessment of the operational system and electric power system based on dependent network,'' in \emph{Proc. PHM}, 2022, pp. 1--8.

\bibitem{cas_fail_power_7}
Y.~Chen, Y.~Li, W.~Li \emph{et~al.}, ``Cascading failure analysis of cyber physical power system with multiple interdependency and control threshold,'' \emph{IEEE Access}, vol.~6, pp. 39\,353--39\,362, 2018.

\bibitem{Component_power_1}
Q.~Lin and H.~Wu, ``Interconnect reliability modeling and index failure analysis for power amplifier,'' in \emph{Proc. ITNEC}, 2017, pp. 1188--1191.

\bibitem{cps_compon_fail_1}
W.~Zhu and J.~V. Milanović, ``Cyber-physical system failure analysis based on complex network theory,'' in \emph{Proc. IEEE EUROCON ST}, 2017, pp. 571--575.

\bibitem{Vargas2023}
P.~Vargas and I.~Tien, ``Impacts of {5G} on cyber-physical risks for interdependent connected smart critical infrastructure systems,'' \emph{Int. J. Crit. Infrastruct. Prot.}, vol.~42, 2023.

\bibitem{Qu2023}
X.~Qu and H.~Wang, ``Emergency task offloading strategy based on cloud-edge-end collaboration for smart factories,'' \emph{Comput. Netw.}, vol. 234, 2023.

\bibitem{Prado20231845}
A.~Prado, F.~Stockeler, F.~Mehmeti \emph{et~al.}, ``Enabling proportionally-fair mobility management with reinforcement learning in {5G} networks,'' \emph{IEEE J. Sel. Areas Commun.}, vol.~41, no.~6, pp. 1845 -- 1858, 2023.

\bibitem{Tayyab202299}
M.~Tayyab, N.~Kolehmainen, M.~M. Butt \emph{et~al.}, ``Energy efficient rrm relaxation for reduced capability ues in {5G} networks,'' in \emph{Proc. IEEE GLOBECOM}, 2022.

\bibitem{Zhang202217936}
X.~Zhang, S.~Lin, W.~Feng \emph{et~al.}, ``Cell edge user capacity-coverage reliability tradeoff for {5G-R} systems with overlapped linear coverage,'' \emph{IEEE Trans. Intell. Transp. Syst.}, vol.~23, no.~10, pp. 17\,936 -- 17\,951, 2022.

\bibitem{Haghrah2023}
A.~Haghrah, A.~Haghrah, J.~M. Niya \emph{et~al.}, ``Handover triggering estimation based on fuzzy logic for {LTE-A/5G} networks with ultra-dense small cells,'' \emph{Soft Computing}, 2023.

\bibitem{Gundogan2023}
A.~Gündogan, A.~Badalıoğlu, P.~Spapis \emph{et~al.}, ``On the modelling and performance analysis of lower layer mobility in {5G}-advanced,'' in \emph{Proc. IEEE WCNC}, 2023.

\bibitem{deSouza20231}
A.~B. de~Souza, P.~A.~L. Rego, V.~Chamola \emph{et~al.}, ``A bee colony-based algorithm for task offloading in vehicular edge computing,'' \emph{IEEE Syst. J.}, pp. 1--12, 2023.

\bibitem{network_vehicle_1}
J.~Xu, F.~Luo, and H.~Shao, ``Failure rate analysis for time-sensitive networking,'' in \emph{Proc. ICMIA}, 2016/11.

\bibitem{cas_fail_power_8}
J.~Huang, Q.~Wang, Z.~Sang \emph{et~al.}, ``The dc power flow model based analysis on failure of power grid communication network,'' in \emph{Proc. IFEEA}, 2022, pp. 683--686.

\bibitem{cas_fail_power_9}
X.~Gao, M.~Peng, and C.~K. Tse, ``Cascading failure analysis of cyber–physical power systems considering routing strategy,'' \emph{IEEE Trans. Circuits Syst. II Express Briefs}, vol.~70, no.~1, pp. 136--140, 2023.

\bibitem{cas_fail_power_10}
A.~Salehpour, I.~Al-Anbagi, K.-C. Yow \emph{et~al.}, ``Modeling cascading failures in coupled smart grid networks,'' \emph{IEEE Access}, vol.~10, pp. 81\,054--81\,070, 2022.

\bibitem{Chandran2022561}
D.~Chandran and Y.~U. Lee, ``Color visible light communication technique based on machine learning compensation map for {6G} communication service,'' \emph{Int. J. Microw. Opt. Technol.}, vol.~17, no.~5, pp. 561 -- 573, 2022.

\bibitem{Adeogun2021959}
R.~Adeogun, G.~Berardinelli, and P.~Mogensen, ``Learning to dynamically allocate radio resources in mobile {6G} in-{X} subnetworks,'' in \emph{Proc. IEEE PIMRC}, vol. 2021-September, 2021.

\bibitem{Khan20222726}
A.~U. Khan, G.~Abbas, Z.~H. Abbas \emph{et~al.}, ``Reliability analysis of cognitive radio networks with reserved spectrum for {6G}-{I}o{T},'' \emph{IEEE Trans. Netw. Serv. Manag.}, vol.~19, no.~3, pp. 2726 -- 2737, 2022.

\bibitem{Abbas20227151}
G.~Abbas, A.~U. Khan, Z.~H. Abbas \emph{et~al.}, ``{FMCPR}: {F}lexible multiparameter-based channel prediction and ranking for {CR}-enabled massive {I}o{T},'' \emph{IEEE Internet Things J.}, vol.~9, no.~10, pp. 7151 -- 7165, 2022.

\bibitem{Lv20222831}
X.~Lv, H.~Rui, and J.~Xu, ``Double layers flexible radio access network: {U}ser cluster centric architecture towards {6G},'' in \emph{Proc. IEEE GLOBECOM}, 2022.

\bibitem{rca_fail_network_2}
M.~Choi, T.~Kim, J.~p. Lee \emph{et~al.}, ``An empirical study on root cause analysis and prediction of network failure using deep learning,'' in \emph{Proc. ICTC}, 2021, pp. 741--746.

\bibitem{Nishio202176}
T.~Nishio, Y.~Koda, J.~Park \emph{et~al.}, ``When wireless communications meet computer vision in beyond {5G},'' \emph{IEEE Communications Standards Magazine}, vol.~5, no.~2, pp. 76 -- 83, 2021.

\bibitem{Basu20216885}
D.~Basu, A.~Jain, U.~Ghosh \emph{et~al.}, ``A reverse path-flow mechanism for latency aware controller placement in v{SDN} enabled {5G} network,'' \emph{IEEE Trans. Ind. Informat.}, vol.~17, no.~10, pp. 6885 -- 6893, 2021.

\bibitem{Khan2022}
S.~Khan, S.~Khan, Y.~Ali \emph{et~al.}, ``Highly accurate and reliable wireless network slicing in 5th generation networks: {A} hybrid deep learning approach,'' \emph{J. Netw. Comput. Appl.}, vol.~30, no.~2, 2022.

\bibitem{Khan202231}
S.~Khan, A.~Hussain, S.~Nazir \emph{et~al.}, ``Efficient and reliable hybrid deep learning-enabled model for congestion control in {5G/6G} networks,'' \emph{Comput. Commun.}, vol. 182, pp. 31 -- 40, 2022.

\bibitem{Majumdar20222321}
S.~Majumdar, R.~Trivisonno, and G.~Carle, ``Scalability of distributed intelligence architecture for {6G} network automation,'' in \emph{Proc. IEEE ICC}, 2022, pp. 2321 -- 2326.

\bibitem{Shakeel2021969}
P.~M. Shakeel, S.~Baskar, H.~Fouad \emph{et~al.}, ``Creating collision-free communication in {I}o{T} with {6G} using multiple machine access learning collision avoidance protocol,'' \emph{Mobile Networks and Applications}, vol.~26, no.~3, pp. 969 -- 980, 2021.

\bibitem{Manogaran202214644}
G.~Manogaran, T.~Baabdullah, D.~B. Rawat \emph{et~al.}, ``Ai-assisted service virtualization and flow management framework for {6G}-enabled cloud-software-defined network-based {I}o{T},'' \emph{IEEE Internet Things J.}, vol.~9, no.~16, pp. 14\,644 -- 14\,654, 2022.

\bibitem{Ortin20221287}
J.~Ortin, P.~Serrano, J.~Garcia-Reinoso \emph{et~al.}, ``Analysis of scaling policies for nfv providing {5G/6G} reliability levels with fallible servers,'' \emph{IEEE Trans. Netw. Serv. Manag.}, vol.~19, no.~2, pp. 1287 -- 1305, 2022.

\bibitem{Sarkar2022367}
S.~Sarkar, S.~Vittal, and A.~Antony~Franklin, ``{LOCOMOTIVE 5G} core for {6G} ready resilient and highly available network slices and {SFC}s,'' in \emph{Proc. CNSM}, 2022.

\bibitem{Sun202012240}
W.~Sun, H.~Zhang, R.~Wang \emph{et~al.}, ``Reducing offloading latency for digital twin edge networks in {6G},'' \emph{IEEE Trans. Veh. Technol.}, vol.~69, no.~10, pp. 12\,240 -- 12\,251, 2020.

\bibitem{Mogyorosi20222453}
F.~Mogyorosi, P.~Babarczi, J.~Zerwas \emph{et~al.}, ``Resilient control plane design for virtualized {6G} core networks,'' \emph{IEEE Trans. Netw. Serv. Manag.}, vol.~19, no.~3, pp. 2453 -- 2467, 2022.

\bibitem{Saravanan2022851}
J.~Saravanan, A.~k. Tamilarasan, R.~Rajendran \emph{et~al.}, ``Performance analysis of digital twin edge network implementing bandwidth optimization algorithm,'' \emph{Int. J. comput. Digit. Syst.}, vol.~12, no.~1, pp. 851 -- 858, 2022.

\bibitem{Ganjalizadeh20224208}
M.~Ganjalizadeh, H.~S. Ghadikolaei, J.~Haraldson \emph{et~al.}, ``Interplay between distributed {AI} workflow and {URLLC},'' in \emph{Proc. IEEE GLOBECOM}, 2022.

\bibitem{phy_vir_mixed_failure_1}
R.~Birke, I.~Giurgiu, L.~Y. Chen \emph{et~al.}, ``Failure analysis of virtual and physical machines: {P}atterns, causes and characteristics,'' in \emph{Proc. IEEE/IFIP DSN}, 2014, pp. 1--12.

\bibitem{rca_fail_network_1}
Y.~Shan, L.~Bin, Z.~Zheng \emph{et~al.}, ``Root cause analysis of failures for power communication network based on {CNN},'' in \emph{Proc. ICCSN}, 2020, pp. 100--105.

\bibitem{rca_fail_network_3}
J.~M.~N. Gonzalez, J.~A. Jimenez, J.~C.~D. Lopez \emph{et~al.}, ``Root cause analysis of network failures using machine learning and summarization techniques,'' \emph{IEEE Commun. Mag.}, vol.~55, no.~9, pp. 126--131, 2017.

\bibitem{10234427}
Y.~Cui, T.~Lv, W.~Ni \emph{et~al.}, ``Digital twin-aided learning for managing reconfigurable intelligent surface-assisted, uplink, user-centric cell-free systems,'' \emph{IEEE Journal on Selected Areas in Communications}, vol.~41, no.~10, pp. 3175--3190, 2023.

\bibitem{Munilla2021}
J.~Munilla, M.~Burmester, and R.~Barco, ``An enhanced symmetric-key based {5G-AKA} protocol,'' \emph{Comput. Netw.}, vol. 198, 2021.

\bibitem{Yan2022}
X.~Yan and M.~Ma, ``A privacy-preserving handover authentication protocol for a group of {MTC} devices in {5G} networks,'' \emph{Comput. Secur.}, vol. 116, 2022.

\bibitem{BenSaad20231612}
S.~Ben~Saad, B.~Brik, and A.~Ksentini, ``Toward securing federated learning against poisoning attacks in zero touch {B5G} networks,'' \emph{IEEE Trans. Netw. Serv. Manag.}, vol.~20, no.~2, pp. 1612--1624, 2023.

\bibitem{Shi202361}
S.~Shi, Y.~Xiao, C.~Du \emph{et~al.}, ``{MS-PTP}: {P}rotecting network timing from byzantine attacks,'' in \emph{Proc. ACM WiSec}, 2023.

\bibitem{Yan20221678}
X.~Yan, M.~Ma, and R.~Su, ``A certificateless efficient and secure group handover authentication protocol in {5G} enabled vehicular networks,'' in \emph{Proc. IEEE ICC}, 2022.

\bibitem{Xu20226368}
D.~Xu, K.~Yu, and J.~A. Ritcey, ``Cross-layer device authentication with quantum encryption for {5G} enabled {II}o{T} in {I}ndustry 4.0,'' \emph{IEEE Trans. Ind. Informat.}, vol.~18, no.~9, pp. 6368 -- 6378, 2022.

\bibitem{Feng20226224}
C.~Feng, B.~Liu, Z.~Guo \emph{et~al.}, ``Blockchain-based cross-domain authentication for intelligent {5G}-enabled internet of drones,'' \emph{IEEE Internet Things J.}, vol.~9, no.~8, pp. 6224 -- 6238, 2022.

\bibitem{Yang202313959}
Y.~Yang, J.~Cao, R.~Ma \emph{et~al.}, ``{FHAP}: {F}ast handover authentication protocol for high-speed mobile terminals in {5G} satellite-terrestrial-integrated networks,'' \emph{IEEE Internet Things J.}, vol.~10, no.~15, pp. 13\,959--13\,973, 2023.

\bibitem{Ming2022}
Z.~Ming, X.~Li, C.~Sun \emph{et~al.}, ``Sleeping cell detection for resiliency enhancements in {5G/B5G} mobile edge-cloud computing networks,'' \emph{ACM Trans.Sens. Netw.}, vol.~18, no.~3, 2022.

\bibitem{Xu20212429}
D.~Xu and P.~Ren, ``Quantum learning based nonrandom superimposed coding for secure wireless access in {5G URLLC},'' \emph{IEEE Trans. Inf. Forensics Secur.}, vol.~16, pp. 2429 -- 2444, 2021.

\bibitem{Wijethilaka2022915}
S.~Wijethilaka and M.~Liyanage, ``A federated learning approach for improving security in network slicing,'' in \emph{Proc. IEEE GLOBECOM}, 2022.

\bibitem{Hewa20227174}
T.~Hewa, A.~Braeken, M.~Liyanage \emph{et~al.}, ``Fog computing and blockchain-based security service architecture for {5G} industrial {I}o{T}-enabled cloud manufacturing,'' \emph{IEEE Trans. Ind. Informat.}, vol.~18, no.~10, pp. 7174 -- 7185, 2022.

\bibitem{hu20}
S.~Hu, Q.~Wu, and X.~Wang, ``Energy management and trajectory optimization for {UAV}-enabled legitimate monitoring systems,'' \emph{IEEE Trans. Wireless Commun.}, vol.~20, no.~1, pp. 142--155, Jan. 2021.

\bibitem{7931680}
X.~Zha, W.~Ni, K.~Zheng \emph{et~al.}, ``Collaborative authentication in decentralized dense mobile networks with key predistribution,'' \emph{IEEE Trans. Inf. Forensics Secur.}, vol.~12, no.~10, pp. 2261--2275, 2017.

\bibitem{braeken2020symmetric}
A.~Braeken, ``Symmetric key based 5g aka authentication protocol satisfying anonymity and unlinkability,'' \emph{Comput. Netw.}, vol. 181, p. 107424, 2020.

\bibitem{Taskou20222328}
S.~K. Taskou, M.~Rasti, and P.~H.~J. Nardelli, ``Energy and cost efficient resource allocation for blockchain-enabled {NFV},'' \emph{IEEE Trans. Serv. Comput.}, vol.~15, no.~4, pp. 2328 -- 2341, 2022.

\bibitem{8989788}
G.~Yu, X.~Zha, X.~Wang \emph{et~al.}, ``Enabling attribute revocation for fine-grained access control in blockchain-{IoT} systems,'' \emph{IEEE Trans Eng. Manag.}, vol.~67, no.~4, pp. 1213--1230, 2020.

\bibitem{10.1016/j.ipm.2021.102492}
\BIBentryALTinterwordspacing
G.~Yu, L.~Zhang, X.~Wang \emph{et~al.}, ``A novel dual-blockchained structure for contract-theoretic {LoRa}-based information systems,'' \emph{Inf. Process. Manage.}, vol.~58, no.~3, may 2021. [Online]. Available: \url{https://doi.org/10.1016/j.ipm.2021.102492}
\BIBentrySTDinterwordspacing

\bibitem{DUAN2023102897}
L.~Duan, W.~Xu, W.~Ni, and W.~Wang, ``{BSAF: A} blockchain-based secure access framework with privacy protection for cloud-device service collaborations,'' \emph{J. Syst. Archit.}, vol. 140, p. 102897, 2023.

\bibitem{MAKHDOOM2019251}
\BIBentryALTinterwordspacing
I.~Makhdoom, M.~Abolhasan, H.~Abbas \emph{et~al.}, ``Blockchain's adoption in {IoT}: {T}he challenges, and a way forward,'' \emph{J. Netw. Comput. Appl.}, vol. 125, pp. 251--279, 2019. [Online]. Available: \url{https://www.sciencedirect.com/science/article/pii/S1084804518303473}
\BIBentrySTDinterwordspacing

\bibitem{zheng2021federated}
J.~Zheng, K.~Li, E.~Tovar \emph{et~al.}, ``Federated learning for energy-balanced client selection in mobile edge computing,'' in \emph{Proc. IWCMC}, 2021.

\bibitem{zheng2022exploring}
J.~Zheng, K.~Li, N.~Mhaisen \emph{et~al.}, ``Exploring deep-reinforcement-learning-assisted federated learning for online resource allocation in privacy-preserving edgeiot,'' \emph{IEEE Internet Things J.}, vol.~9, no.~21, pp. 21\,099--21\,110, 2022.

\bibitem{Yan20233104}
X.~Yan, M.~Ma, and R.~Su, ``Efficient group handover authentication for secure 5g-based communications in platoons,'' \emph{IEEE Trans. Intell. Transp. Syst.}, vol.~24, no.~3, pp. 3104--3116, 2023.

\bibitem{Li2023889}
X.~Li, X.~Hu, R.~Zhang \emph{et~al.}, ``A model-driven security analysis approach for {5G} communications in industrial systems,'' \emph{IEEE Trans. Wireless Commun.}, vol.~22, no.~2, pp. 889--902, 2023.

\bibitem{shahriar2014phy}
C.~Shahriar, M.~La~Pan, M.~Lichtman \emph{et~al.}, ``Phy-layer resiliency in ofdm communications: A tutorial,'' \emph{IEEE Commun. Surv. Tut.}, vol.~17, no.~1, pp. 292--314, 2014.

\bibitem{Thiruvasagam20212502}
P.~K. Thiruvasagam, A.~Chakraborty, and C.~S.~R. Murthy, ``Resilient and latency-aware orchestration of network slices using multi-connectivity in {MEC}-enabled {5G} networks,'' \emph{IEEE Trans. Netw. Serv. Manag.}, vol.~18, no.~3, pp. 2502 -- 2514, 2021.

\bibitem{Thiruvasagam20211491}
P.~K. Thiruvasagam, A.~Chakraborty, A.~Mathew \emph{et~al.}, ``Reliable placement of service function chains and virtual monitoring functions with minimal cost in softwarized {5G} networks,'' \emph{IEEE Trans. Netw. Serv. Manag.}, vol.~18, no.~2, pp. 1491 -- 1507, 2021.

\bibitem{Liu202024}
Y.~Liu, J.~Peng, J.~Kang \emph{et~al.}, ``A secure federated learning framework for {5G} networks,'' \emph{IEEE Wireless Commun.}, vol.~27, no.~4, pp. 24 -- 31, 2020.

\bibitem{cps_secu_fail_3}
V.~Belenko, V.~Chernenko, V.~Krundyshev \emph{et~al.}, ``Data-driven failure analysis for the cyber physical infrastructures,'' in \emph{Proc. ICPS}, 2019, pp. 1--5.

\bibitem{bi_tifs_2024}
S.~Bi, K.~Li, S.~Hu \emph{et~al.}, ``Detection and mitigation of position spoofing attacks on cooperative {UAV} swarm formations,'' \emph{IEEE Trans. Inf. Forensics Secur.}, vol.~19, pp. 1883--1895, 2024.

\bibitem{kurunathan2023machine}
H.~Kurunathan, H.~Huang, K.~Li \emph{et~al.}, ``Machine learning-aided operations and communications of unmanned aerial vehicles: {A} contemporary survey,'' \emph{IEEE Commun. Surv. Tut.}, 2023.

\bibitem{li2020joint}
K.~Li, W.~Ni, E.~Tovar \emph{et~al.}, ``Joint flight cruise control and data collection in {UAV}-aided internet of things: An onboard deep reinforcement learning approach,'' \emph{IEEE Internet Things J.}, vol.~8, no.~12, pp. 9787--9799, 2020.

\bibitem{Bastami20215018}
H.~Bastami, M.~Letafati, M.~Moradikia \emph{et~al.}, ``On the physical layer security of the cooperative rate-splitting-aided downlink in {UAV} networks,'' \emph{IEEE Trans. Inf. Forensics Secur.}, vol.~16, pp. 5018 -- 5033, 2021.

\bibitem{li2019energy}
K.~Li, R.~C. Voicu, S.~S. Kanhere \emph{et~al.}, ``Energy efficient legitimate wireless surveillance of {UAV} communications,'' \emph{IEEE Trans. Veh. Technol.}, vol.~68, no.~3, pp. 2283--2293, 2019.

\bibitem{yuan18capacity}
X.~Yuan, Z.~Feng, W.~Xu \emph{et~al.}, ``Capacity analysis of {UAV} communications: {C}ases of random trajectories,'' \emph{IEEE Trans. Veh. Technol.}, vol.~67, no.~8, pp. 7564--7576, 2018.

\bibitem{hu21}
S.~{Hu}, W.~{Ni}, X.~{Wang} \emph{et~al.}, ``Joint optimization of trajectory, propulsion, and thrust powers for covert {UAV-on-UAV} video tracking and surveillance,'' \emph{IEEE Trans. Inf. Forensics Secur.}, vol.~16, pp. 1959--1972, Jan. 2021.

\bibitem{Qiu2020Multiple}
C.~Qiu, Z.~Wei, X.~Yuan \emph{et~al.}, ``Multiple {UAV}-mounted base station placement and user association with joint fronthaul and backhaul optimization,'' \emph{IEEE Trans. Commun.}, vol.~68, no.~9, pp. 5864--5877, 2020.

\bibitem{Chen2020Performance}
X.~Chen, Z.~Feng, Z.~Wei \emph{et~al.}, ``Performance of joint sensing-communication cooperative sensing {UAV} network,'' \emph{IEEE Trans. Veh. Technol.}, vol.~69, no.~12, pp. 15\,545--15\,556, 2020.

\bibitem{li2021bloothair}
K.~Li, N.~Lu, J.~Zheng \emph{et~al.}, ``Bloothair: A secure aerial relay system using bluetooth connected autonomous drones,'' \emph{ACM Transactions on Cyber-Physical Systems}, vol.~5, no.~3, pp. 1--22, 2021.

\bibitem{Gupta20211712}
R.~Gupta, M.~M. Patel, S.~Tanwar \emph{et~al.}, ``Blockchain-based data dissemination scheme for {5G}-enabled softwarized {UAV} networks,'' \emph{IEEE Trans. Green Commun. Netw.}, vol.~5, no.~4, pp. 1712 -- 1721, 2021.

\bibitem{Xue20225284}
K.~Xue, X.~Luo, Y.~Ma \emph{et~al.}, ``A distributed authentication scheme based on smart contract for roaming service in mobile vehicular networks,'' \emph{IEEE Trans. Veh. Technol.}, vol.~71, no.~5, pp. 5284 -- 5297, 2022.

\bibitem{Abdulqadder2021866}
I.~H. Abdulqadder, D.~Zou, I.~T. Aziz \emph{et~al.}, ``Deployment of robust security scheme in {SDN} based {5G} network over {NFV} enabled cloud environment,'' \emph{IEEE Trans. Emerg. Topics Comput.}, vol.~9, no.~2, pp. 866 -- 877, 2021.

\bibitem{yuan23tcom}
X.~Yuan, S.~Hu, W.~Ni \emph{et~al.}, ``Joint user, channel, modulation-coding selection, and {RIS} configuration for jamming resistance in multiuser {OFDMA} systems,'' \emph{IEEE Trans. Commun.}, vol.~71, no.~3, pp. 1631--1645, Mar. 2023.

\bibitem{hu23ris}
S.~Hu, X.~Yuan, W.~Ni \emph{et~al.}, ``{RIS}-assisted jamming rejection and path planning for {UAV-Borne IoT} platform: {A} new deep reinforcement learning framework,'' \emph{IEEE Internet Things J.}, pp. 1--1, 2023.

\bibitem{yuan23tifs}
X.~Yuan, S.~Hu, W.~Ni \emph{et~al.}, ``Deep reinforcement learning-driven reconfigurable intelligent surface-assisted radio surveillance with a fixed-wing {UAV},'' \emph{IEEE Trans. Inf. Forensics Secur.}, vol.~18, pp. 4546--4560, 2023.

\bibitem{Garzon2022}
S.~R. Garzon, H.~Yildiz, and A.~Kupper, ``Towards decentralized identity management in multi-stakeholder {6G} networks,'' in \emph{Proc. 6GNet}, 2022.

\bibitem{Fang20232091}
H.~Fang, Z.~Xiao, X.~Wang \emph{et~al.}, ``Collaborative authentication for {6G} networks: {A}n edge intelligence based autonomous approach,'' \emph{IEEE Trans. Inf. Forensics Secur.}, vol.~18, pp. 2091 -- 2103, 2023.

\bibitem{Zhang2020SDSFSDFSD}
H.~Zhang, S.~Leng, and H.~Chai, ``A blockchain enhanced dynamic spectrum sharing model based on proof-of-strategy,'' in \emph{Proc. IEEE ICC}, 2020.

\bibitem{Chai20224620}
H.~Chai, S.~Leng, J.~He \emph{et~al.}, ``{C}yber{C}hain: {C}ybertwin empowered blockchain for lightweight and privacy-preserving authentication in internet of vehicles,'' \emph{IEEE Trans. Veh. Technol.}, vol.~71, no.~5, pp. 4620 -- 4631, 2022.

\bibitem{Ghourab2022}
E.~M. Ghourab, L.~Bariah, S.~Muhaidat \emph{et~al.}, ``Blockchain-enabled moving target defense for secure {CR} networks,'' in \emph{Proc. ITC}, 2022.

\bibitem{Lin20217204}
H.~Lin, S.~Garg, J.~Hu \emph{et~al.}, ``A blockchain-based secure data aggregation strategy using sixth generation enabled {N}etwork-in-{B}ox for industrial applications,'' \emph{IEEE Trans. Ind. Informat.}, vol.~17, no.~10, pp. 7204 -- 7212, 2021.

\bibitem{Tanwar2022}
S.~Tanwar, U.~Bodkhe, M.~D. Alshehri \emph{et~al.}, ``Blockchain-assisted industrial automation beyond {5G} networks,'' \emph{Compu. Ind. Eng.}, vol. 169, 2022.

\bibitem{Chandwani2020}
A.~Chandwani, S.~Dey, and A.~Mallik, ``Cybersecurity of onboard charging systems for electric vehicles-review, challenges and countermeasures,'' \emph{IEEE Access}, 2020.

\bibitem{wlan_network_fail_1}
R.~Manohar, R.~K. Sreenilayam, and V.~Pandey, ``Functional approach for reliability evaluation of wlan communication networks,'' in \emph{Proc. ICRCICN}, 2020, pp. 125--129.

\bibitem{8954616}
G.~Yu, X.~Wang, K.~Yu \emph{et~al.}, ``Survey: {S}harding in blockchains,'' \emph{IEEE Access}, vol.~8, pp. 14\,155--14\,181, 2020.

\bibitem{10201805}
G.~Yu, X.~Wang, W.~Ni \emph{et~al.}, ``Adaptive resource scheduling in permissionless sharded-blockchains: {A} decentralized multiagent deep reinforcement learning approach,'' \emph{IEEE Trans. Syst. Man Cybern. Syst.}, vol.~53, no.~11, pp. 7256--7268, 2023.

\bibitem{security_power_1}
Y.~Wang, C.~Feng, Y.~Li \emph{et~al.}, ``Expected failure method and its analysis for safety evaluation in a cyber-physical power system,'' \emph{IEEE Access}, vol.~10, pp. 133\,348--133\,356, 2022.

\bibitem{cas_fail_secu_power_4}
R.~Atat, M.~Ismail, S.~S. Refaat \emph{et~al.}, ``Cascading failure vulnerability analysis in interdependent power communication networks,'' \emph{IEEE Syst. J.}, vol.~16, no.~3, pp. 3500--3511, 2022.

\bibitem{Cai20231}
\BIBentryALTinterwordspacing
G.~Cai, B.~Fan, Y.~Dong \emph{et~al.}, ``Task-efficiency oriented {V2X} communications: {D}igital twin meets mobile edge computing,'' \emph{IEEE Wireless Commun.}, pp. 1--8, 2023. [Online]. Available: \url{6G-security-IOV}
\BIBentrySTDinterwordspacing

\bibitem{Mezair2022164}
T.~Mezair, Y.~Djenouri, A.~Belhadi \emph{et~al.}, ``A sustainable deep learning framework for fault detection in {6G I}ndustry 4.0 heterogeneous data environments,'' \emph{Comput. Commun.}, vol. 187, pp. 164 -- 171, 2022.

\bibitem{Jayaweera2022}
N.~Jayaweera, D.~Marasinghe, N.~Rajatheva \emph{et~al.}, ``{L}i{DAR} aided wireless networks - {L}o{S} detection and prediction based on static maps,'' in \emph{Proc. IEEE VTC}, 2022.

\bibitem{Rathinavel2022594}
G.~Rathinavel, N.~Muralidhar, N.~Ramakrishnan \emph{et~al.}, ``Efficient generative wireless anomaly detection for next generation networks,'' in \emph{Proc. MILCOM}, 2022.

\bibitem{Anand2022283}
R.~Anand and T.~Isobe, ``Differential fault attack on {R}occa,'' \emph{Lecture Notes in Computer Science (including subseries Lecture Notes in Artificial Intelligence and Lecture Notes in Bioinformatics)}, vol. 13218 LNCS, pp. 283 -- 295, 2022.

\bibitem{Fernandez-Fernandez202328}
A.~Fernandez-Fernandez, E.~Coronado, A.~Erspamer \emph{et~al.}, ``Unlocking the path toward intelligent telecom marketplaces for beyond {5G} and {6G} networks,'' \emph{IEEE Commun. Mag.}, vol.~61, no.~3, pp. 28 -- 34, 2023.

\bibitem{Faisal202380}
T.~Faisal, J.~A.~O. Lucena, D.~R. Lopez \emph{et~al.}, ``How to design autonomous service level agreements for {6G},'' \emph{IEEE Commun. Mag.}, vol.~61, no.~3, pp. 80 -- 85, 2023.

\bibitem{cps_fail_2}
H.~Xiao-Jie, C.~Yu-Lei, B.~Ting \emph{et~al.}, ``Analysis and research on vehicle-ground communication failure of {CBTC} system,'' in \emph{Proc. ICCSN}, 2019, pp. 437--441.

\bibitem{Prathiba202119}
S.~B. Prathiba, G.~Raja, S.~Anbalagan \emph{et~al.}, ``{SOSC}hain: Self optimizing streamchain for last-mile {6G UAV}-truck networks,'' in \emph{Proc. ACM MobiCom}, 2021.

\bibitem{Fernandes20221580}
A.~L.~P. Fernandes, D.~D. Souza, D.~B. Da~Costa \emph{et~al.}, ``Cell-free massive {MIMO} with segmented fronthaul: {R}eliability and protection aspects,'' \emph{IEEE Wireless Commun. Lett.}, vol.~11, no.~8, pp. 1580 -- 1584, 2022.

\bibitem{Khare2023}
S.~Khare, A.~K. Garg, and V.~Janyani, ``40 {G}bps hybrid fiber optic over {FSO} based latency aware ring architecture with intra-{ODN} transmission capability,'' \emph{Opt. Quantum Electron.}, vol.~55, no.~4, 2023.

\bibitem{Ferreira20235178}
T.~Ferreira, A.~Figueiredo, D.~Raposo \emph{et~al.}, ``Millimeter-wave feasibility in {5G} backhaul: {A} cross-layer analysis of blockage impact,'' \emph{IEEE Access}, vol.~11, pp. 5178 -- 5192, 2023.

\bibitem{Yaghoubi2022616}
F.~Yaghoubi, M.~Furdek, A.~Rostami \emph{et~al.}, ``Design and reliability performance of wireless backhaul networks under weather-induced correlated failures,'' \emph{IEEE Trans. Rel.}, vol.~71, no.~2, pp. 616 -- 629, 2022.

\bibitem{Yasin20225}
Q.~Yasin, Z.~Iqbal, M.~A. Khan \emph{et~al.}, ``Reliable multipath flow for link failure recovery in {5G} networks using {SDN} paradigm,'' \emph{Inf. Technol. Control.}, vol.~51, no.~1, pp. 5 -- 17, 2022.

\bibitem{Shi2022}
L.~Shi, C.~Wang, X.~Cai \emph{et~al.}, ``{G}a{A}s {RF} amplifier field failure analysis and reliability prediction in {5G} {AAU} system,'' in \emph{Proc. IPFA}, 2022.

\bibitem{Xu202277}
X.~Xu, X.~Deng, B.~Li \emph{et~al.}, ``An analysis of {5G D2D} network in power grids' automation,'' in \emph{Proc. ICCIS}, 2022.

\bibitem{Wang2023118}
Q.~Wang, M.~Chen, J.~Zhao \emph{et~al.}, ``Intelligent dynamic test system for vehicle engine based on {5G} internet of things,'' \emph{IEEE Consum. Electron. Mag.}, vol.~12, no.~2, pp. 118 -- 126, 2023.

\bibitem{Latry2022}
O.~Latry, N.~Moultif, E.~Joubert \emph{et~al.}, ``A time to failure evaluation of {A}l{G}a{N}/{G}a{N} {HEMT} transistors for {RF} applications,'' \emph{e-Prime - Advances in Electrical Engineering, Electronics and Energy}, vol.~2, 2022.

\bibitem{Wu2022304}
R.~Wu, X.~Liu, W.~Huang \emph{et~al.}, ``Calculation and analysis of communication network survivability in {SCADA} communication system,'' in \emph{Proc. IEEE IMCEC}, 2022.

\bibitem{Cherrared20212515}
S.~Cherrared, S.~Imadali, E.~Fabre \emph{et~al.}, ``{SFC} self-modeling and active diagnosis,'' \emph{IEEE Trans. Netw. Serv. Manag.}, vol.~18, no.~3, pp. 2515 -- 2530, 2021.

\bibitem{Bai20232811}
J.~Bai, X.~Chang, F.~Machida \emph{et~al.}, ``Impact of service function aging on the dependability for {MEC} service function chain,'' \emph{IEEE Trans. Dependable Secure Comput.}, vol.~20, no.~4, pp. 2811 -- 2824, 2023.

\bibitem{Chantre20222478}
H.~D. Chantre and N.~L. S.~D. Fonseca, ``Design of {5G} mec-based networks with {1:N:K} protection scheme,'' \emph{IEEE Trans. Netw. Serv. Manag.}, vol.~19, no.~3, pp. 2478 -- 2491, 2022.

\bibitem{Zhu202248}
H.~Zhu, J.~Li, J.~Hu \emph{et~al.}, ``Failure-aware and automated disaster backup in the {5G} core network,'' in \emph{Proc. CECCC}, 2022.

\bibitem{Ahmad2023754}
M.~Ahmad, S.~M.~N. Ali, M.~T. Tariq \emph{et~al.}, ``Neutrino: {A} fast and consistent edge-based cellular control plane,'' \emph{IEEE/ACM Trans. Netw.}, vol.~31, no.~2, pp. 754 -- 769, 2023.

\bibitem{10146517}
D.~Xiao, J.~A. Zhang, X.~Liu \emph{et~al.}, ``A two-stage {GCN}-based deep reinforcement learning framework for {SFC} embedding in multi-datacenter networks,'' \emph{IEEE Transactions on Network and Service Management}, vol.~20, no.~4, pp. 4297--4312, 2023.

\bibitem{Baldvinsson2022621}
J.~R. Baldvinsson, M.~Ganjalizadeh, A.~Alabbasi \emph{et~al.}, ``{IL-GAN}: {R}are sample generation via incremental learning in {GAN}s,'' in \emph{Proc. IEEE GLOBECOM}, 2022.

\bibitem{Mandal2022}
P.~Mandal, N.~Sarkar, S.~Santra \emph{et~al.}, ``Hybrid {WDM-FSO-PON} with integrated {SMF/FSO} link for transportation of rayleigh backscattering noise mitigated wired/wireless information in long-reach,'' \emph{Opt. Commun.}, vol. 507, 2022.

\bibitem{Wu2022}
Z.~Wu, W.~Wang, Z.~Wang \emph{et~al.}, ``Investigation on the thermo-mechanical reliability of bump with different solder shape in ultra-large size {FCBGA} package,'' in \emph{Proc. ICEPT}, 2022.

\bibitem{Marozau2022}
I.~Marozau, S.~Unterhofer, M.~Berry \emph{et~al.}, ``Reliability assessment of miniaturised electromechanical {RF} relays for space applications,'' \emph{Microelectron. Reliab.}, vol. 138, 2022.

\bibitem{Nielsen20225044}
M.~H. Nielsen, Y.~Zhang, C.~Xue \emph{et~al.}, ``Robust and efficient fault diagnosis of mm-{W}ave active phased arrays using baseband signal,'' \emph{IEEE Trans. Antennas Propagat.}, vol.~70, no.~7, pp. 5044 -- 5053, 2022.

\bibitem{Demir2022}
Y.~I. Demir, M.~S.~J. Solaija, and H.~Arslan, ``On the performance of handover mechanisms for non-terrestrial networks,'' in \emph{Proc. IEEE VTC}, 2022.

\bibitem{Maiwada2022486}
U.~D. Maiwada, K.~Usman~Danyaro, and A.~B. Sarlan, ``An improved mobility state detection mechanism for femtocells in {LTE} networks,'' in \emph{Proc. DASA}, 2022.

\bibitem{Taniguchi20223899}
Y.~Taniguchi, Y.~Ikegami, H.~Fujikawa \emph{et~al.}, ``Counseling (ro)bot as a use case for {5G/6G},'' \emph{Complex and Intell. Syst.}, vol.~8, no.~5, pp. 3899 -- 3917, 2022.

\bibitem{Linsalata20221}
F.~Linsalata, S.~Mura, M.~Mizmizi \emph{et~al.}, ``{L}o{S}-map construction for proactive relay of opportunity selection in {6G V2X} systems,'' \emph{IEEE Trans. Veh. Technol.}, pp. 1--15, 2022.

\bibitem{Morandi2021}
F.~Morandi, F.~Linsalata, M.~Brambilla \emph{et~al.}, ``A probabilistic codebook technique for fast initial access in {6G} vehicle-to-vehicle communications,'' in \emph{Proc. IEEE ICC Workshops}, 2021.

\bibitem{Li20221292}
J.~Li, K.~Zhu, and Y.~Zhang, ``Knowledge-assisted few-shot fault diagnosis in cellular networks,'' in \emph{Proc. IEEE GLOBECOM}, 2022.

\bibitem{Wang2021sasasas}
Y.~Wang, Y.~Ruan, and Y.~Tang, ``Intelligent fault diagnosis method for mobile cellular networks,'' in \emph{Proc. IEEE Globecom Workshops}, 2021.

\bibitem{Zheng2021sasasasas}
T.~Zheng, Z.~Meng, Q.~Gu \emph{et~al.}, ``A preliminary prototype based on biological mimicry for hardware data acquisition,'' in \emph{Proc. IEEE ETFA}, 2021.

\bibitem{Mohamed2020}
A.~Mohamed, H.~Ruan, M.~H.~H. Abdelwahab \emph{et~al.}, ``An inter-disciplinary modelling approach in industrial {5G/6G} and machine learning era,'' in \emph{Proc. IEEE ICC Workshops}, 2020.

\bibitem{Lvssssssss20217}
Z.~L{\"u}, M.~Yuan, Y.~L{\"u} \emph{et~al.}, ``An overlapping architecture of large-size wide-body aircraft based on cloud sea computing in 5g ogce,'' \emph{Smart Innovation, Systems and Technologies}, vol. 191, pp. 7 -- 13, 2021.

\bibitem{Zhong2022}
Y.~Zhong, G.~Yang, H.~Xu \emph{et~al.}, ``Learning-based health prediction method for airborne dme receiver with signal processing techniques in {6G} networks,'' \emph{J. Circuit. Syst. Comput.}, vol.~31, no.~12, 2022.

\bibitem{Malakoutian2022}
M.~Malakoutian, X.~Zheng, K.~Woo \emph{et~al.}, ``Low thermal budget growth of near-isotropic diamond grains for heat spreading in semiconductor devices,'' \emph{Adv. Funct. Mater.}, vol.~32, no.~47, 2022.

\bibitem{Huang2023}
G.~Huang, Z.~Wan, S.~Yang \emph{et~al.}, ``Mechanism investigation of micro-drill fracture in {PCB} large aspect ratio micro-hole drilling,'' \emph{J. Mater. Process. Technol.}, vol. 316, 2023.

\bibitem{Shaghaghi2022437}
A.~Shaghaghi, A.~Zakeri, N.~Mokari \emph{et~al.}, ``Proactive and {A}o{I}-aware failure recovery for stateful {NFV}-enabled zero-touch {6G} networks: Model-free {DRL} approach,'' \emph{IEEE Trans. Netw. Serv. Manag.}, vol.~19, no.~1, pp. 437 -- 451, 2022.

\bibitem{Mismar20213330}
F.~B. Mismar and J.~Hoydis, ``Unsupervised learning in next-generation networks: {R}eal-time performance self-diagnosis,'' \emph{IEEE Communications Letters}, vol.~25, no.~10, pp. 3330 -- 3334, 2021.

\bibitem{Reddy202243}
R.~Reddy, S.~Baradie, M.~Gundall \emph{et~al.}, ``{CPU} resource resilience in wireless mobile communications: {D}esign and evaluation on {COTS} virtual distributed platform,'' in \emph{Proc. IEEE BlackSeaCom}, 2022.

\bibitem{Babou2022295}
C.~S.~M. Babou, Y.~Owada, M.~Inoue \emph{et~al.}, ``{HEC}-{N}erve{N}et: {A} resilient edge cloud architecture for beyond {5G} networks,'' in \emph{Proc. IEEE ICC Workshops}, 2022.

\bibitem{Shayesteh2021217}
B.~Shayesteh, C.~Fu, A.~Ebrahimzadeh \emph{et~al.}, ``Auto-adaptive fault prediction system for edge cloud environments in the presence of concept drift,'' in \emph{Proc. IEEE IC2E}, 2021.

\bibitem{Sethi2020}
A.~Sethi and M.~Bisht, ``Intelligent and stable next generation hetnets with self-aggregated framework,'' in \emph{Proc. IOPCS}, no.~1, 2020.

\bibitem{3gpp2018}
\BIBentryALTinterwordspacing
``{3GPP TS} 23.527 version 15.0.0 {R}elease-15,'' 2018, accessed on 2024-1-7. [Online]. Available: \url{https://www.etsi.org/deliver/etsi_ts/123500_123599/123527/15.00.00_60/ts_123527v150000p.pdf}
\BIBentrySTDinterwordspacing

\bibitem{3gpp2020}
\BIBentryALTinterwordspacing
``How to tackle fast recovery from radio link failure,'' 2020, accessed on 2024-1-7. [Online]. Available: \url{https://www.ericsson.com/en/blog/2020/9/fast-recovery-from-radio-link-failure}
\BIBentrySTDinterwordspacing

\bibitem{3gpp2023}
\BIBentryALTinterwordspacing
``{3GPP TS} 28.532 version 17.5.2 {Release-17},'' 2023, accessed on 2024-1-7. [Online]. Available: \url{https://www.etsi.org/deliver/etsi_ts/128500_128599/128532/17.05.02_60/ts_128532v170502p.pdf}
\BIBentrySTDinterwordspacing

\bibitem{3gpp20231}
\BIBentryALTinterwordspacing
``{ETSI TS} 151 010-2 v13.15.0,'' 2023, accessed on 2024-1-7. [Online]. Available: \url{https://www.etsi.org/deliver/etsi_ts/151000_151099/15101002/13.15.00_60/ts_15101002v131500p.pdf}
\BIBentrySTDinterwordspacing

\bibitem{3gpp20232}
\BIBentryALTinterwordspacing
``{3GPP TS 38.473 version 17.3.0 Release 17},'' 2023, accessed on 2024-1-7. [Online]. Available: \url{https://www.etsi.org/deliver/etsi_ts/138400_138499/138473/17.03.00_60/ts_138473v170300p.pdf}
\BIBentrySTDinterwordspacing

\bibitem{3gpp20233}
\BIBentryALTinterwordspacing
``{3GPP TS 38.523-3 version 17.7.0 Release 17},'' 2023, accessed on 2024-1-7. [Online]. Available: \url{https://www.etsi.org/deliver/etsi_ts/138500_138599/13852303/17.07.00_60/ts_13852303v170700p.pdf}
\BIBentrySTDinterwordspacing

\bibitem{3gpp20234}
\BIBentryALTinterwordspacing
``{3GPP TS 38.522 version 17.9.0 Release 17},'' 2023, accessed on 2024-1-7. [Online]. Available: \url{https://www.etsi.org/deliver/etsi_ts/138500_138599/138522/17.09.00_60/ts_138522v170900p.pdf}
\BIBentrySTDinterwordspacing

\bibitem{3gpp20236}
\BIBentryALTinterwordspacing
``{3GPP Rel-18: 5G-Advanced RAN Features},'' 2023, accessed on 2024-1-7. [Online]. Available: \url{https://rimedolabs.com/blog/3gpp-rel-18/}
\BIBentrySTDinterwordspacing

\bibitem{ieee1}
\BIBentryALTinterwordspacing
``{IEEE} topic on failure-analysis,'' 2023, accessed on 2024-1-7. [Online]. Available: \url{https://technav.ieee.org/topic/failure-analysis}
\BIBentrySTDinterwordspacing

\bibitem{ieee2}
\BIBentryALTinterwordspacing
``{ATS}2021,'' 2023, accessed on 2024-1-7. [Online]. Available: \url{https://www.ats2021.info.hiroshima-cu.ac.jp/}
\BIBentrySTDinterwordspacing

\bibitem{ieee3}
\BIBentryALTinterwordspacing
``{IPFA}2021,'' 2023, accessed on 2024-1-7. [Online]. Available: \url{https://ipfa-ieee.org/2021/}
\BIBentrySTDinterwordspacing

\bibitem{ngmn2}
\BIBentryALTinterwordspacing
``{NGMN SUSTAINABLE TRUST},'' 2023, accessed on 2024-1-7. [Online]. Available: \url{https://www.ngmn.org/wp-content/uploads/210726-NGMN-Sustainable-Trust-V1.0.pdf}
\BIBentrySTDinterwordspacing

\bibitem{wwrf1}
\BIBentryALTinterwordspacing
``{WWRF Outlook30},'' 2023, accessed on 2024-1-7. [Online]. Available: \url{https://wwrf.ch/wp-content/publications/outlook/Outlook30.pdf}
\BIBentrySTDinterwordspacing

\bibitem{one6g1}
\BIBentryALTinterwordspacing
``{O}ne{6G} {6G Vertical Use Cases},'' 2023, accessed on 2024-1-7. [Online]. Available: \url{https://one6g.org/download/2027/}
\BIBentrySTDinterwordspacing

\bibitem{one6g2}
\BIBentryALTinterwordspacing
``{O}ne{6G} {6G} technology overview,'' 2023, accessed on 2024-1-7. [Online]. Available: \url{https://one6g.org/download/2699/}
\BIBentrySTDinterwordspacing

\bibitem{one6g3}
\BIBentryALTinterwordspacing
``{O}ne{6G} {6G} \& {Robotics},'' 2023, accessed on 2024-1-7. [Online]. Available: \url{https://one6g.org/download/2991/}
\BIBentrySTDinterwordspacing

\bibitem{nist}
\BIBentryALTinterwordspacing
``The framework on artificial intelligence risk management,'' 2023, accessed on 2024-1-7. [Online]. Available: \url{https://nvlpubs.nist.gov/nistpubs/ai/NIST.AI.100-1.pdf}
\BIBentrySTDinterwordspacing

\bibitem{eu1}
\BIBentryALTinterwordspacing
``{E}uropean {U}nion {W}hite {P}aper,'' 2023, accessed on 2024-1-7. [Online]. Available: \url{https://eur-lex.europa.eu/legal-content/EN/TXT/PDF/?uri=CELEX:52020DC0065}
\BIBentrySTDinterwordspacing

\bibitem{eu2}
\BIBentryALTinterwordspacing
``European union proposal on regulating and harmonizing the development of {AI},'' 2023, accessed on 2024-1-7. [Online]. Available: \url{https://artificialintelligenceact.eu/the-act/}
\BIBentrySTDinterwordspacing

\bibitem{eu3}
\BIBentryALTinterwordspacing
``European parliament adopted its negotiating position on the {AI} act,'' 2023, accessed on 2024-1-7. [Online]. Available: \url{https://www.europarl.europa.eu/news/en/headlines/society/20230601STO93804/eu-ai-act-first-regulation-on-artificial-intelligence/}
\BIBentrySTDinterwordspacing

\bibitem{uk1}
\BIBentryALTinterwordspacing
``The parliament of {UK AI} regulation,'' 2023, accessed on 2024-1-7. [Online]. Available: \url{https://www.gov.uk/government/publications/ai-regulation-a-pro-innovation-approach/white-paper}
\BIBentrySTDinterwordspacing

\bibitem{cd1}
\BIBentryALTinterwordspacing
``The government of {C}anada {AI} paper,'' 2023, accessed on 2024-1-7. [Online]. Available: \url{https://www.tbs-sct.canada.ca/pol/doc-eng.aspx?id=32592}
\BIBentrySTDinterwordspacing

\bibitem{cd2}
\BIBentryALTinterwordspacing
``The government of {C}anada {AI} guide,'' 2023, accessed on 2024-1-7. [Online]. Available: \url{https://www.canada.ca/en/government/system/digital-government/digital-government-innovations/responsible-use-ai/guide-use-generative-ai.html}
\BIBentrySTDinterwordspacing

\bibitem{au1}
\BIBentryALTinterwordspacing
``The government of {A}ustralia {AI} paper,'' 2023, accessed on 2024-1-7. [Online]. Available: \url{https://www.minister.industry.gov.au/ministers/husic/media-releases/safe-and-responsible-ai}
\BIBentrySTDinterwordspacing

\bibitem{9810299}
S.~Wang, C.~Yuen, W.~Ni \emph{et~al.}, ``Multiagent deep reinforcement learning for cost- and delay-sensitive virtual network function placement and routing,'' \emph{IEEE Trans. Commun.}, vol.~70, no.~8, pp. 5208--5224, 2022.

\bibitem{10123399}
W.~Li, T.~Lv, Y.~Cao \emph{et~al.}, ``Multi-carrier {NOMA}-empowered wireless federated learning with optimal power and bandwidth allocation,'' \emph{IEEE Trans. Wirel. Commun.}, vol.~22, no.~12, pp. 9762--9777, 2023.

\bibitem{yang_tut}
W.~Yang, H.~Du, Z.~Q. Liew \emph{et~al.}, ``Semantic communications for future internet: {F}undamentals, applications, and challenges,'' \emph{IEEE Commun. Surv. Tut.}, vol.~25, no.~1, pp. 213--250, 2023.

\bibitem{Pradhan2023807}
\BIBentryALTinterwordspacing
M.~R. Pradhan, B.~Mago, and K.~Ateeq, ``A classification-based sensor data processing method for the {I}nternet of {T}hings assimilated wearable sensor technology,'' \emph{Cluster Computing}, vol.~26, no.~1, pp. 807 -- 822, 2023. [Online]. Available: \url{6G-resource-IOT}
\BIBentrySTDinterwordspacing

\bibitem{het4}
I.~Alqerm and J.~Pan, ``{DeepEdge}: {A} new {QoE}-based resource allocation framework using deep reinforcement learning for future heterogeneous edge-{IoT} applications,'' \emph{IEEE Trans. Netw. Service Manag.}, vol.~18, no.~4, pp. 3942--3954, 2021.

\bibitem{modal1}
C.~Wang, X.~Yu, L.~Xu \emph{et~al.}, ``Energy-efficient task scheduling based on traffic mapping in heterogeneous mobile-edge computing: {A} green {IoT} perspective,'' \emph{IEEE Trans. Green Commun. Netw.}, vol.~7, no.~2, pp. 972--982, 2023.

\bibitem{BI2023107172}
S.~Bi, C.~Wang, B.~Wu \emph{et~al.}, ``A comprehensive survey on applications of {AI} technologies to failure analysis of industrial systems,'' \emph{Eng. Fail. Anal.}, vol. 148, p. 107172, 2023.

\bibitem{Bi_iot0222}
S.~Bi, J.~Cui, W.~Ni \emph{et~al.}, ``Three-dimensional cooperative positioning for {Internet of Things} provenance,'' \emph{IEEE Internet Things J.}, vol.~9, no.~20, pp. 19\,945--19\,958, 2022.

\bibitem{modal2}
G.~Qiu, G.~Tang, C.~Li \emph{et~al.}, ``A complete and comprehensive semantic perception of mobile travelling for mobile communication services,'' \emph{IEEE Internet Things J.}, pp. 1--1, 2023.

\bibitem{modal3}
P.~Wang, L.~T. Yang, J.~Li \emph{et~al.}, ``{MMDP}: {A} mobile-{IoT} based multi-modal reinforcement learning service framework,'' \emph{IEEE Trans. Serv. Comput.}, vol.~13, no.~4, pp. 675--684, 2020.

\bibitem{modal4}
Q.~Yu, L.~Hu, B.~Alzahrani \emph{et~al.}, ``Intelligent visual-{IoT}-enabled real-time {3D} visualization for autonomous crowd management,'' \emph{IEEE Wireless Commun.}, vol.~28, no.~4, pp. 34--41, 2021.

\bibitem{het1}
D.~Gao, L.~Wang, and B.~Hu, ``Spectrum efficient communication for heterogeneous {IoT} networks,'' \emph{IEEE Trans. Network Sci. Eng.}, vol.~9, no.~6, pp. 3945--3955, 2022.

\bibitem{het2}
K.~Fizza, P.~P. Jayaraman, A.~Banerjee \emph{et~al.}, ``{IoT-QWatch}: {A} novel framework to support the development of quality-aware autonomic {IoT} applications,'' \emph{IEEE Internet Things J.}, vol.~10, no.~20, pp. 17\,666--17\,679, 2023.

\bibitem{het3}
Z.~Zhou, S.~Yu, W.~Chen \emph{et~al.}, ``{CE-IoT}: {C}ost-effective cloud-edge resource provisioning for heterogeneous {IoT} applications,'' \emph{IEEE Internet Things J.}, vol.~7, no.~9, pp. 8600--8614, 2020.

\bibitem{het5}
T.~Kim, H.~Park, Y.~Jin \emph{et~al.}, ``Partition placement and resource allocation for multiple {DNN}-based applications in heterogeneous {IoT} environments,'' \emph{IEEE Internet Things J.}, vol.~10, no.~11, pp. 9836--9848, 2023.

\bibitem{het6}
I.~Behnke and H.~Austad, ``Real-time performance of industrial {IoT} communication technologies: {A} review,'' \emph{IEEE Internet Things J.}, pp. 1--1, 2023.

\bibitem{hu21dml}
S.~Hu, X.~Chen, W.~Ni \emph{et~al.}, ``Distributed machine learning for wireless communication networks: {T}echniques, architectures, and applications,'' \emph{IEEE Commun. Surveys Tuts.}, vol.~23, no.~3, pp. 1458--1493, 3rd Quart., 2021.

\bibitem{Zhangsasasasas2020}
W.~Zhang, X.~Li, X.-D. Jia \emph{et~al.}, ``Machinery fault diagnosis with imbalanced data using deep generative adversarial networks,'' \emph{Measurement: Journal of the International Measurement Confederation}, vol. 152, 2020.

\bibitem{Jia2020120974}
F.~Jia, S.~Li, H.~Zuo \emph{et~al.}, ``Deep neural network ensemble for the intelligent fault diagnosis of machines under imbalanced data,'' \emph{IEEE Access}, vol.~8, pp. 120\,974--120\,982, 2020.

\bibitem{Xudsfsdfds202177}
J.~Xu, Y.~Li, F.~Meng \emph{et~al.}, ``Fault diagnosis on imbalanced data using an adaptive cost-sensitive multiscale attention network,'' in \emph{Proc. ICITES}, 2021.

\bibitem{Ansdffsdfs2021}
Z.~An, X.~Jiang, J.~Cao \emph{et~al.}, ``Self-learning transferable neural network for intelligent fault diagnosis of rotating machinery with unlabeled and imbalanced data,'' \emph{Knowl. Based Syst.}, vol. 230, 2021.

\bibitem{Liqweqweqweq2019}
Q.~Li, L.~Chen, C.~Shen \emph{et~al.}, ``Enhanced generative adversarial networks for fault diagnosis of rotating machinery with imbalanced data,'' \emph{Meas. Sci. Technol.}, vol.~30, no.~11, 2019.

\bibitem{Sunewqewqewqewq2021}
Y.~Sun, T.~Zhao, Z.~Zou \emph{et~al.}, ``Imbalanced data fault diagnosis of hydrogen sensors using deep convolutional generative adversarial network with convolutional neural network,'' \emph{Rev. Sci. Instrum.}, vol.~92, no.~9, 2021.

\bibitem{Haovxdsfsdfs2020}
W.~Hao and F.~Liu, ``Imbalanced data fault diagnosis based on an evolutionary online sequential extreme learning machine,'' \emph{Symmetry}, vol.~12, no.~8, 2020.

\bibitem{fewshot1}
Z.~Xu, G.~Tang, and B.~Pang, ``An infrared thermal image few-shot learning method based on {CAPNet} and its application to induction motor fault diagnosis,'' \emph{IEEE Sensors J.}, vol.~22, no.~16, pp. 16\,440--16\,450, 2022.

\bibitem{fewshot2}
Y.~Hu, R.~Liu, X.~Li \emph{et~al.}, ``Task-sequencing meta learning for intelligent few-shot fault diagnosis with limited data,'' \emph{IEEE Trans. Ind. Informat.}, vol.~18, no.~6, pp. 3894--3904, 2022.

\bibitem{fewshot3}
C.~Yang, J.~Liu, Q.~Xu \emph{et~al.}, ``A generalized graph contrastive learning framework for few-shot machine fault diagnosis,'' \emph{IEEE Trans. Ind. Informat.}, pp. 1--10, early access, Jul. 27, 2023.

\bibitem{zeroshot1}
B.~Zhao, T.~Li, W.~Dai \emph{et~al.}, ``Fault diagnosis of rotating machinery based on {FMEA} and zero-shot learning,'' in \emph{Proc. PHM}, 2022.

\bibitem{zeroshot2}
J.~Xu and K.~Li, ``Generative zero-shot learning compound fault diagnosis of bearings,'' in \emph{Proc. ICSMD}, 2021.

\bibitem{zeroshot3}
P.~A. Traganitis and E.~G. Strangas, ``Perspectives of transfer learning on the diagnosis of faults in electrical machines, power electronics, and drives,'' in \emph{Proc. IEEE SDEMPED}, 2023.

\bibitem{meta1}
Z.~Zhao, R.~Zhao, Y.~Xu \emph{et~al.}, ``Task-generalization-based graph convolutional network for fault diagnosis of rod-fastened rotor system,'' \emph{IEEE Trans. Ind. Informat.}, pp. 1--11, 2023.

\bibitem{meta2}
P.~Lyu, X.~Li, W.~Yu \emph{et~al.}, ``A novel fault diagnosis method based on feature fusion and model agnostic meta-learning,'' in \emph{Proc. IEEE CASE}, 2023.

\bibitem{meta3}
Z.~Lao, D.~He, Z.~Jin \emph{et~al.}, ``Few-shot fault diagnosis of turnout switch machine based on semi-supervised weighted prototypical network,'' \emph{Knowl. Based Syst.}, vol. 274, 2023.

\bibitem{meta4}
J.~Hu, W.~Li, A.~Wu \emph{et~al.}, ``Novel joint transfer fine-grained metric network for cross-domain few-shot fault diagnosis,'' \emph{Knowl. Based Syst.}, vol. 279, 2023.

\bibitem{meta5}
Y.~Zhang, D.~Han, J.~Tian \emph{et~al.}, ``Domain adaptation meta-learning network with discard-supplement module for few-shot cross-domain rotating machinery fault diagnosis,'' \emph{Knowl. Based Syst.}, vol. 268, 2023.

\bibitem{meta6}
Z.~Lei, P.~Zhang, Y.~Chen \emph{et~al.}, ``Prior knowledge-embedded meta-transfer learning for few-shot fault diagnosis under variable operating conditions,'' \emph{Mech. Syst. Signal Process.}, vol. 200, 2023.

\bibitem{meta7}
Z.~Zhao, R.~Zhao, X.~Wu \emph{et~al.}, ``A meta-learning network with anti-interference for few-shot fault diagnosis,'' \emph{Neurocomputing}, vol. 552, 2023.

\bibitem{meta8}
J.~Lin, H.~Shao, X.~Zhou \emph{et~al.}, ``Generalized {MAML} for few-shot cross-domain fault diagnosis of bearing driven by heterogeneous signals,'' \emph{Expert Syst. Appl.}, vol. 230, 2023.

\bibitem{subservice1}
J.~Bai, Y.~Li, X.~Chang \emph{et~al.}, ``Understanding {NFV}-enabled vehicle platooning application: {A} dependability view,'' \emph{IEEE Trans. Cloud Comput.}, pp. 1--14, 2023.

\bibitem{subservice2}
P.~Abdisarabshali, M.~Liwang, A.~Rajabzadeh \emph{et~al.}, ``Decomposition theory meets reliability analysis: {P}rocessing of computation-intensive dependent tasks over vehicular clouds with dynamic resources,'' \emph{IEEE/ACM Trans. Netw.}, pp. 1--16, 2023.

\bibitem{subservice3}
L.~Cai, X.~Wei, C.~Xing \emph{et~al.}, ``Failure-resilient {DAG} task scheduling in edge computing,'' \emph{Comput. Netw.}, vol. 198, 2021.

\bibitem{cas_fail_power__ml_review_1}
N.~M. Sami and M.~A. Naeini, ``Machine learning applications in cascading failure analysis in power systems: {A} review,'' \emph{ArXiv}, vol. abs/2305.19390, 2023.

\bibitem{10263803}
Y.~Wang, T.~Sun, S.~Li \emph{et~al.}, ``Adversarial attacks and defenses in machine learning-empowered communication systems and networks: {A} contemporary survey,'' \emph{IEEE Commun. Surveys Tuts.}, vol.~25, no.~4, pp. 2245--2298, 2023.

\bibitem{10129254}
Y.~Wang, T.~Li, S.~Li \emph{et~al.}, ``New adversarial image detection based on sentiment analysis,'' \emph{IEEE Trans Neural Netw. Learn. Syst.}, pp. 1--15, 2023, early access, May 19, 2023.

\end{thebibliography}

\end{document}